\documentclass[aps,prl,reprint,nofootinbib]{revtex4-1}
\usepackage{natbib}
\usepackage{graphicx}
\usepackage{blindtext}
\usepackage{braket}
\usepackage{mathtools}
\usepackage{amsmath}
\usepackage{amssymb}
\usepackage{cancel}
\usepackage{empheq}
\usepackage[strict]{changepage}
\usepackage{xcolor,soul}
\sethlcolor{yellow}
\usepackage[scr]{rsfso}
\usepackage[colorlinks=true]{hyperref}
\usepackage{etoolbox}
\newcommand*\widefbox[1]{\fbox{\hspace{2em}#1\hspace{2em}}}

\begin{document}
\title{Energy Spectrum of Lost Alpha Particles in Magnetic Mirror Confinement}
\author{Alejandro Mesa Dame$^\dagger$$^{1,2}$}
\footnotetext{$^{\dagger}$\,Corresponding author: amesadam@pppl.gov}
\author{Ian E. Ochs$^{1,2}$}
\author{Nathaniel J. Fisch$^{1,2}$}
\affiliation{$^{1}$Department of Astrophysical Sciences, Princeton University, Princeton NJ 08544\\
$^{2}$Princeton Plasma Physics Laboratory, Princeton University, Princeton, NJ 08540}
\begin{abstract}

In a magnetic mirror fusion reactor, capturing the energy of fusion-produced alpha particles is essential to sustaining the reaction. However, since alpha particles are born at energies much higher than the confining potential, a substantial fraction are lost due to pitch-angle scattering before they can transfer their energy to the plasma via drag. The energy of lost alpha particles can still be captured through direct conversion, but designing an effective mechanism requires a description of the energies and times at which they become deconfined. Here we present analytical solutions for the loss velocity, energy, and time distributions of alpha particles in a magnetic mirror. After obtaining the Fokker-Planck collision operator, we asymptotically solve for the eigenfunctions of the Legendre operator to reveal a closed-form solution. Our framework applies to any high-energy species, for any applied potential and mirror ratio $R > 1$, making this work broadly applicable to mirror devices.\\
   
\end{abstract}

\maketitle
\vspace{-1pt}
\section{I. Introduction}
\label{sec:intro}
\vspace{-4pt}
Magnetic mirrors have experienced a recent resurgence due to the introduction of sheared-flow stabilization~\cite{Forest}, new high-field superconductors~\cite{Whyte,Carlson}, and more efficient methods of sustaining electron temperatures in tandem mirror end-plugs~\cite{Simonen}. These advancements have paved the way for a new generation of axisymmetric mirror experiments including multiple-mirror (MM) traps~\cite{Burdakov-MM,Burdakov-GOL3,Skovorodin, Miller}, the Centrifugal Mirror Fusion eXperiment (CMFX)~\cite{Carson,Schwartz}, and the Wisconsin High-field Axisymmetric Mirror (WHAM)~\cite{Endrizzi,Egedal}.\\
\indent Magnetic mirrors represent a promising alternative to more conventional magnetic confinement schemes like tokamaks due to their relative simplicity, steady-state operation, lack of driven current, differential confinement features, and high $\beta$ values~\cite{Post,Li}. Unlike tokamaks, however, mirrors have an open field line configuration and thus rely on conservation of the first adiabatic invariant $\mathcal{M} = W_{\perp}/B$~\cite{Heidbrink}, along with a combination of applied and ambipolar potential differences $\Phi$ between the midplane and ends to keep particles confined~\cite{Munirov}. Large end losses, together with loss-cone instabilities from the resulting population inversion~\cite{Cordey}, call into question whether the temperature of the plasma and fusion power density can be sustained on useful timescales.\\
\indent Centrifugal mirrors~\cite{Longmire,Bekhtenev_1980,Lehnert_1971,Lehnert_1974,White,Ellis,Romero,Abel} have addressed some of these doubts. These experiments apply a radial electric field to the plasma via end electrodes or alternative methods~\cite{Kolmes_Nature}, inducing an $\vec{E}\times\vec{B}$ drift around the central axis of the mirror~\cite{Gueroult, Ohkawa, Oiler, Liziakin}. The resulting rotation can reach supersonic speeds and produces a centrifugal potential well centered at the midplane~\cite{Lehnert_1971,Lehnert_1974}. This applied potential, in addition to demonstrated suppression of loss-cone instabilities~\cite{Kolmes,White}, makes centrifugal mirrors an exciting prospect for magnetic confinement fusion. Nevertheless, a careful examination of energetic particle losses is necessary to evaluate the viability of magnetic mirror confinement.\\
\indent Alpha particles, or He-4 nuclei, are produced in both deuterium-tritium (DT) and proton-boron-11 (p-B11)~\cite{Putvinski} fusion reactions, carrying a significant fraction of the fusion power in each case: 20\% and 100\%, respectively. A substantial population of highly energetic alpha particles is expected to greatly influence plasma temperature, fusion power density, transport, and stability~\cite{Ochs}. The role of alpha particles in fusion power balance involves two competing effects. On the one hand, accumulation of alpha particles is typically deleterious to fusion reactions. This is because they initially provide free energy that drives instabilities~\cite{Mizuno,Cai}, then transfer that energy to electrons enhancing bremsstrahlung radiation~\cite{England,Fisch}, and ultimately force reactant species to share the available pressure. These effects serve to undermine confinement, hemorrhage free energy, and decrease the effective fusion cross-section~\cite{Ochs}. This is especially undesirable for aneutronic reactions which have naturally smaller cross-sections~\cite{Kolmes_Alpha,Ochs_Alpha2,Ochs_Alpha}. On the other hand, if alpha particles are allowed to leave with their birth energy, it might be difficult to energetically sustain the reaction. A proposed solution to simultaneously address these concerns is alpha channeling~\cite{Fisch_Effect,Fisch_PRL,Zhmoginov,Fisch,Rax}, where wave interactions are leveraged to expel alpha particles while transferring their energy to fuel ions. Any residual energy could be captured via direct conversion~\cite{Yasaka} as they leave. Describing the loss energy and time spectra of alpha particles is the first step towards enacting such measures.\\
\indent Energetic particle losses in a magnetic mirror have been previously examined by Ochs et al.~\cite{Munirov}, who presented a covariant Fokker-Planck equation and associated timescales for relativistic tail electrons, by Pastukhov~\cite{Pastukhov}, who obtained expressions for collisional losses of electrons in a two-component plasma in the large potential limit $\Phi \!>\!>\! k_{B}T$, and by Najmabadi et al.~\cite{Najmabadi}, who derived a collision operator and similar expressions in the modest potential limit $\Phi \!<\! k_{B}T$. However, these efforts do not consider fast ion species like energetic alpha particles. Fokker-Planck codes have been used to describe fast-ion losses~\cite{Matsuda-Stewart,Tsuji,Peigney}, but these studies are entirely numerical. Killeen et al.~\cite{Killeen}, and more recently, Egedal et al.~\cite{Egedal} have considered semi-analytical solutions of fast-ion losses in the low-collisionality limit. They obtained results through spectral decomposition in terms of the unknown angular eigenfunctions of the Fokker-Planck model, which were computed numerically for the changing boundary conditions. However, understanding which system parameters define key features of alpha particle loss behavior and gaining insight into how we might use them to our advantage requires robust scaling relations and closed-form solutions. \\
\indent We derive here a fully analytical closed-form solution describing the loss of alpha particles in a magnetic mirror with arbitrary applied potential, in the low-collisionality and modest potential limits, for any mirror ratio $R \!>\! 1$. We employ the Wentzel–Kramers–Brillouin (WKB) approximation~\cite{Bender_Orszag} and Fourier analysis to asymptotically solve for Killeen and Egedal's unknown angular eigenfunctions and eigenvalues, compose from them a solution, and demonstrate agreement with a Monte-Carlo simulation.\\
\indent This solution condenses all of the information regarding alpha particle losses into a single expression, the Green's function solution for the distribution function. From this function, one can derive the probability distribution functions (PDFs) for the loss velocities, energies, and times for any initial distribution of alpha particles. These PDFs shed light on many difficult questions including optimizing direct conversion efficiencies, tuning device parameters such as mirror ratio and potential strength, and determining the various fates that particles meet, such as deconfinement via pitch-angle scattering or thermalization via drag.\\
\indent We find that the nature of our loss spectra is determined by a handful of dimensionless parameters with readily interpretable physical meanings, which can be manipulated externally through choice of device parameters. Our results show that for both DT and p-B11 scenarios, even a modest rotational potential succeeds in trapping and thermalizing a substantial fraction of the initially confined alpha particles. This suggests that differential confinement is an important consideration when designing a reactor based on centrifugal mirror confinement, and that even a small applied potential affects the power balance of the reactor significantly.\\
\indent The outline of the paper is as follows. We begin in Sec.\,\hyperref[sec:trapping]{II} by reviewing the definition of standard mirror coordinates and deriving the hyperbolic trapping boundary for a mirror with potential $\Phi$. In Sec.\,\hyperref[sec:fractions]{III}, we discuss the various fates that alpha particles meet and provide approximate expressions for their relative proportions or ``loss fractions" $F_{l}^{(i-iii)}$. In Sec.\,\hyperref[sec:collision]{IV}, we present the classical non-relativistic Rosenbluth formulation~\cite{NRL} of the Landau collision operator~\cite{landau} and derive from it a Fokker-Planck partial differential equation (PDE) in mirror coordinates to describe all nontrivial loss behavior. In Sec.\,\hyperref[sec:montecarlo]{V}, we find a numerical solution to this PDE via Monte-Carlo methods. In Sec.\hyperref[sec:pde]{VI}, we propose an ordering of timescales that allows us to solve our PDE by separation of variables and obtain an analytical Green’s function solution for the distribution function $f(t,x,\mu)$. In Sec.\hyperref[sec:eigenmodes]{VII}, we integrate this distribution function to obtain the remaining number density as a sum of eigenmode populations with characteristic decay rates. In Sec.\,\hyperref[sec:pdfs]{VIII}, we derive the loss velocity and time PDFs from the number density solution and generalize our Green's function solution to any initial velocity distribution. In Sec.\,\hyperref[sec:energyntime]{IX}, we present expressions for the mean loss energy and time due to pitch-angle scattering. In Sec.\,\hyperref[sec:fractions2]{X}, we generalize our loss fractions for any initial velocity distribution. In Sec.\,\hyperref[sec:steadystate]{XI}, we compute the equilibrium distribution for a steady-state delta source. In Sec.\,\hyperref[sec:relativity]{XII}, we outline the application of our model to relativistic alpha particles and tail electrons. In Sec.\,\hyperref[sec:conclusions]{XIII}, we present our conclusions.
\vspace{-8pt}
\section{II. Trapping Condition \& Phase Space}
\label{sec:trapping}
\begin{figure}
\centering
\noindent\includegraphics[width=\columnwidth]{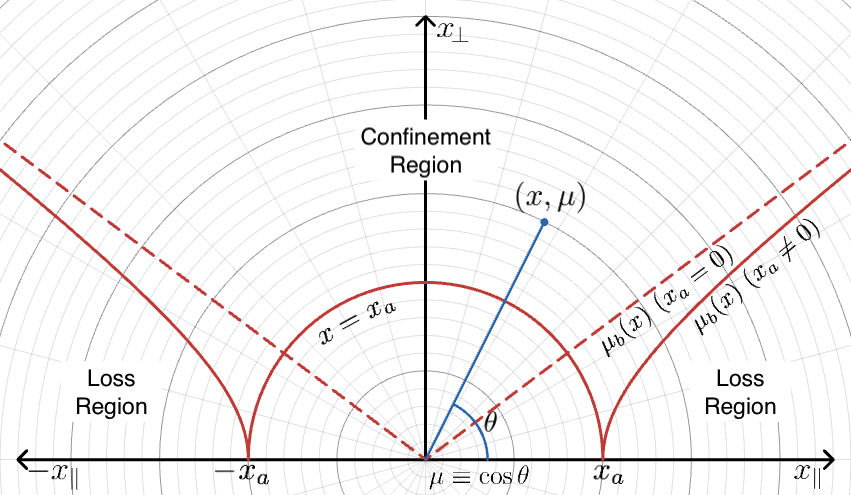}
\caption{Mirror Phase Space. Here we see the projection of velocity phase space into the $(x_{\parallel},x_{\perp}) \equiv (x\mu,x\sqrt{1\!-\!\mu^{2}})$ plane. The regions on the left and right represent the loss cone, which in the presence of a confining potential becomes the hyperboloid described by $|\mu|\geq\mu_{b}(x)$ (Eq.~(\ref{Eq:hyperbola})\!). The upper region given by $|\mu| < \mu_{b}(x)$ corresponds to confined particles and the circular region around the origin corresponds to potential-trapped, thermalized particles. The bottom half of the picture is perfectly symmetric and has been omitted for clarity.\label{Fig:hyperbola}}
\end{figure}

\vspace{-8pt}
\indent Consider a fast ion species `$a$' confined in an axisymmetric centrifugal magnetic mirror with mirror ratio $R \equiv B_{e}/B_{m}$ subject to applied potential $\Phi_{a}\!\equiv\!\Phi_{e}\!-\!\Phi_{m}$, where subscripts $m$ and $e$ denote the midplane and ends, respectively. This potential will typically be a combination of centrifugal and ambipolar electric potentials. Then conservation of energy between the midplane and ends of the device can be used to derive a trapping condition as follows. We choose to work in standard mirror coordinates~\cite{Munirov}, which are isomorphic to spherical coordinates with azimuthal symmetry corresponding to gyrotropicity,
\begin{align}x \equiv v/v_{_{th,a}}, \,\,\mu \equiv v_{z}/v,\,\,\text{tan}\,\varphi \equiv v_{y}/v_{x} \end{align}
where normalized velocity $x$ is the radial coordinate, pitch-angle $\mu \!\equiv\! \cos\theta$ corresponds to polar angle $\theta$ with $v_{z}\!\equiv\! v_{\parallel}$, $\varphi$ is the gyrophase angle, and $v_{th,a}\!\equiv\!\sqrt{2k_{B}T_{a}/m_{a}}$ is the alpha thermal velocity.
Let us denote the kinetic and potential energies $W_{\perp,\parallel},\Phi$. Then the boundary between trapped and untrapped occurs precisely when $W_{e\parallel} = 0$.
\begin{equation}
\begin{aligned}
&W_{m\perp}+W_{m\parallel} + \Phi_{m} = W_{e\perp}+\cancel{W_{e\parallel}} + \Phi_{e}\end{aligned}
\end{equation}
We can use conservation of the first adiabatic invariant to substitute $R = B_{e}/B_{m} = W_{e\perp}/W_{m\perp}$ and obtain,
\begin{align}
\label{Eq:hyperbola}
&x_{m}^{2} \!\!=\!Rx_{m}^{2}(1\!-\!\mu^{2})\!+\!\frac{\Phi_{a}}{k_{B}T_{a}\!}\!\rightarrow\mu_{b}(x) \!\equiv\! \sqrt{1\!-\!\frac{1}{R}\!\!\left(\!1\!-\!\frac{x_{a}^{2}}{x^{2}}\!\right)}\!\!
\end{align}
where we have defined $x_{a} \!\equiv\! \sqrt{\Phi_{a}/k_{B}T_{a}}$. This trapping boundary in midplane coordinates corresponds to a hyperboloid in $\mathbb{R}^{3}$, which we can further simplify by realizing gyrophase angle is irrelevant to confinement and taking a 2D cross-section at $\varphi = 0$ as shown in Figure~\ref{Fig:hyperbola}. \\
\indent Throughout our analysis, we will consider all particles of this fast ion species to experience the same constant potential $\Phi_{a}$. In the centrifugal case, this corresponds to neglecting shear in the rotational velocity profile of the reactor, which presents an alternative loss channel where particles diffuse radially and then exit in a region with lower effective confining potential \cite{White}. This assumption is valid for a sufficient combination of the reactor radius being many times the Larmor radius and drag dominating over pitch-angle scattering, as shown in Appendix~\hyperref[App:J]{J}.
\vspace{-20pt}
\section{III. Loss Fractions}
\label{sec:fractions}
\begin{figure}
\centering
\noindent\includegraphics[height = 0.75\columnwidth,width=\columnwidth]{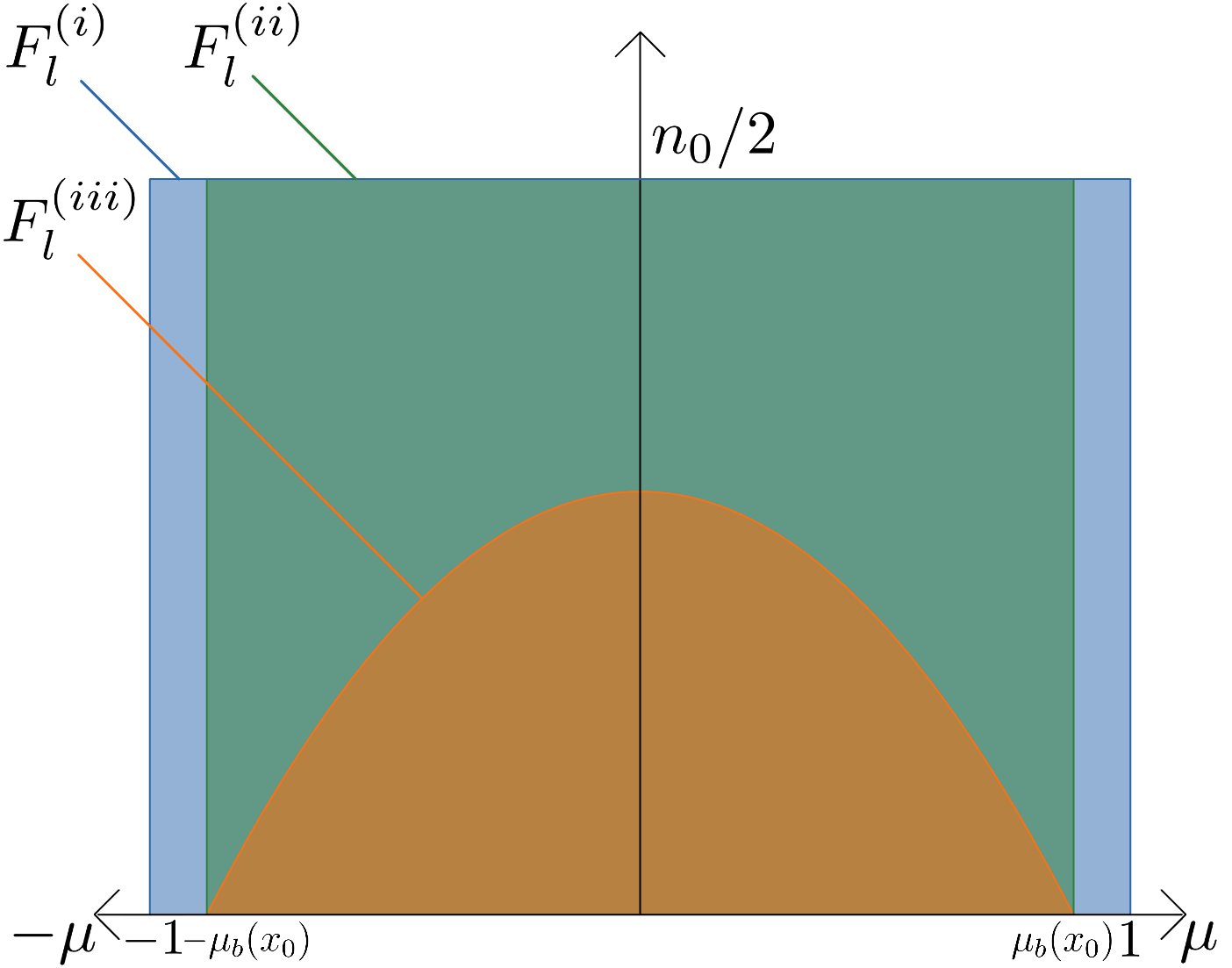}
\caption{Loss Fractions. Shown here for $n_{0}$ particles are the fractions of the initial uniform distribution in $\mu$ that are born deconfined (blue), leave gradually via pitch-angle scattering (green), and end up confined as part of the resulting thermal distribution (orange).\label{Fig:fractions}}
\end{figure}

\vspace{-6pt}
Let us consider an isotropic population of $n_{0}$ alpha particles born at a single velocity $x=x_{0}$ as shown in Figure~\ref{Fig:DensityShellModel}.
This corresponds to a uniform distribution across pitch-angles $\mu \in [-1,1]$ as shown in Figure~\ref{Fig:fractions}, where we can estimate the fractions of particles that are lost from the mirror in various ways over time. Let us denote,
\begin{align*}&i. \rightarrow \text{never confined: born in loss cone with $|\mu| \geq \mu_{b}(x_{0})$}\\
&ii. \!\rightarrow \text{gradually lost: deconfined via pitch-angle scattering}\\
&iii. \!\rightarrow \text{retained: slowed and trapped by the potential}\end{align*}
Suppose now that we knew how many of our initially confined particles remained in our distribution at a given value of $x$, described by our number density $n(x|x_{0})$ (Eq.~(\ref{Eq:n(x|x0)})\!) normalized to $\hat{n}_{0}\!\equiv\!n_{0}\mu_{b}(x_{0})$. Then~we~could~immediately~estimate~loss~fractions,
\begin{align}\label{Eq:lossfractions}&F_{l}^{(i)}\!(x_{0}) = \frac{n_{0}-\hat{n}_{0}}{n_{0}} = 1-\mu_{b}(x_{0})\nonumber\\
&\!F_{l}^{(ii)}\!(x_{0}) = \frac{\hat{n}_{0}-n(x_{a}|x_{0})}{n_{0}} = \mu_{b}(x_{0})\!\biggl(\!1\!-\!\frac{\!n(x_{a}|x_{0})}{\hat{n}_{0}}\!\biggr)\\
&F_{l}^{(iii)}\!(x_{0}) = \frac{n(x_{a}|x_{0})}{n_{0}} = \mu_{b}(x_{0})\frac{n(x_{a}|x_{0})}{\hat{n}_{0}}\nonumber\end{align}
We know the loss energy and time of particles in $F_{l}^{(i)}$ since they roughly leave the mirror at their birth energy and time, and we can approximate for weak parallel diffusion that those particles making it to $F_{l}^{(iii)}$ never leave the mirror at all since they are confined by the potential. What happens to those particles in $F_{l}^{(ii)}$ which are gradually lost between birth and when they would otherwise be trapped by the potential remains to be seen.\\ \\
\vspace{-29pt}
\section{IV. Landau Collision Operator}
\vspace{-4pt}
\label{sec:collision}
Making the conventional assumption that all collisions occur at the midplane, or equivalently that our mirror potential is a square-well~\cite{Pastukhov,Post}, we model our distribution function as living in the confined region illustrated in Figure~\ref{Fig:hyperbola} with homogeneous boundary conditions and evolving solely due to collisions such that $df_{a}/dt = \partial f_{a}/\partial t = \left(\partial f_{a}/\partial t\right)_{coll}$.\\
\indent We present the non-relativistic Landau collision operator~\cite{landau} for a highly energetic ion species (alpha particles) colliding with Maxwellian distributions of other less energetic, slower ions (deuterium, tritium/protons, boron-11) and less energetic, albeit faster, electrons. We obtain the following exact tensor PDE in Cartesian velocity coordinates from the Rosenbluth formulation~\cite{NRL} in Appendix \hyperref[App:A]{A} using the identities in Appendix \hyperref[App:B]{B},
\begin{align}\frac{\partial f_{a}}{\partial t} = \frac{\partial}{\partial\vec{v}}\cdot\!\left(\!^{\vec{v}}\!\vec{A}_{a}f_{a}+\!^{\vec{v}}\overline{\overline{D}}_{a}\cdot\frac{\partial f_{a}}{\partial\vec{v}}\!\right)\end{align}
where the advection vector $^{\vec{v}}\!\vec{A}_{a}$, and diffusion tensor $^{\vec{v}}\overline{\overline{D}}_{a}$ are given for $x_{b}\equiv v/v_{th,b}$, error function $\text{erf}(x)$, and special function $\text{slp}(x)\!\equiv \text{erf}(x)\!-\!x\,\text{erf}\,'\!(x)$~\cite{chandrasekhar} by,
\begin{align}&\!^{\vec{v}}\!\vec{A}_{a} \!=\! \sum_{b}\!C_{ab}\!\!\left\{\!\frac{m_{a}\vec{v}}{m_{b}v^{3}}\!\right\}\!\text{slp}\!\left(x_{b}\right)\!, \,C_{ab} \!\equiv\! \frac{4\pi n_{b}}{m_{a}^{2}}\!\!\left(\!\frac{Z_{a}Z_{b}e^{2}}{4\pi\epsilon_{0}}\!\right)^{\!\!2}\!\!\!\lambda_{ab}\nonumber\\
&\!\!^{\vec{v}}\overline{\overline{D}}_{a} \!\!=\!\! \sum_{b}\!\frac{C_{ab}}{2}\!\!\left\{\!\!\frac{v^{2}\overline{\overline{1}}\!-\!\vec{v}\vec{v}}{v^{3}}\!\!\left(\!\!\text{erf}(x_{b})\!-\!\frac{\text{slp}(x_{b})}{2x_{b}^{2}}\!\right)\!\!+\!\frac{\vec{v}\vec{v}}{v^{3}}\frac{\text{slp}(x_{b})}{x_{b}^{2}}\!\!\right\}\!\!\!\!\end{align}
Transforming to standard mirror coordinates $(v_{x},v_{y},v_{z})\rightarrow(x,\mu,\varphi)$ as shown in Appendix \hyperref[App:C]{C}, and converting the resulting tensor equation into a scalar one as shown in Appendix \hyperref[App:D]{D} we obtain the following PDE,
\begin{equation}
\begin{aligned}&\!\!\frac{\partial f_{a}}{\partial t}\!\!=\!\!\frac{1}{x^{2}}\!\frac{\partial}{\partial x}\!\!\!\left[\!\!\left\{\!\!\sum_{b}\!\frac{C_{ab}}{v_{th,a}^{3}\!}\!\frac{m_{a}}{m_{b}}\text{slp}(x_{b})\!\!\right\}\!\!f_{a}\!\!+\!\!\left\{\!\!\sum_{b}\!\frac{\!C_{ab}x\!}{2v_{th,a}^{3}}\frac{\text{slp}(x_{b})\!}{x_{b}^{2}}\!\!\right\}\!\!\frac{\partial f_{a}}{\partial x}\!\right]\!\!\\
&\!\!\!\!+\!\!\frac{1}{x^{3}}\!\!\left\{\!\!\sum_{b}\!\frac{\!C_{ab}\!}{2v_{th,a}^{3}\!\!}\!\!\left(\!\!\text{erf}(x_{b})\!-\!\frac{\text{slp}(x_{b})\!}{2x_{b}^{2}}\!\right)\!\!\!\right\}\!\!\!\left[\!\frac{\partial}{\partial\mu}\!\!\!\left[\!(1\!\!-\!\!\mu^{\!2})\!\frac{\partial f_{a}}{\partial\mu}\!\right]\!\!\!+\!\!\frac{1}{1\!\!-\!\!\mu^{2}}\!\frac{\partial^{2}f_{a}\!}{\partial\varphi^{2}\!}\!\right]\!\!\end{aligned}
\end{equation}
where symmetry in gyrophase makes the last term vanish. In the case of alpha particles `$a$' colliding with species $b\in\{i,e\}$, we consider the limits $x_{i}\!>\!>\!1$ and $x_{e} \!<\!<\! 1$, subject to constraints $v_{th,i} \!<\!<\! v_{th,a}x \!<\!<\! v_{th,e}$. For a DT scenario (Table \hyperref[Tab:transport]{1}) this corresponds to validity regime $0.1\lesssim x \lesssim 5.6$, so we shall restrict $x_{a}\in[0.1,x_{0}]$. Expanding in the appropriate limit for each species and considering trace alphas $n_{a} \!<\!<\! n_{i},n_{e}$ we have,
\begin{align}\label{Eq:scalarcollisionoperator}&\tau_{0}^{i}\frac{\partial f_{a}}{\partial t} \!=\! \frac{1}{x^{2}}\frac{\partial}{\partial x}\!\left[\!\left(\!Z_{\parallel}^{i}\!+\!Z_{\parallel}^{e}x^{3}\!\right)\!\!f_{a}\!+\!\frac{1}{2}\!\left(\!\frac{1}{x}\!+\!Z_{\perp}^{e}x^{2}\!\!\right)\!\!\frac{\partial f_{a}}{\partial x}\!\right]\\
&\!+\!\frac{1}{2x^{3}}\!\!\left(\!\!Z_{\perp}^{i}\!\!+\!Z_{\perp}^{e}x\!-\!\frac{1}{2x^{2}}\!\!\right)\!\!\frac{\partial}{\partial\mu}\!\!\left[\!(1\!-\!\mu^{2})\frac{\partial f_{a}}{\partial\mu}\!\right]\nonumber\end{align}
where our transport coefficients are given by,
\begin{align}&
(\tau_{0}^{i})\,^{\!\!-1}\!\!\equiv\!\frac{4\pi e^{4}}{(4\pi\epsilon_{0})^{2}}\!\sum_{b\neq e}\!\frac{n_{b}Z_{a}^{2}Z_{b}^{2}\lambda_{ab}}{m_{a}^{2}v_{th,a}^{3}}\frac{m_{a}T_{b}}{m_{b}T_{a}},\,\,\,\tau_{s}\equiv\frac{\tau_{0}^{i}}{Z_{\parallel}^{e}}\nonumber\\
&(\tau_{0}^{e})\,^{\!\!-1}\!\!\equiv\!\frac{4\pi e^{4}}{(4\pi\epsilon_{0})^{2}}\frac{\frac{4}{3\sqrt{\pi}}n_{e}Z_{a}^{2}\lambda_{ae}}{m_{a}^{2}v_{th,a}^{3}}\!\!\left(\!\frac{m_{e}T_{a}}{m_{a}T_{e}}\!\right)^{\!\!1/2} \!\!\!\!\!\!= \frac{Z_{\perp}^{e}}{\tau_{0}^{i}}\\
&Z_{\parallel}^{i} \!\equiv\! \frac{\sum_{b\neq e}n_{b}Z_{b}^{2}\lambda_{ab}T_{a}/m_{b}}{\sum_{b\neq e}n_{b}Z_{b}^{2}\lambda_{ab}T_{b}/m_{b}},Z_{\parallel}^{e} \!\equiv\! \frac{\frac{4}{3\sqrt{\pi}}n_{e}\lambda_{ae}\left(\!\frac{m_{e}T_{a}^{5}}{m_{a}^{3}T_{e}^{3}}\!\right)^{\!1/2}}{\sum_{b\neq e}n_{b}Z_{b}^{2}\lambda_{ab}T_{b}/m_{b}}\nonumber\\
&Z_{\!\perp}^{i} \!\!\equiv\! \frac{\sum_{b\neq e}n_{b}Z_{b}^{2}\lambda_{ab}T_{a}/m_{a}}{\sum_{b\neq e}n_{b}Z_{b}^{2}\lambda_{ab}T_{b}/m_{b}},Z_{\!\perp}^{e}\!\!\equiv\! \frac{\!\frac{4}{3\sqrt{\pi}}n_{e}\lambda_{ae}\!\!\left(\!\frac{m_{e}T_{a}^{3}}{m_{a}^{3}T_{e}}\!\right)^{\!1/2}\!\!\!\!}{\sum_{b\neq e}\!n_{b}Z_{b}^{2}\lambda_{ab}T_{b}/m_{b}\!}\nonumber\end{align}
\vspace{-1pt}
\noindent Their approximate values are summarized in Table 1 for DT and p-B11 fusion scenarios, along with the projected reactor parameters and Coulomb logarithms (Appendix \hyperref[App:E]{E}) used to compute them. Parallel ($\parallel$) and perpendicular ($\perp$) here refer to velocity and pitch-angle, respectively, not orientation relative to the magnetic field. The given ion and electron timescales $\tau_{0}^{i,e}$ are for parallel diffusion due to each species, with $\tau_{s}$ being the slowing down time.\\
\begin{table}
\centering
Table 1. Projected Values for DT \& p-B11 Scenarios
\begin{tabular}{ |p{0.6cm}|p{2.2cm}|p{0.55cm}|p{1.75cm}|p{0.5cm}|p{1.75cm}|  }
\hline
\multicolumn{6}{|c|}{DT Parameters \& Transport Coefficients} \\
\hline
$n_{a}$ & $0.3\cdot 10^{13}\text{cm}^{-3}$ & $T_{a}$ & $3.5$MeV& $\tau_{0}^{i}$ & $2114.3$ s\\
\hline
$n_{d}$ & $4.85\cdot 10^{13}\text{cm}^{-3}$ & $T_{d}$&$15$keV& $\tau_{0}^{e}$ & $129.16$ s\\
\hline
$n_{t}$ & $4.85\cdot 10^{13}\text{cm}^{-3}$ & $T_{t}$ & $15$keV& $Z_{\parallel}^{i}$ & 233.33\\
\hline
$n_{e}$ & $1.03\cdot 10^{14}\text{cm}^{-3}$ & $T_{e}$ & $15$keV&$Z_{\parallel}^{e}$ & 3819.5\\
\hline
$\lambda_{ad}$ & 24.686 & $\lambda_{ae}$ & 20.355 & $Z_{\perp}^{i}$ & 140.14\\
\hline
$\lambda_{at}$ & 24.934 & $\lambda_{ee}$ & 21.017
 &$Z_{\perp}^{e}$ & 16.369\\
\hline
\end{tabular}
\begin{tabular}{ |p{0.6cm}|p{2.2cm}|p{0.55cm}|p{1.75cm}|p{0.5cm}|p{1.75cm}|  }
\hline
\multicolumn{6}{|c|}{p-B11 Parameters \& Transport Coefficients} \\
\hline
$n_{a}$ & $0.3\cdot10^{13}\text{cm}^{-3}$ & $T_{a}$ & $2.9$MeV& $\tau_{0}^{i}$ & $22.953$ s\\
\hline
$n_{p}$ & $8.25\cdot10^{13}\text{cm}^{-3}$ & $T_{p}$&$300$keV& $\tau_{0}^{e}$ & $187.90$ s\\
\hline
$n_{B}$ & $1.45\cdot10^{13}\text{cm}^{-3}$ & $T_{B}$ & $300$keV& $Z_{\parallel}^{i}$ & 9.6667\\
\hline
$n_{e}$ & $1.61\cdot 10^{14}\text{cm}^{-3}$ & $T_{e}$ & $150$keV&$Z_{\parallel}^{e}$ & 2.3616\\
\hline
$\lambda_{ap}$ & 25.150 & $\lambda_{ae}$ & 23.454 & $Z_{\perp}^{i}$ & 9.1924\\
\hline
$\lambda_{aB}$ & 24.530 & $\lambda_{ee}$& 24.144 &$Z_{\perp}^{e}$ & 0.1222\\
\hline
\end{tabular}\label{Tab:transport}\end{table}
\indent Examining our coefficients, we observe that at high energies, electron drag and ion pitch-angle scattering are the dominant processes for DT, while for p-B11, both drag and pitch-angle scattering are mostly due to ions.\\
\indent Considering $x_{0} \!\sim\! O(1)$ for the overwhelming majority of particles, parallel diffusion can be neglected in both cases since it is the 2nd order process in $x$ and has a small coefficient. Let us further define $Z_{\parallel}(x) \!\equiv\! Z_{\parallel}^{i}\!+\!Z_{\parallel}^{e}x^{3}, \, Z_{\!\perp}\!(x) \!\equiv\! Z_{\perp}^{i}\!\!+\!Z_{\perp}^{e}x\!-\!1/2x^{2}$ to obtain the Fokker-Planck advection-diffusion equation,
\begin{align}\label{Eq:advection-diffusion}\!\!\!\!\!\frac{\partial f_{a}}{\partial t} \!+\! \frac{1}{x^{2}}\frac{\partial}{\partial x}\!\left(x^{2}v(x)f_{a}\right)=\nu(x)\frac{\partial}{\partial\mu}\!\!\left[\!(1\!-\!\mu^{2})\frac{\partial f_{a}}{\partial\mu}\!\right]\end{align}
with parallel advection velocity $v(x)\!\equiv\! -Z_{\parallel}(x)/\tau_{0}^{i}x^{2}$ corresponding to frictional drag, and collision frequency $\nu(x)\!\equiv\! Z_{\!\perp}\!(x)/2\tau_{0}^{i}x^{3}$ characterizing pitch-angle diffusion due to successive scatterings.\\
\indent Our problem has now been reduced to an initial value problem of a scalar PDE in (2+1) dimensions where we must solve for the distribution function $f_{a}(t,x,\mu)$ subject to initial condition $f_{a}(0,x,\mu)$. 
\vspace{-9pt}
\section{V. Numerical Verification}
\label{sec:montecarlo}
\begin{figure}
\centering
\noindent\includegraphics[height = 0.63\columnwidth,width=\columnwidth]{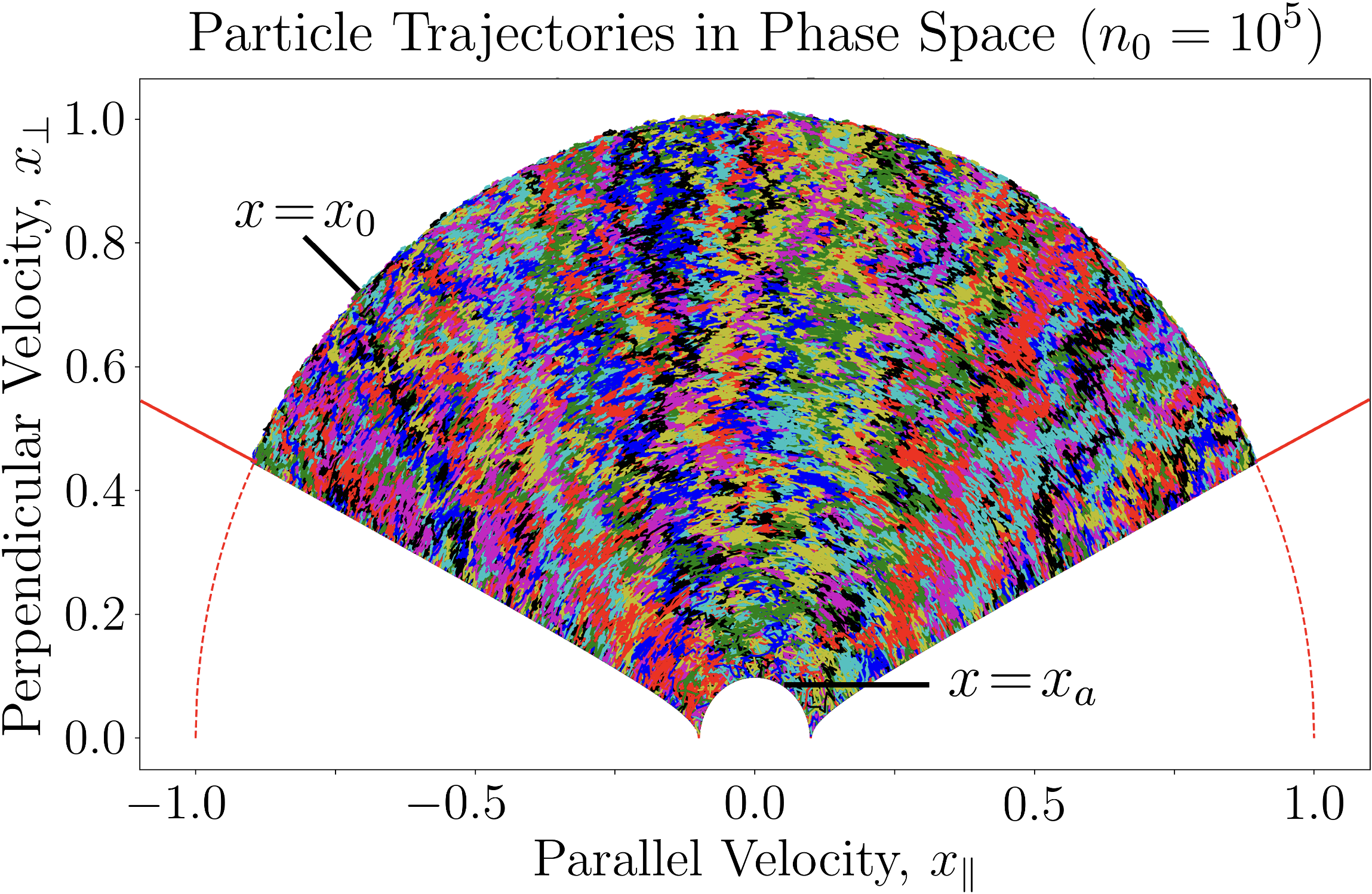}
\caption{Monte-Carlo Particle Trajectories. Shown here are the trajectories of $n_{0} = 10^{5}$ alpha particles initially isotropically distributed at $x = x_{0}$, undergoing deterministic advection and stochastic diffusion as dictated by the Fokker-Planck Model in Eqs.~(\ref{Eq:advection-diffusion},\ref{Eq:MC-evolution}) for a DT scenario with $x_{0} = 1, x_{a} = 0.1,\, R \!=\! 5$.\label{Fig:montecarlo}}
\end{figure}

\vspace{-5pt}
\indent Before we begin our analytical calculation, we should first find a numerical solution of the advection-diffusion equation (Eq.~(\ref{Eq:advection-diffusion})\!) which can be used to benchmark our model. To this end, we employ Monte-Carlo simulations comprising $n_{0} = 10^{5}$ particles sampled uniformly between $\mu\in[-1,1]$ at $x = x_{0}$, in line with the setup described in Sec.\,\hyperref[sec:fractions]{III}. This will also allow us to better understand the physical dynamics that Eq.~(\ref{Eq:advection-diffusion}) represents.\\
\indent Since Eq.~(\ref{Eq:advection-diffusion}) is already in conservative form, the time-evolution rules for an individual particle can be directly obtained, and since $\hat{x}$ and $\hat{\mu}$ are orthogonal, we can further treat these behaviors independently from each particle's point of view. In $\hat{x}$ we have deterministic advection,
\begin{align}\frac{\partial f_{a}}{\partial t} + \vec{\nabla}_{\vec{x}}\cdot(f_{a}v(x)\hat{x}) = 0 \rightarrow \frac{dx}{dt} = v(x)\end{align}
However, in $\hat{\mu}$ we have stochastic diffusion. For $\mu\equiv\cos\theta$,
\begin{align}\frac{\partial f_{a}}{\partial t} = \nu(x)\frac{\partial}{\partial\mu}\!\!\left[\!(1\!-\!\mu^{2})\frac{\partial f_{a}}{\partial\mu}\!\right]\rightarrow \frac{\partial f_{a}}{\partial t} = \nu(x)\nabla_{\theta}^{2}f_{a}\end{align}
which is the classic heat equation with constant (in $\theta$) diffusion coefficient. The response to a single particle, or delta-distribution $\delta(\theta-\theta_{0})$, is given by the heat kernel,
\begin{align}f(0,\theta) = \delta(\theta\!-\!\theta_{0}) \rightarrow f(t,\theta) \!=\! \frac{1}{\sqrt{4\pi\nu(x)t}}e^{-\frac{(\theta-\theta_{0})^{2}}{4\nu(x)t}}\!\!\end{align}
Then we can think of $d\theta \equiv \theta-\theta_{0}$ as a random variable distributed as the heat kernel at each time step \cite{Killeen}. Our evolution rules are therefore,
\begin{equation}\label{Eq:MC-evolution}\begin{aligned}dx_{n} = v(x_{n})dt \rightarrow x_{n+1} = x_{n}+dx_{n}\\
d\theta_{n}\sim \mathcal{N}(0,2\nu(x_{n})dt) \rightarrow \theta_{n+1} = \theta_{n} + d\theta_{n}\end{aligned}\end{equation}
The individual particle trajectories generated by this random motion are shown in Figure~\ref{Fig:montecarlo}. Monte-Carlo results will be given for comparison with our analytical model for various parameters in a DT scenario in Figures~\ref{Fig:n(x|x0)},\ref{Fig:n(t|x0)},\ref{Fig:p(x|x0)},\ref{Fig:p(t|x0)}. Now let us proceed with our analytical calculation.
\vspace{-5pt}
\section{VI. Fokker-Planck Solution}
\label{sec:pde}
\begin{figure}
\centering
\noindent\includegraphics[width=\columnwidth]{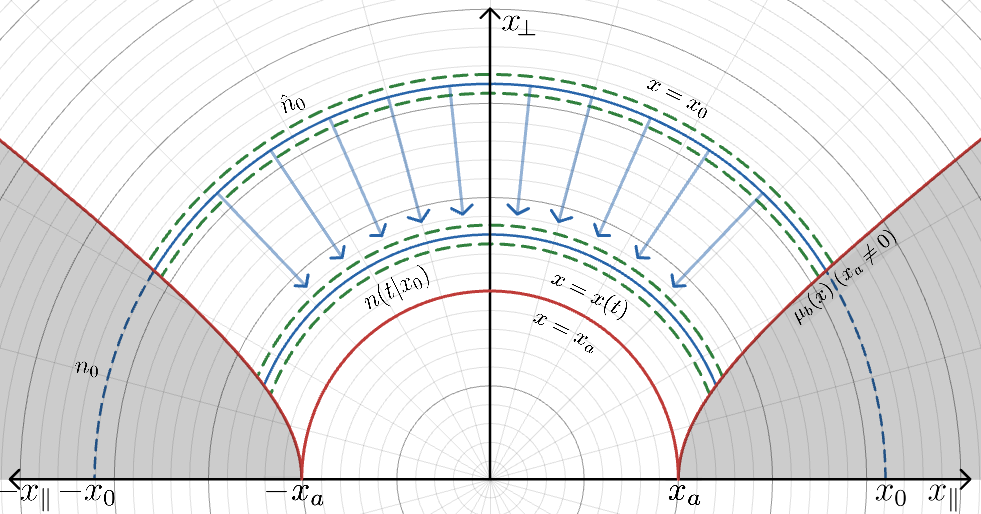}
\caption{Density Shell Model. Here we see a thin shell of initially confined density $\hat{n}_{0}$ advecting inwards from initial radial coordinate $x = x_{0}$ while diffusing in angle and leaking through the trapping boundary into the loss cone (hyperboloid). Near the origin, we see the circular trapped region established by the rotational potential, where a (possibly vanishing) fraction of the distribution becomes confined.\label{Fig:DensityShellModel}}
\end{figure}

\vspace{-6pt}
Considering precisely the same setup described in Sec.\,\hyperref[sec:fractions]{III}, we have as our initial condition the infinitesimally thin shell of density $\hat{n}_{0}$, starting at $x = x_{0}$ as shown in Figure~\ref{Fig:DensityShellModel}. Since we neglect parallel diffusion, particles comprising the shell should advect towards the origin together, with the entire confined distribution at radius $x = x(t)$ at any given time. This also implies that in a more general situation, particles beginning at different values of $x = x_{0}$ evolve independently of one another. Then we can understand that the response to this infinitesimally thin shell is essentially the Green's function solution to our problem. We solve for this Green's function solution with the understanding that it can later be integrated over any initial velocity distribution to yield fully general results for both the loss fractions and loss distributions (Sec.\,\hyperref[sec:pdfs]{VIII-X}).\\
\indent We limit our search to approximately separable solutions $f_{a}(t,x,\mu) = g(t,x)h(\mu|\mu_{b}(x)\!)$ where we use the fact that our bounding hyperbola $\mu_{b}(x)$ varies slowly relative to pitch-angle diffusion to categorize the dependence of $h(\mu|\mu_{b}(x)\!)$ on $x$ as weak. Then Eq.~(\ref{Eq:advection-diffusion}) becomes,
\begin{align}&\!\!\!\!\frac{1}{\nu(x)g}\!\!\left[\frac{\partial g}{\partial t}\!+\!\frac{\!1\!}{x^{2}\!}\frac{\partial}{\partial x}(x^{2}v(x)g)\!\right] \!\!=\! \frac{1}{h}\frac{\partial}{\partial\mu}\!\!\left[\!(1\!-\!\mu^{2})\frac{\partial h}{\partial\mu}\!\right] \!\!=\! -\lambda\!\!\!\end{align}
where we've defined separation ``constant" $-\lambda(\mu_{b}(x)\!)$ to obtain the following pair of equations: advection and pitch-angle scattering loss,
\begin{align}&\frac{\partial g}{\partial t} + \frac{1}{x^{2}}\frac{\partial}{\partial x}\!\left(x^{2}v(x)g\right) + \lambda \nu(x)g = 0\end{align}
and shaping of the distribution in pitch-angle by the diffusive character of scattering and the moving boundary conditions,
\begin{align}\label{Eq:Legendre}&\frac{\partial}{\partial\mu}\!\!\left[(1\!-\!\mu^{2})\frac{\partial h}{\partial \mu}\right] \!=\! -\lambda h\end{align}
From the advection equation we easily obtain $g(t,x)$ via method of characteristics as shown in Appendix \hyperref[App:F]{F}. We find that for the initial condition of a delta shell distribution at $x = x_{0}$, namely $g(0,x) = \delta(x\!-\!x_{0})/x_{0}^{2}$,
\begin{align}\label{Eq:advection-soln}g(t,x) = \frac{\delta(x\!-\!x(t)\!)}{x(t)^{2}}\exp\!\left(\!\!-\!\!\int\limits_{0}^{t}\!\!\lambda(\mu_{b}(x(t)\!)\!)\nu(x(t)\!)\,dt\!\!\right)\!\!\end{align}
where our characteristic advection path is given by,
\begin{align}\label{Eq:x(t)}x(t)\!=\!\left[(x_{0}^{3}+\eta^{3})e^{-3t/\tau_{s}}-\eta^{3}\right]^{1/3}\!\!\!\!\!\!\!\!,\,\,\,\,\eta\equiv(Z_{\parallel}^{i}/Z_{\parallel}^{e})^{1/3}\end{align}
In $h(\mu|\mu_{b})$ we have an eigenvalue problem of the Legendre operator (Eq.~(\ref{Eq:Legendre})\!). Its prototypical eigenfunctions,
the Legendre functions of the first and second kind, do
not satisfy our moving homogeneous boundary conditions
so we instead employ the eikonal WKB approximation~\cite{Bender_Orszag},
\begin{align}&h(\mu|\mu_{b}) \sim e^{is/\!\sqrt{\epsilon}}, \,\,\,\, s(\mu|\mu_{b}) \equiv \sum\limits_{n=0}^{\infty}\epsilon^{n/2}s_{n}(\mu|\mu_{b})\end{align}
which when substituted into Eq.~(\ref{Eq:Legendre}) yields,
\begin{align}\label{Eq:sequation}-2\mu\frac{is'}{\sqrt{\epsilon}}+(1\!-\!\mu^{2})\frac{is''}{\sqrt{\epsilon}}-(1\!-\!\mu^{2})\frac{s'^{2}}{\epsilon} = -\lambda\end{align}
As shown in Eq.~(\ref{Eq:advection-soln}), higher eigenvalues correspond to those eigenmodes with higher decay rates, which naturally dominate our loss spectra. If we suppose then that our eigenvalue contributes at the highest order, $\lambda \sim \mathcal{O}(1/\epsilon)$, our dominant balance is,
\begin{align}\label{Eq:dominantbalance}(1\!-\!\mu^{2})\frac{s_{0}'^{2}}{\epsilon}\!\sim\!\lambda \,\rightarrow\, s_{0}\!\sim\!\pm\sqrt{\epsilon\lambda}\sin^{\!-1}\!\!\mu\end{align}
And at $\mathcal{O}(1/\sqrt{\epsilon})$,
\begin{align}\label{Eq:subdominantbalance}\!\frac{-i\mu}{1\!-\!\mu^{2}}\!+\!\frac{is_{0}''}{2s_{0}'}\!\sim\!s_{1}'\,\rightarrow\, s_{1} \!\sim\!\frac{i}{4}\!\ln(1\!-\!\mu^{2})\end{align}
Let us first find a solution in the Geometrical Optics (GO) approximation~\cite{hojlund,Bender_Orszag}, $h \!\!\sim\!\! e^{is_{0}/\sqrt{\epsilon}}$. The Physical Optics (PO) approximation $h \!\sim\! e^{is_{0}/\!\sqrt{\epsilon}+s_{1}}$ shown in Appendix \hyperref[App:G]{G} represents a negligible correction. Then,
\begin{align}h(\mu|\mu_{b})\!\sim\!a(\mu_{b})\cos(\!\sqrt{\lambda}\sin^{\!-1}\!\!\mu)\!+\!b(\mu_{b})\sin(\!\sqrt{\lambda}\sin^{\!-1}\!\!\mu)\!\!\end{align}
Our boundary conditions $h(\pm\mu_{b}|\mu_{b}) = 0$ allow two nontrivial solutions,
\begin{equation}\label{Eq:eigenvalues}\begin{aligned}a\!\neq\!0,b\!=\!0 \rightarrow\! \lambda_{k}^{(i)} \!\!=\! \frac{\!(k\!+\!1/2)^{2}\pi^{2}\!\!\!}{\!(\sin^{\!-1}\!\!\mu_{b})^{2}\!},\,\!h_{k}^{\!(i)}\!\!\!\sim\! \cos\!\left(\!\!\sqrt{\!\lambda_{k}^{(i)}}\!\sin^{\!-1}\!\!\mu\!\!\right)\!\!\!\!\!\\
a\!=\!0,b\!\neq\!0 \!\rightarrow\!\lambda_{k}^{(ii)} \!\!=\! \frac{\!k^{2}\pi^{2}\!\!}{\!(\sin^{\!-1}\!\!\mu_{b})^{2}\!},\,\!h_{k}^{\!(ii)}\!\!\!\sim\! \sin\!\!\left(\!\!\sqrt{\!\lambda_{k}^{(ii)}}\!\sin^{\!-1}\!\!\mu\!\!\right)\!\!\!\!\end{aligned}\end{equation}
It is worth noting that we have found an orthogonal set of eigenfunctions with associated eigenvalues which can be used to construct asymptotic solutions to any eigenvalue problem of the Legendre operator with homogeneous boundary conditions on the interval $\mu\in[-\mu_{b},\mu_{b}]$ (or $\mu\in[0,\mu_{b}]$), for any $\mu_{b} < 1$. Then, our general solution for $f(t,x,\mu)$ is a linear combination of these two solutions,
\begin{align}\label{Eq:eigendistribution}&\!f(t,x,\mu)\!\sim\!\!\frac{\delta(x\!-\!x(t)\!)}{x(t)^{2}}\!\sum\limits_{k=0}^{\infty}\biggl[a_{k}h_{k}^{\!(i)}\!e^{\!\!-\!\!\int\limits_{0}^{t}\!\!\lambda_{k}^{\!(i)}\!\nu dt}\!\!\!\!\!\!\!\!\!\!\!+\!b_{k}h_{k}^{\!(ii)}\!e^{\!\!-\!\!\int\limits_{0}^{t}\!\!\lambda_{k}^{\!(ii)}\! \nu dt}\!\biggr]\!\!\!\!\end{align}
All that remains now is to determine our coefficient functions $a_{k}(\mu_{b})$ and $b_{k}(\mu_{b})$. If we start with a confined delta shell in velocity, uniformly distributed in angle, then our initial condition is,
\begin{align}\label{Eq:initial-condition}
f(0,x,\mu) \!=\! \hat{n}_{0}\frac{\delta(x\!-\!x_{0})\!}{x_{0}^{2}}\frac{H(\mu_{b}\!-\!|\mu|)}{2\mu_{b}}
\end{align}
where $H(x)$ is the Heaviside step function. However, if we blindly match our coefficients with Eq.~(\ref{Eq:initial-condition}), we find that our solution does not conserve particles. This is because as $\mu_{b}(x)$ changes, the eigenfunctions themselves evolve and their coefficients must evolve with them. Our $a_{k}$ and $b_{k}$ are functions of $\mu_{b}(x)$ and require more information to be fully determined. If we further impose that, in the absence of pitch-angle scattering, particles are conserved and the distribution remains uniform, we have,
\begin{align}\label{Eq:conservation}\lim_{\nu\rightarrow0}f(t,x,\mu) = \hat{n}_{0}\frac{\delta(x\!-\!x(t)\!)\!}{x(t)^{2}}\frac{H(\mu_{b}\!-\!|\mu|)}{2\mu_{b}}\end{align}
The difference between fixing our coefficients with Eq.~(\ref{Eq:initial-condition}) and letting them evolve according to Eq.~(\ref{Eq:conservation}) is a factor of $\mu_{b}(x_{0})/\mu_{b}(x)$, precisely diluting the distribution to compensate for expansion of our domain in $\mu$, thereby preventing the spurious creation of particles.\\
\indent Then, defining $z \!\equiv\! \sin^{\!-1}\!\!\mu/\sin^{\!-1}\!\!\mu_{b}$ and renormalizing $a_{k},b_{k} \rightarrow \hat{a}_{k},\hat{b}_{k}$ by $\hat{n}_{0}/2\mu_{b}$ we can match our coefficients between Eq.~(\ref{Eq:eigendistribution}) and Eq.~(\ref{Eq:conservation}) via,
\begin{align}\label{Eq:matching}\sum_{k=0}^{\infty}\left[\hat{a}_{k}\cos\!\left(\!(k\!+\!1/2)\pi z\right)+\hat{b}_{k}\sin\!\left(k\pi z\right)\right] = \phi(z)\end{align}
where since we do not have a constant term available for matching on the LHS, we periodically extend the RHS beyond our interval to be the even square wave $\phi(z)$ with average value zero,
\begin{align}
\phi(z)\!\equiv\! \begin{cases}1, \,z\!\in\![-1,1]\quad \phi(z\!+\!4n) = \phi(z)\\
\!-1, \,z\!\in\!(1,3]\quad\quad\quad\forall\,\,n\in\mathbb{Z}\end{cases}\!\!\!\!\!\!\!\!\!\!\!\!\!\!\!\!\!\!\!\!\!\!\!\!\!\!\!\!\!\!\!\!\!\!\!\!\!\!\!\!\!\!\!\!\!\!\!\!\!\!\!\!\!\!,\quad\quad\quad\quad\quad\quad\quad
\end{align}
Since our $\phi(z)$ is now even, we can immediately discard all the $b_{k}$. The $a_{k}$ can be found via Fourier's trick,
\begin{align}\label{Eq:coefficients}a_{k} =\! \frac{\hat{n}_{0}}{2\mu_{b}}\!\int_{0}^{2}\!\!\!\phi(z)\cos\left(\!(k\!+\!1/2)\pi z\right)dz,\,\,\,\, b_{k} = 0\end{align} Since only $a\neq 0$, our eigenvalues and eigenfunctions of interest are,
\begin{align}\label{Eq:eigenfunctions}\lambda_{k} \equiv \frac{\!(k\!+\!1/2)^{2}\pi^{2}\!\!\!}{(\sin^{\!-1}\!\!\mu_{b})^{2}},\quad\!h_{k}(\mu|\mu_{b}) \!\sim\! \cos\!\left(\!\!\sqrt{\!\lambda_{k}}\sin^{\!-1}\!\!\mu\!\right)\end{align}
\begin{figure}
\centering
\noindent\includegraphics[width=\columnwidth,height=0.7\columnwidth]{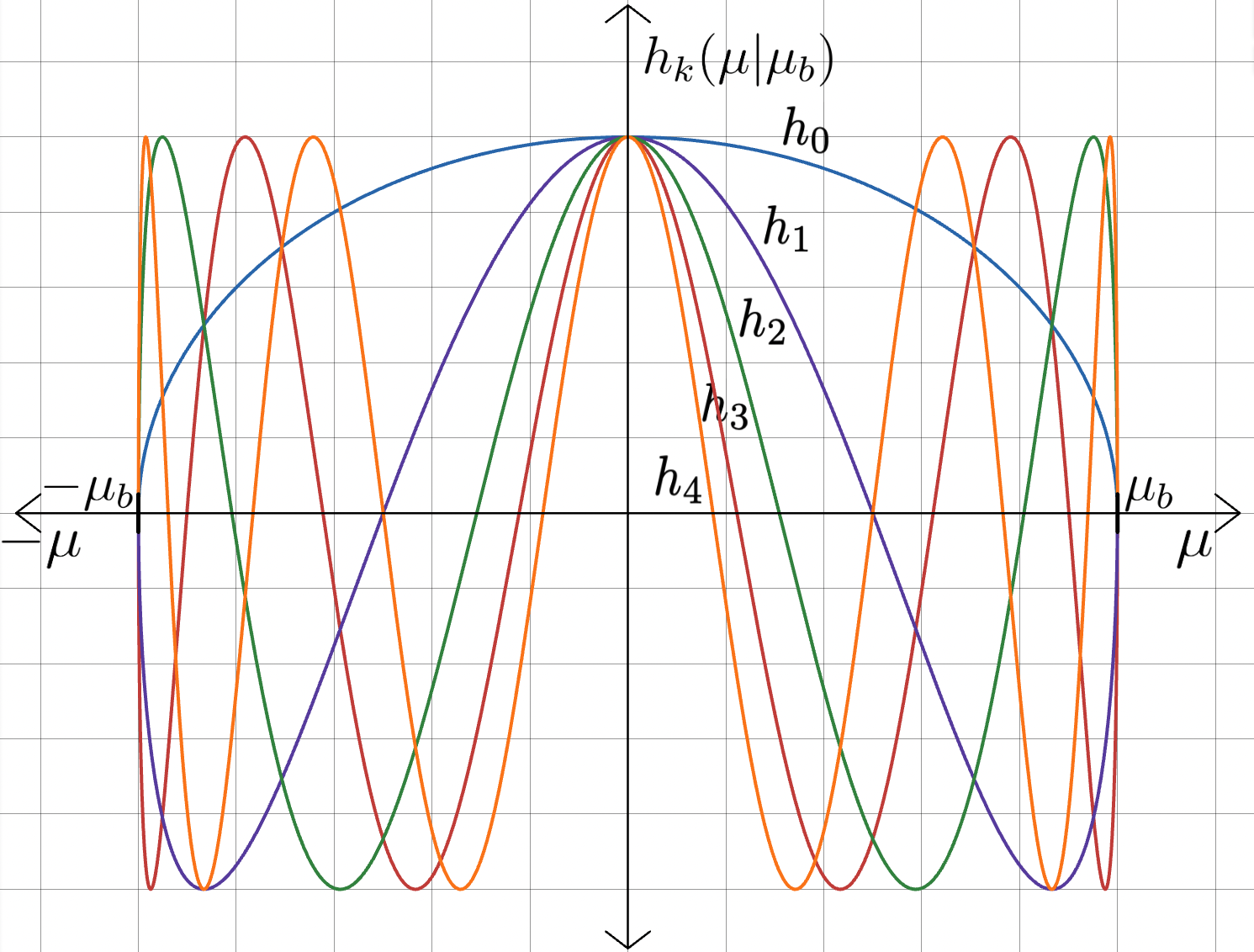}
\caption{Asymptotic Eigenfunctions of the Legendre Operator. Shown here are the first five even orthogonal eigenfunctions of the Legendre operator with homogeneous boundary conditions in the Geometrical Optics WKB approximation (Eq.(\ref{Eq:eigenfunctions})\!).\label{Fig:eigenfunctions}}
\end{figure}
\noindent \!\!as shown in Figure~\ref{Fig:eigenfunctions}. Substituting our eigenfunctions (Eq.~(\ref{Eq:eigenfunctions})\!) and coefficients (Eq.~(\ref{Eq:coefficients})\!) into our spectral decomposition (Eq.~(\ref{Eq:eigendistribution})\!), we obtain the Green's function solution for the distribution function,
\begin{align}&\label{Eq:f(t,x,mu)}f(t,x,\mu)\!\sim\!\hat{n}_{0}\frac{\delta(x\!-\!x(t)\!)}{x(t)^{2}}\frac{H(\mu_{b}\!-\!|\mu|)}{2\mu_{b}}\!\sum\limits_{k=0}^{\infty}\!\frac{2(-1)^{k}}{(k\!+\!1/2)\pi}\nonumber\\
&\cdot\cos\!\left(\!\!\sqrt{\!\lambda_{k}}\sin^{\!\!-1}\!\!\mu\!\right)\exp\!\left(\!\!-\!\!\int\limits_{0}^{t}\!\!\lambda_{k}(\mu_{b}(x(t)\!)\!)\nu(x(t)\!)dt\!\!\right)\end{align}
which neatly encapsulates all the information about alpha particle losses. Its time evolution is shown in Figure~\ref{Fig:f(t,mu)}. It is worth noting that the above procedure could be repeated for any desired initial distribution in pitch-angle. One has only to rematch the Fourier coefficients $\hat{a}_{k}$ and $\hat{b}_{k}$ to the new function $\phi(z)$ in Eq.~(\ref{Eq:matching}).
\begin{figure}
\centering
\noindent\includegraphics[width=\columnwidth]{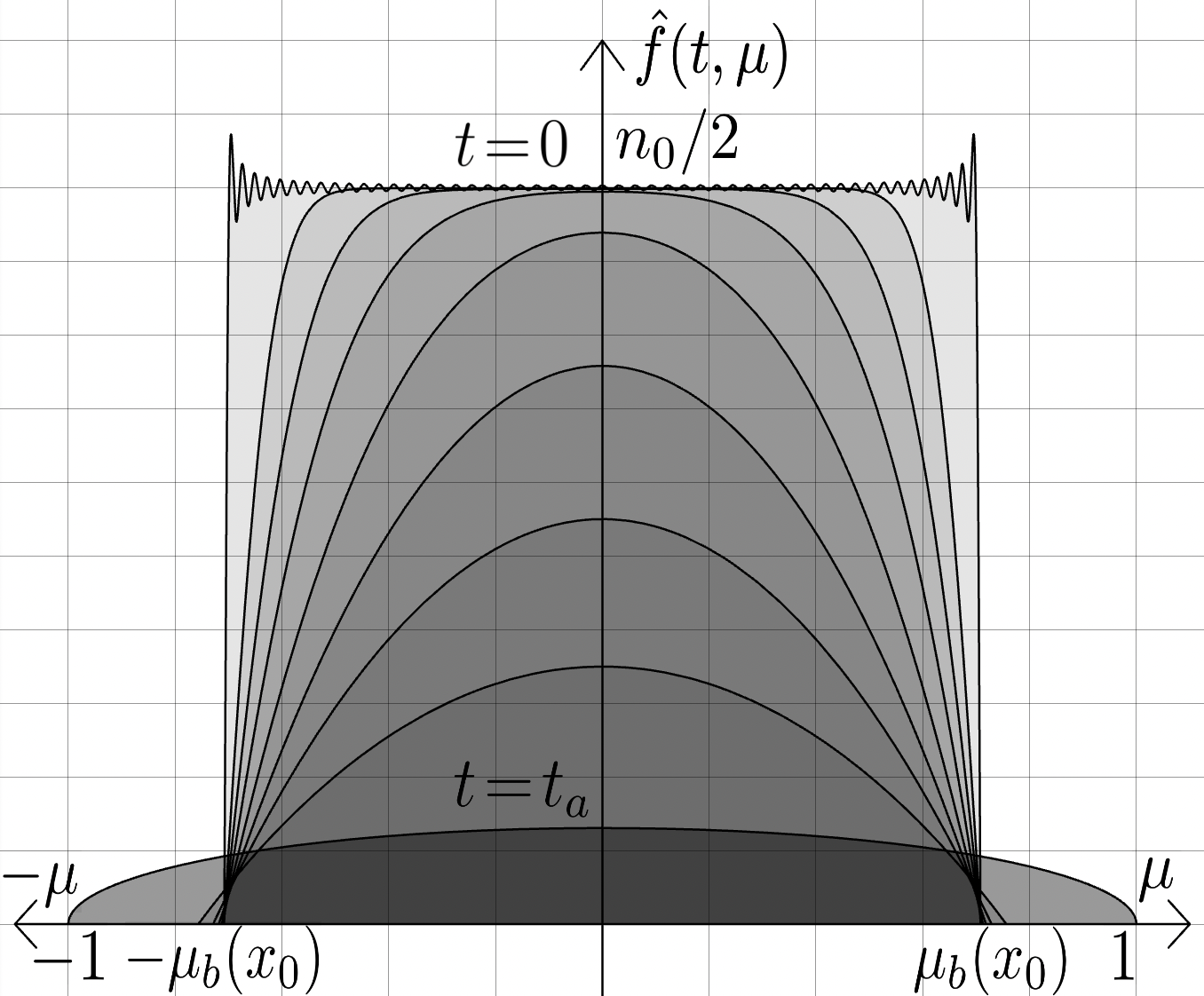}
\caption{Evolution of the Angular Distribution $\hat{f}(t,\mu)\equiv x^{2}f(t,x,\mu)/\delta(x-x(t)\!)$ vs $\mu$. Here we see the uniform distribution normalized to $\hat{n}_{0}$ diffuse in pitch-angle while obeying homogeneous boundary conditions at $\mu \!=\! \pm\mu_{b}(x(t)\!)$ over times $t\!\in\![0,t_{a}]$ where $x(t_{a})\equiv x_{a}$ (Eq.~(\ref{Eq:ta})\!), for $x_{0} = 1, x_{a}=0.1$, $R=2$.\label{Fig:f(t,mu)}}
\end{figure}
\vspace{-5pt}
\section{VII. Number Density Eigenmodes}
\begin{figure}
\centering
\noindent\includegraphics[height = 0.78\columnwidth,width=\columnwidth]{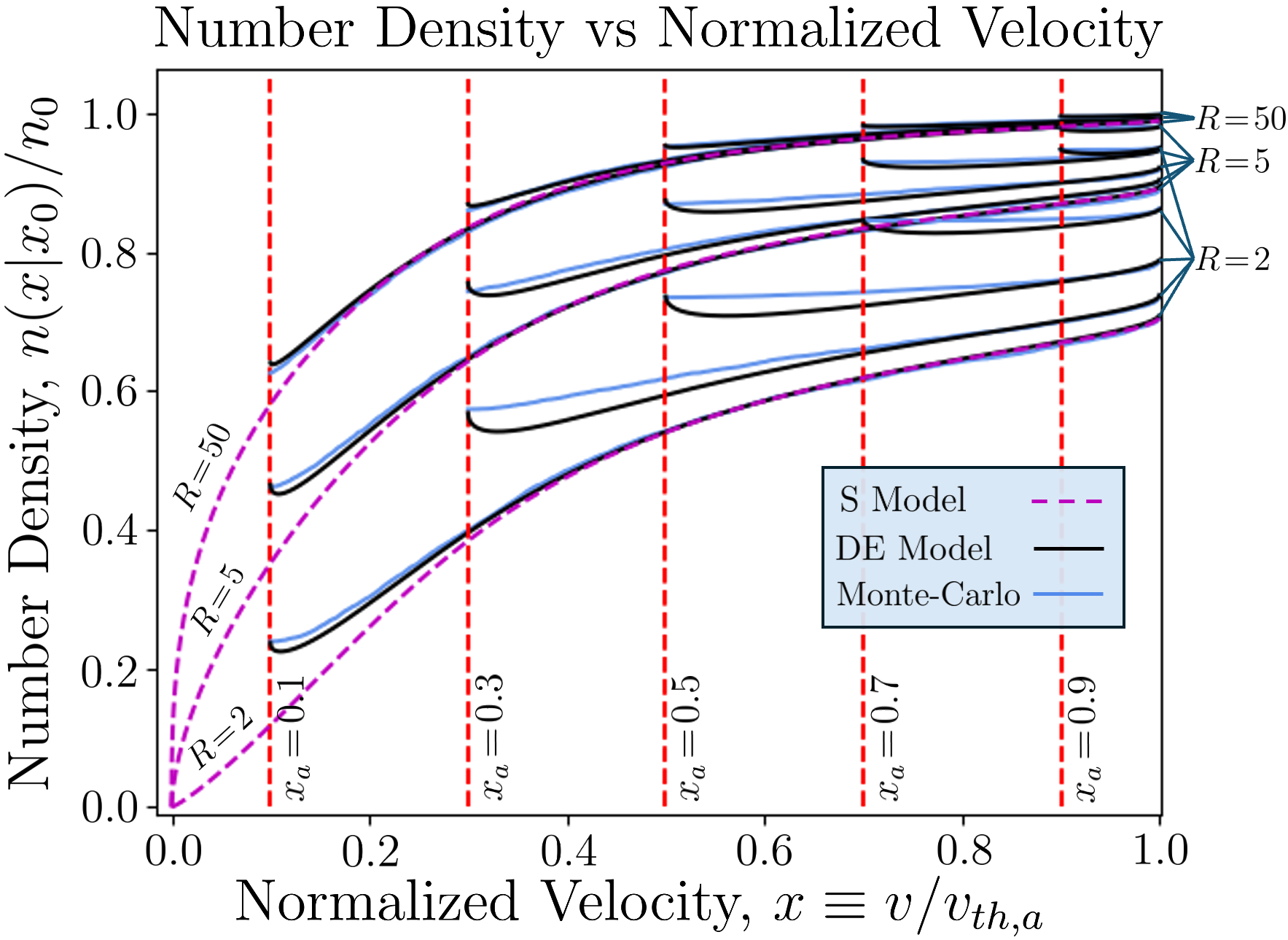}
\caption{Shell Number Density vs Normalized Velocity. Shown here are the remaining number densities $n(x|x_{0})$ for an isotropic delta shell distribution starting at $x = x_{0}$ as projected by the S and DE models to $500$ terms together with the Monte-Carlo simulation results for a DT scenario with $n_{0} \!=\! 10^{5}, x_{0} \!=\! 1$ for $(x_{a},R) \in \{0.1,0.3,0.5,0.7,0.9\}\times\{2,5,50\}$.\label{Fig:n(x|x0)}}
\end{figure}

\begin{figure}
\centering
\noindent\includegraphics[height = 0.78\columnwidth,width=\columnwidth]{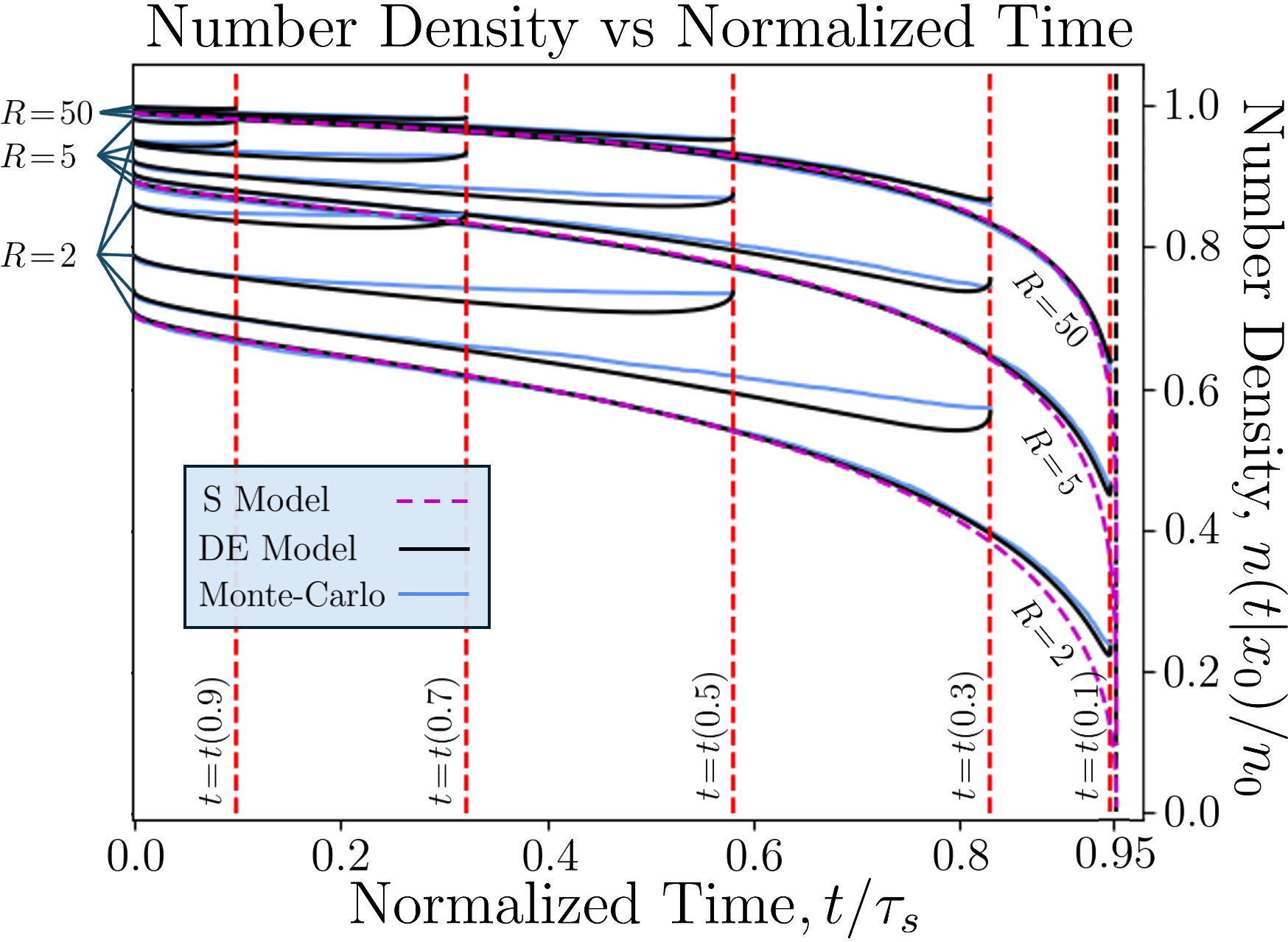}
\caption{Shell Number Density vs Normalized Time. Shown here are the remaining number densities $n(t|x_{0})$ for an isotropic delta shell distribution starting at $x = x_{0}$ as projected by the S and DE models to $500$ terms together with the Monte-Carlo simulation results for a DT scenario with $n_{0} \!=\! 10^{5}, x_{0} \!=\! 1$ for $(x_{a},R) \in \{0.1,0.3,0.5,0.7,0.9\}\times\{2,5,50\}$.\label{Fig:n(t|x0)}}
\end{figure}

\vspace{-1pt}
\label{sec:eigenmodes}
At any given time, our remaining density is simply,
\begin{align}n(t) \!\equiv\! \!\smashoperator[r]{\int\limits_{0}^{\infty}}\!\!\!\smashoperator[r]{\int\limits_{\,-\mu_{b}(x)}^{\,\mu_{b}(x)}} \!\!f(t,x,\mu)x^{2}d\mu\,dx
\end{align}
which yields,
\begin{align}\label{Eq:n(t|x0)}n(t|x_{0}) \!\sim\!\!\sum\limits_{k=0}^{\infty}n_{0}^{\!(k)}e^{\!\!-\!\int\limits_{0}^{t}\!\!\lambda_{k}\nu dt}\!,\,\,n_{0}^{\!(k)}\!\equiv\!\frac{\hat{n}_{0}}{\lambda_{k}\!\!-\!1}\frac{2\sqrt{1\!-\!\mu_{b}^{2}}}{\mu_{b}\sin^{\!-1}\!\!\mu_{b}}\!\!\!\!\end{align}
where we've written $x\equiv x(t), \mu_{b}\equiv\mu_{b}(x)$, $\lambda_{k}\equiv\lambda_{k}(\mu_{b})$, $\nu \equiv \nu(x)$ for brevity.
Then we can interpret our result as the distribution being composed of an infinite series of eigenmode populations $n_{0}^{(k)}\!(\mu_{b})$ with `decay' or loss rates equal to the time-integrated product of their corresponding eigenvalue and the pitch-angle scattering frequency. If we now use our characteristic bijection $t(x) \!\leftrightarrow\! x(t)$ (Eq.~(\ref{Eq:x(t)})\!), we obtain,
\begin{align}\label{Eq:n(x|x0)}n(x|x_{0}) \!\sim \!\sum\limits_{k=0}^{\infty}n_{0}^{\!(k)}\!\exp\!\!\left(\!\!-\!\!\int\limits_{x}^{x_{0}}\!\!\lambda_{k}\frac{Z_{\perp}\!(x)}{Z_{\parallel}(x)}\frac{dx}{2x}\!\!\right)\end{align}
In the zero potential limit $\Phi_{a} = 0$, our trapping boundary (Eq.~(\ref{Eq:hyperbola})\!) becomes $\mu_{b}(x) = \mu_{b0} \equiv \sqrt{1\!-\!1/R}$,
and in the case where $Z_{\perp}(x)\sim Z_{\perp}^{i}$ which is almost always true, we obtain the simple scaling relation,
\begin{align}\label{Eq:n(x|x0)Phi=0} n(x|x_{0}) \!\sim\!\!\sum\limits_{k=0}^{\infty}\!n_{0}^{\!(k)}\!(\mu_{b0}\!)\!\!\left(\!\frac{x^{3}}{x_{0}^{3}}\frac{x_{0}^{3}\!+\!\eta^{3}}{x^{3}\!+\!\eta^{3}}\!\right)^{\!\beta_{k}/6}\!\!\!\!\!\!\!\!\!\!\!,\,\beta_{k}\!\equiv\!\lambda_{k}(\mu_{b0})\frac{Z_{\perp}^{i}}{Z_{\parallel}^{i}}\!\!\end{align}
similar to that obtained by Egedal et al.~\cite{Egedal} Eq.~(14-15) but now with explicit forms for the eigenmode populations (Eq.~(\ref{Eq:n(t|x0)})\!), eigenvalues (Eq.~(\ref{Eq:eigenfunctions})\!), and full set of critical exponents $\beta_{k}$ (Eq.~(\ref{Eq:n(x|x0)Phi=0})\!). Correspondingly, the remaining density, as a function of time, scales as,
\begin{align}\label{Eq:n(t|x0)Phi=0}n(t|x_{0})\sim \!\sum\limits_{k=0}^{\infty}n_{0}^{\!(k)}\!(\mu_{b0}\!)\!\!\left(\!1\!-\!\frac{\eta^{3}}{x_{0}^{3}}\!\left(e^{3Z_{\parallel}^{e}t/\tau_{0}^{i}}\!-\!1\right)\!\!\right)^{\!\beta_{k}/6}\!\!\end{align}
The relative importance of drag and pitch-angle scattering can therefore be roughly evaluated by examining the `confinement parameter',
\begin{align}\zeta\equiv\frac{Z_{\perp}^{i}/Z_{\parallel}^{i}}{\left(\sin^{\!-1}\!\!\mu_{b0}\right)^{2}}\propto \beta_{k}\end{align}
with higher $\zeta$ corresponding to increasing dominance of scattering over drag. The lower this parameter, the more particles are eventually trapped for a given potential. Our models are thus,
\begin{align*}&\text{Dynamic Eigenmode (DE)} \rightarrow \text{Eqs.~(\ref{Eq:n(t|x0)},\ref{Eq:n(x|x0)})}\\
&\text{Basic Scaling (S)}\!\rightarrow \text{Eqs.~(\ref{Eq:n(x|x0)Phi=0},\ref{Eq:n(t|x0)Phi=0})}\end{align*}
The predicted remaining number densities $n(x|x_{0})$ and $n(t|x_{0})$ are displayed in Figures~\ref{Fig:n(x|x0)},\ref{Fig:n(t|x0)} along with the Monte-Carlo results. We observe outstanding agreement, particularly for $x >> x_{a}$.\\
\indent For the DE model (Eqs.~(\ref{Eq:n(t|x0)},\ref{Eq:n(x|x0)})\!), we observe a small non-physical rise in $n(x|x_{0})$ (Figure~\ref{Fig:n(x|x0)}) as $x \rightarrow x_{a}$ due to brief violation of slowly varying $\mu_{b}(x)$, asymptotic breakdown of our dominant balance (Eq.~(\ref{Eq:dominantbalance})\!) as $\mu_{b} \rightarrow 1$, and at small $x$, excessive $-1/2x^{2}$ contribution to $Z_{\perp}(x)$, representing breakdown of our limit of the Landau collision operator (Eq.~(\ref{Eq:scalarcollisionoperator})\!). There is also a very slight error associated with approximating the infinite sums to a finite number of terms. In any case, the breakdown is only substantial at small $x \lesssim 0.1$ where our limit of the Landau collision operator is no longer valid. The typically minute increase can therefore be dismissed as a normal symptom of an asymptotic model.\\ \indent If one desires a strictly non-increasing function, one can simply infer $n(t|x_{0}),n(x|x_{0})$ to be constant beyond their minimum value. Alternatively, we could naively evaluate our amplitudes at $n_{0}^{(k)}(\mu_{b}(x_{0})\!)$ as for the basic scaling.\\ \\
\vspace{-30pt}
\section{VIII. Loss Probability Distributions}
\label{sec:pdfs}
\begin{figure}
\centering
\noindent\includegraphics[height = 0.8\columnwidth,width=\columnwidth]{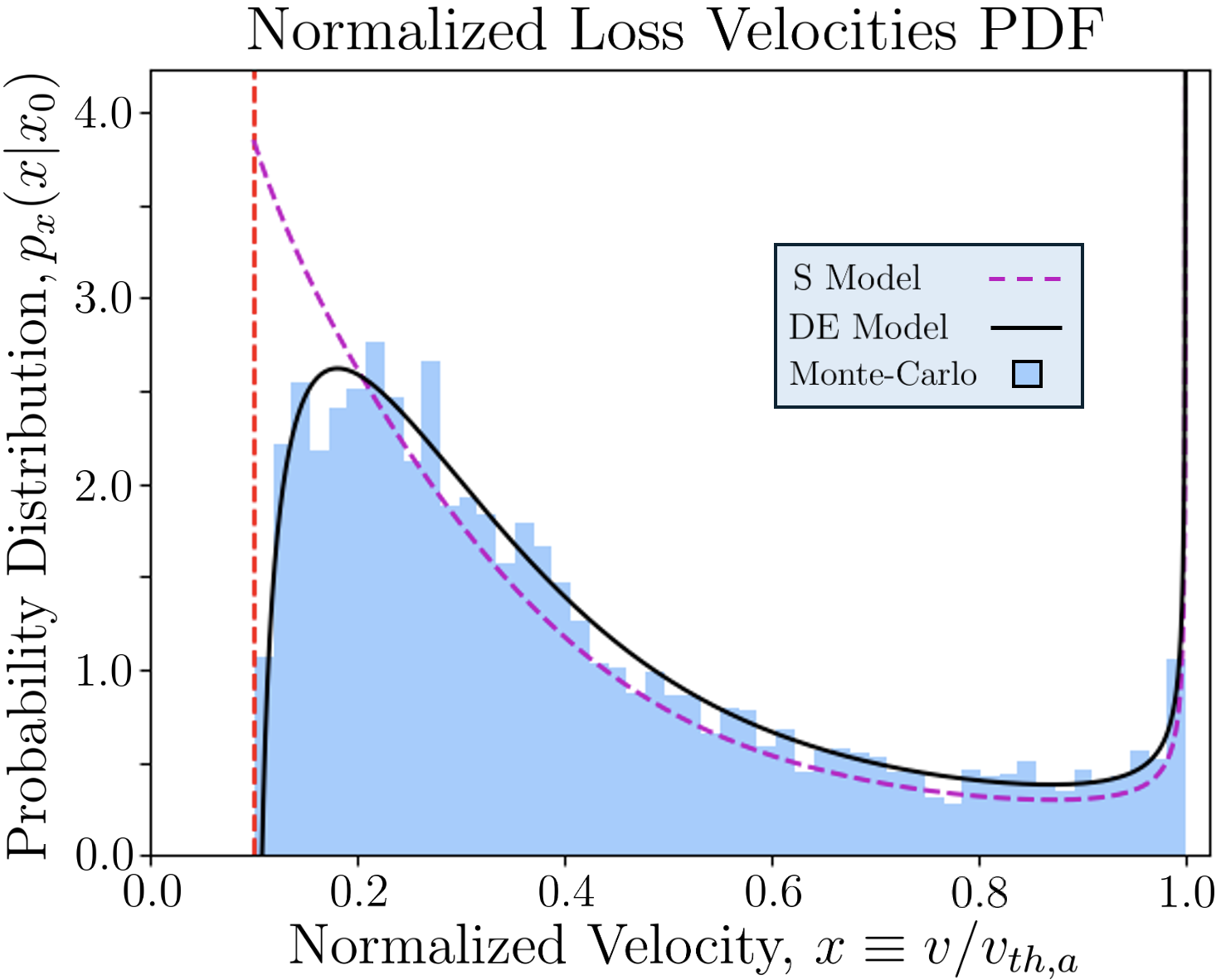}
\caption{Normalized Loss Velocities PDF. Shown here are the normalized loss velocity PDFs $p_{x}(x|x_{0})$ (Eqs.~(\ref{Eq:px-DE},\ref{Eq:px-S})\!) for an isotropic delta shell distribution starting at $x \!=\! x_{0}$ as projected by the S and DE models to 500 terms, normalized on the interval $x\in[x_{a},x_{0}]$, together with the Monte-Carlo simulation results for a DT scenario with $n_{0} \!=\! 10^{5}, x_{0} \!=\! 1, x_{a} \!=\! 0.1, R \!=\! 5$.\label{Fig:p(x|x0)}}
\end{figure}

\begin{figure}
\centering
\noindent\includegraphics[height = 0.8\columnwidth,width=\columnwidth]{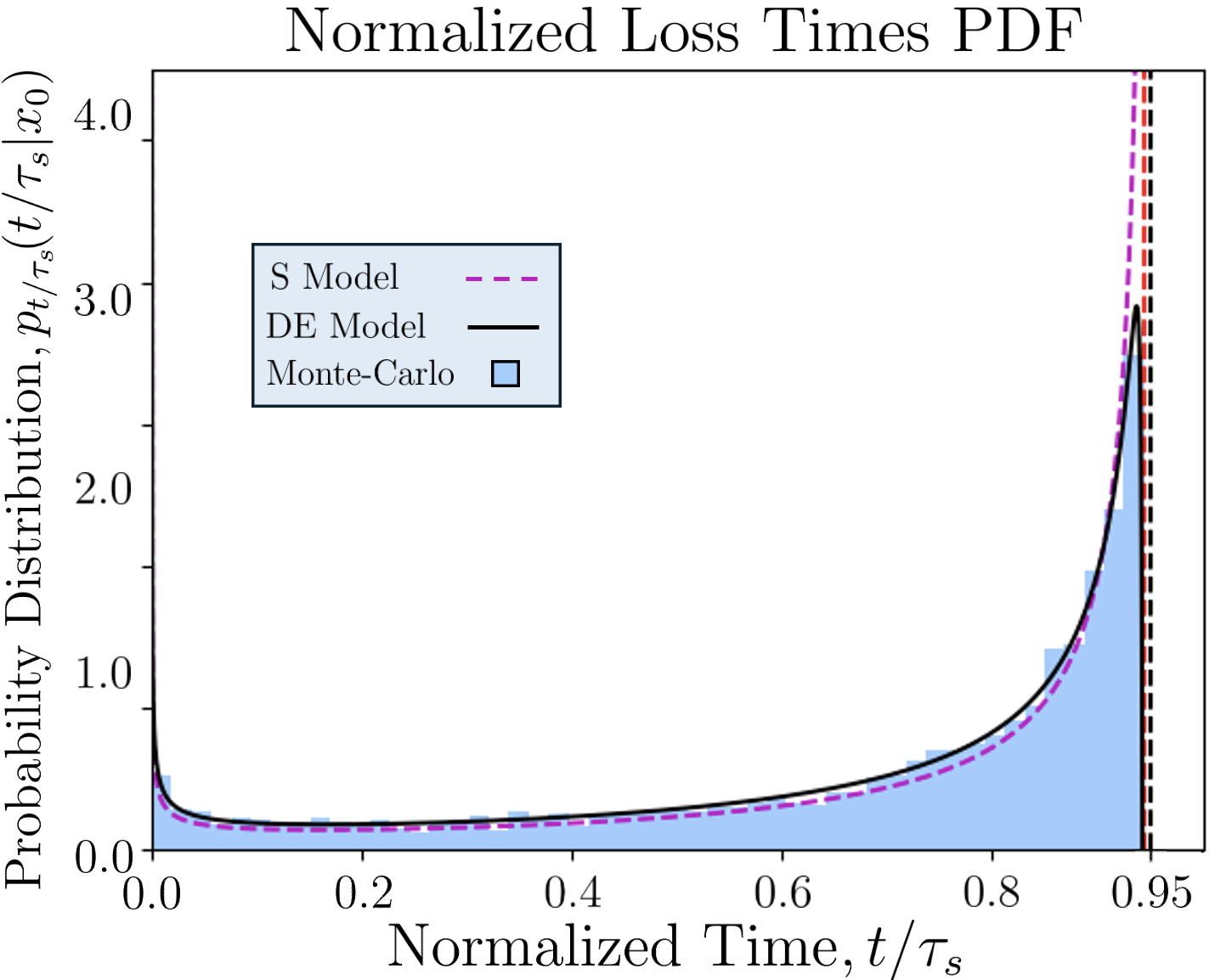}
\caption{Normalized Loss Times PDF. Shown here are the normalized loss time PDFs $p_{t}(t|x_{0})$ (Eq.~(\ref{Eq:PDFs})\!) for an isotropic delta shell distribution starting at $x = x_{0}$ as projected by the S and DE models to 500 terms, normalized on the interval $x\in[x_{a},x_{0}]$, together with the Monte-Carlo simulation results for a DT scenario with $n_{0} = 10^{5}, x_{0} = 1, x_{a} = 0.1, R =\! 5$.\label{Fig:p(t|x0)}}
\end{figure}

We can now compute the probability distribution functions (PDFs) of normalized velocities $x$ at which alpha particles become deconfined. We are concerned here only with those particles that leave gradually via pitch-angle scattering, not the substantial fraction that are born detrapped, nor those that ultimately become trapped by the potential. We therefore consider $x \in [x_{a},x_{0}]$. The distribution of $x$ values at which particles leave is of course proportional to how many particles are leaving when the distribution is at a given value of $x$. We normalize our distribution to unity such that it is a proper~PDF,
\begin{align}\label{Eq:p(x|x0)}p_{x}(x|x_{0}) = \frac{dn/dx}{\hat{n}_{0}\!-\!n(x_{a}|x_{0})}, \,\, x\in[x_{a},x_{0}]\end{align}
For the general model (DE) this becomes,
\begin{align}\label{Eq:px-DE}&p_{x}(x|x_{0}) \!\sim\! \frac{\hat{n}_{0}}{\hat{n}_{0}\!-\!n(x_{a}|x_{0})}\!\sum\limits_{k=0}^{\infty}\!\frac{\lambda_{k}}{\lambda_{k}\!-\!1}\!\Biggl[\!\frac{\sqrt{1\!-\!\mu_{b}^{2}}}{\mu_{b}\sin^{\!-1}\!\!\mu_{b}}\frac{Z_{\perp}\!(x)}{xZ_{\parallel}(x)}\!\!\!\\
&\!\!+\!\!\left(\!\!\frac{\lambda_{k}\!+\!1}{\lambda_{k}\!\!-\!1}\!-\!\frac{\sin^{\!-1}\!\!\mu_{b}}{\mu_{b}\!\sqrt{1\!-\!\mu_{b}^{2}}}\!\right)\!\!\frac{2\mu_{b}'/\mu_{b}}{\pi^{2}(k\!+\!1/2)^{2}}\!\!\Biggr]\!\!\exp\!\!\left(\!\!-\!\!\!\int\limits_{x}^{x_{0}}\!\!\lambda_{k}\frac{Z_{\perp}\!(x)}{Z_{\parallel}(x)}\frac{dx}{2x}\!\!\right)\nonumber\end{align}
while our basic scaling (S) yields,
\begin{align}\label{Eq:px-S}p_{x}(x|x_{0})\!\sim\!\sum_{k=0}^{\infty}\!\frac{n_{0}^{\!(k)}\!(\mu_{b0}\!)}{\hat{n}_{0}}\frac{\beta_{k}\eta^{3}\!/2x}{x^{3}\!+\!\eta^{3}}\!\!\left(\!\frac{x^{3}}{x_{0}^{3}}\frac{x_{0}^{3}\!+\!\eta^{3}}{x^{3}\!+\!\eta^{3}}\!\right)^{\!\beta_{k}/6}\!\!\!\end{align}
Every other PDF is just a change of variables away now via $p_{y}(y) = p_{x}(x(y))|dx/dy|$.
\begin{equation}\begin{aligned}\label{Eq:PDFs}
p_v(v|x_{0}) = \frac{p_{x}(v/v_{th,a}|x_{0})}{v_{th,a}}, \, v \in [v_{a},v_{th,a}x_{0}]\\
p_E(E|x_{0}) = \frac{p_{x}(\sqrt{E/E_{th,a}}|x_{0})}{2\sqrt{EE_{th,a}}}, \, E \in [\Phi_{a},E_{th,a}x_{0}^{2}]\\
p_{t}(t|x_{0}) = \frac{Z_{\parallel}(x)}{\tau_{0}^{i}x^{2}}p_{x}(x|x_{0}), \, t \in [0,t_{a}]
\end{aligned}\end{equation}
where $x = x(t)$ for the PDF in time and,
\begin{align}\label{Eq:ta}x(t_{a})\equiv x_{a}\rightarrow t_{a}(x_{0})\equiv\frac{\tau_{s}}{3}\ln\!\left(\frac{x_{0}^{3}\!+\!\eta^{3}}{x_{a}^{3}\!+\!\eta^{3}}\right).\end{align}
These PDFs are given explicitly for the basic scaling in Appendix \hyperref[App:H]{H} as an example. The normalized velocity distributions $p_{x}(x|x_{0})$ and $p_{t}(t|x_{0})$ are displayed in Figures~\ref{Fig:p(x|x0)},\ref{Fig:p(t|x0)} along with the Monte-Carlo results. We observe outstanding agreement, particularly for $x >> x_{a}$.\\
\indent Now recall that these probability distributions (Eq.~(\ref{Eq:p(x|x0)}-\ref{Eq:PDFs})\!) were derived merely for a shell of particles originating at $x = x_{0}$, or in other words for a given value of $x_{0} \in [x_{a},\infty)$. We can understand then, that these are really conditional probability distributions and extend our definitions to understand what happens in a more general plasma where particles begin with a distribution of $x_{0}$ values. Then we may write,
\begin{align}\label{Eq:jointpdf}p_{x,x_{0}}(x,x_{0}) \!=\! p_{x}(x|x_{0})p_{x_{0}}\!(x_{0})\end{align}
to obtain the joint probability distribution. Furthermore, we can obtain the overall spectra of all lost particles by integrating over our distribution of initial normalized velocities, and compute expectations for any quantity $q(t,x,v,E) = q(x)$.
\begin{align}\label{Eq:expectations}p_{x}(x) \!\equiv \!\!\smashoperator[r]{\int\limits_{0}^{\infty}}\!\!p_{x,x_{0}}(x,x_{0})dx_{0},\langle q(x) \rangle^{\!(ii)} \!\!\equiv\!\!\int\limits_{0}^{\infty}\!\!q(x)p_{x}(x)dx\!\!\!\!\!\end{align}
\indent We have built up a highly generalized framework for analyzing the loss spectra of fast ion species in a magnetic mirror. Once the Green's function solution for the distribution function (Eq.~(\ref{Eq:f(t,x,mu)})\!) has been obtained, one can derive the remaining number densities (Eq.~(\ref{Eq:n(t|x0)},\ref{Eq:n(x|x0)})\!), determine the loss velocity and time distributions (Eqs.~(\ref{Eq:p(x|x0)},\ref{Eq:PDFs})\!) and extract all relevant information about fast ion losses of the plasma in question (Eqs.~(\ref{Eq:jointpdf},\ref{Eq:expectations})\!).
\vspace{-5pt}
\section{IX. Mean Loss Energy \& Time}
\label{sec:energyntime}
\vspace{-3pt}
Now, suppose we wanted to find the mean loss energy. Using Fubini's theorem to interchange the integrals, and omitting regions where $p_{x}(x|x_{0})p_{x_{0}}\!(x_{0}) = 0$,
\begin{align}\label{Eq:<E>ii}\langle E\rangle^{\!(ii)} \!= E_{th,a}\!\!\int\limits_{x_{a}}^{\infty}\!\!\int\limits_{x_{a}}^{\,x_{0}}\!x^{2}p_{x}(x|x_{0})p_{x_{0}}\!(x_{0})dx\,dx_{0}\end{align}
Similarly, we can compute the mean loss time using our bijection between $x(t)\leftrightarrow t(x)$ (Eq.~(\ref{Eq:x(t)})\!),
\begin{align}\label{Eq:<t>ii}&\langle t\rangle^{\!(ii)}\!\!=\!\frac{\tau_{s}}{3}\!\!\int\limits_{x_{a}}^{\infty}\!\!\int\limits_{x_{a}}^{\,x_{0}}\!\ln\!\left(\!\frac{x_{0}^{3}\!+\!\eta^{3}}{x^{3}\!+\!\eta^{3}}\!\right)\!p_{x}(x|x_{0})p_{x_{0}}\!(x_{0})dx\,dx_{0}\!\!\end{align}
The exact integrals in Eqs.~(\ref{Eq:<E>ii},\ref{Eq:<t>ii}) can easily be carried out numerically with a few lines of code to obtain reasonable estimates for the mean loss energy and time for any initial alpha particle distribution. \\
\indent An isotropic delta distribution in velocity space is a good approximation for the strongly peaked alpha particle birth spectrum of a DT fusion reaction. Integrating over solid angle to obtain $p_{v_{0}}\!(v_{0})$ and performing a change of variables $p_{x_{0}}\!(x_{0}) \!=\! p_{v_{0}}\!(v_{0})|dv_{0}/dx_{0}|$ we see that the following distributions are equivalent,
\begin{align}\label{Eq:alphadelta}f(\vec{v}_{0}) = \frac{\delta(v_{0}\!-\!v_{th,a})}{4\pi v_{0}^{2}} \leftrightarrow p_{x_{0}}\!(x_{0}) = \delta(x_{0}\!-\!1)\end{align}
We give plots of $\langle E\rangle^{(ii)}$ and $\langle t\rangle^{(ii)}$ vs key parameters $(x_{a},R)$ for a DT scenario in Figures~\ref{Fig:E vs xa,R},\ref{Fig:t vs xa,R}.
\begin{figure}
\centering
\noindent\includegraphics[height = 0.8\columnwidth,width=\columnwidth]{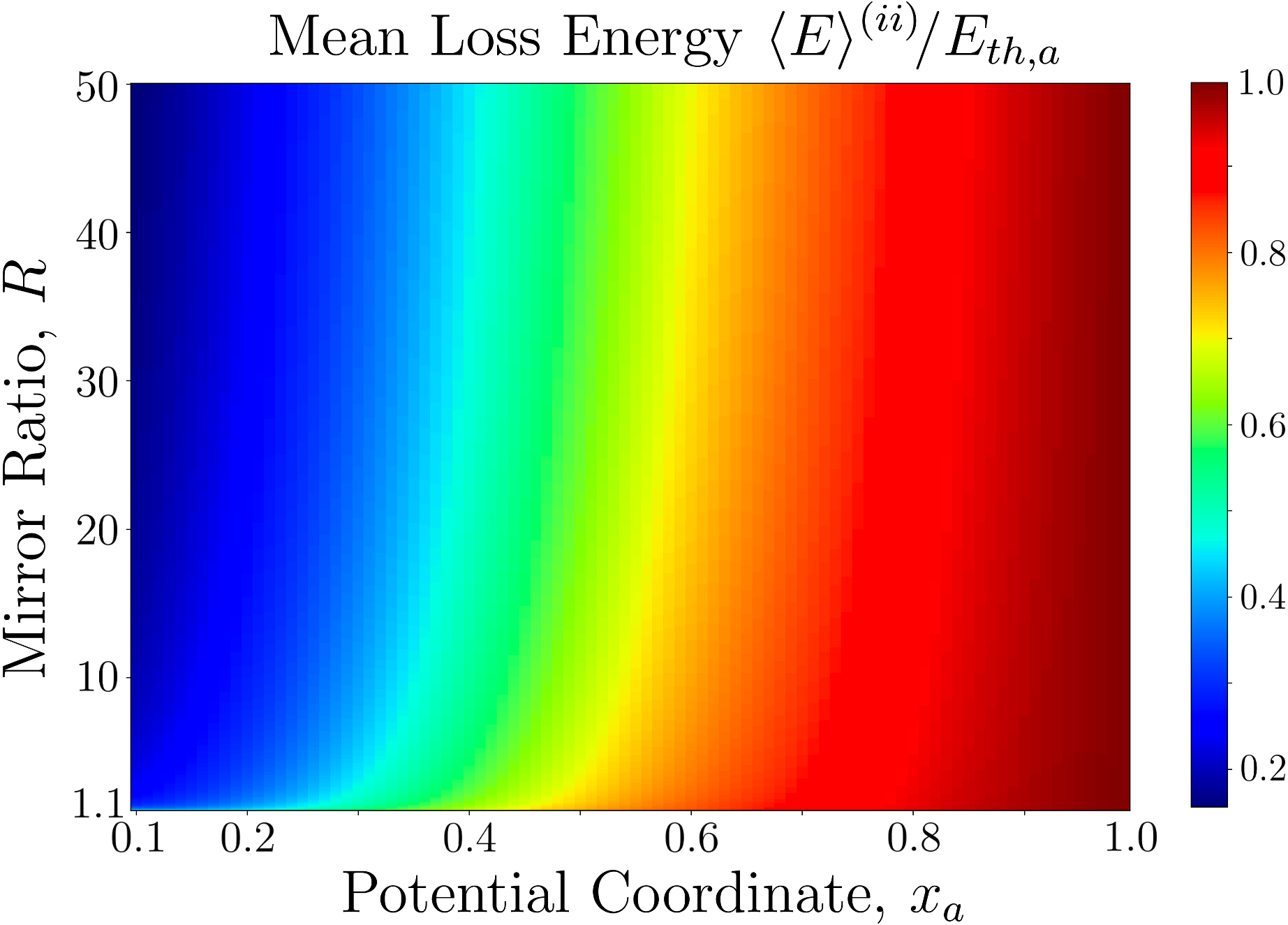}
\caption{Mean Loss Energy $\langle E\rangle^{\!(ii)}$ vs $x_{a},R$. Shown here are numerically integrated estimates of the mean loss energy (Eq.~(\ref{Eq:<E>ii})\!) based on the DE model to 500 terms vs the potential coordinate, $x_{a}$, and the mirror ratio, $R$, for a DT scenario.
\label{Fig:E vs xa,R}}
\end{figure}
\begin{figure}
\centering
\noindent\includegraphics[height = 0.8\columnwidth,width=\columnwidth]{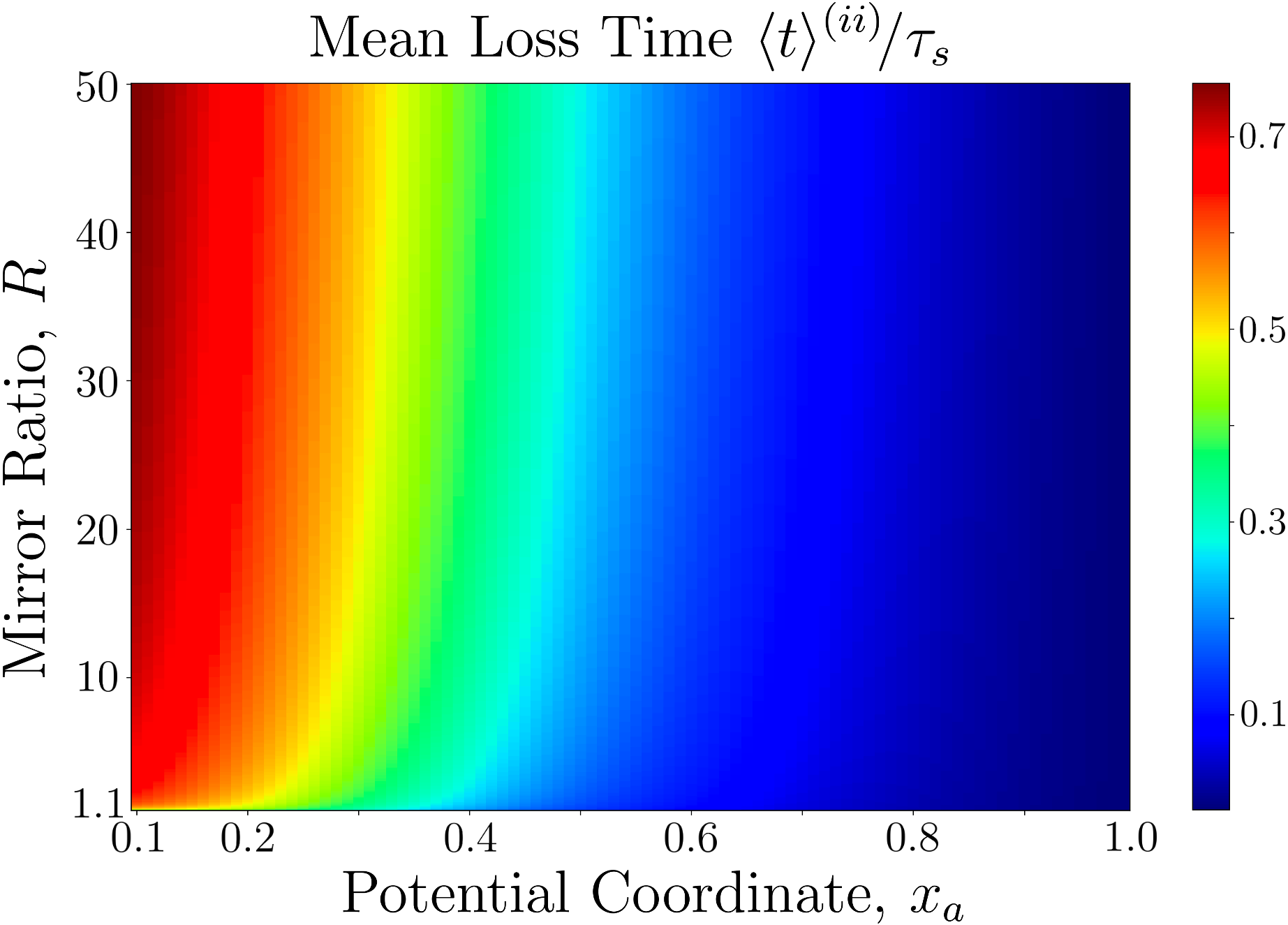}
\caption{Mean Loss Time $\langle t\rangle^{\!(ii)}$ vs $x_{a},R$. Shown here are numerically integrated estimates of the mean loss time (Eq.~(\ref{Eq:<t>ii})\!) based on the DE model to 500 terms vs the potential coordinate, $x_{a}$, and the mirror ratio, $R$, for a DT scenario.
\label{Fig:t vs xa,R}}
\end{figure}
\vspace{-5pt}
\section{X. Generalized Loss Fractions}
\label{sec:fractions2}
Similarly, we can now compute the loss fractions $\langle F_{l}^{(i-iii)}\rangle$ for an isotropically born population of alpha particles which are (i) never confined, (ii) gradually lost, and (iii) retained, not just for the delta shell described in Sec.\,\hyperref[sec:fractions]{III}, but for any initial velocity distribution $p_{x_{0}}(x_{0})$,
\begin{align}\langle F_{l}^{(i-iii)}\rangle \equiv \int\limits_{x_{a}}^{\infty}F_{l}^{(i-iii)}\!(x_{0})\,p_{x_{0}}\!(x_{0})\,dx_{0}\end{align}
\begin{figure}[hbt!]
\centering
\noindent\includegraphics[height = 0.922\columnwidth,width=\columnwidth]{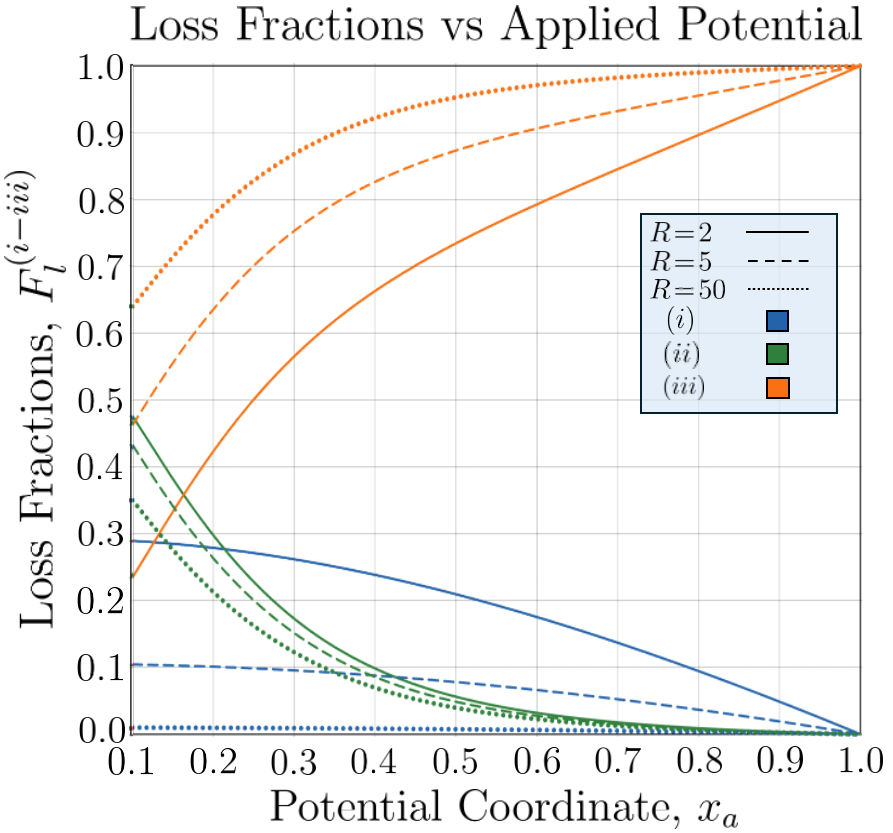}
\caption{Loss Fractions vs Applied Potential. Shown here are the loss fractions $F_{l}^{(i-iii)}$ (Eq.~(\ref{Eq:lossfractions})\!) of particles which are (i) never confined, (ii) gradually lost, and (iii) retained, as estimated by the DE model (Eq.~(\ref{Eq:n(x|x0)})\!) vs the potential coordinate, $x_{a}$, for mirror ratios $R = 2,5,50$, for a DT scenario. 
\label{Fig:lossfractions}}
\end{figure}
Considering again the example of a delta distribution of alphas (Eq.~(\ref{Eq:alphadelta})\!) for a DT scenario, we use the DE model (Eqs.~(\ref{Eq:n(t|x0)},\ref{Eq:n(x|x0)})\!) which seems to predict $n(x_{a}|x_{0})$ extremely well as shown in Figure~\ref{Fig:n(x|x0)} to obtain the predicted loss fractions $F_{l}^{(i-iii)}$ (Eq.~(\ref{Eq:lossfractions})\!) shown in Figure~\ref{Fig:lossfractions}. It is worth noting that producing simple plots like Figures~\ref{Fig:E vs xa,R},\ref{Fig:t vs xa,R},\ref{Fig:lossfractions} with numerical solutions would involve exceedingly high time-complexity. Each point $(x_{a},R)$ would require a full Monte-Carlo simulation with $n_{0}\sim 10^{5}$ particles. Using the DE model (Eqs.~(\ref{Eq:n(t|x0)},\ref{Eq:n(x|x0)})\!), even to 500 terms, is several orders of magnitude faster and significantly more straightforward.\\
\vspace{-15pt}
\section{XI. Steady-State Distribution}
\label{sec:steadystate}
\begin{figure}
\centering
\noindent\includegraphics[height = 0.7\columnwidth,width=\columnwidth]{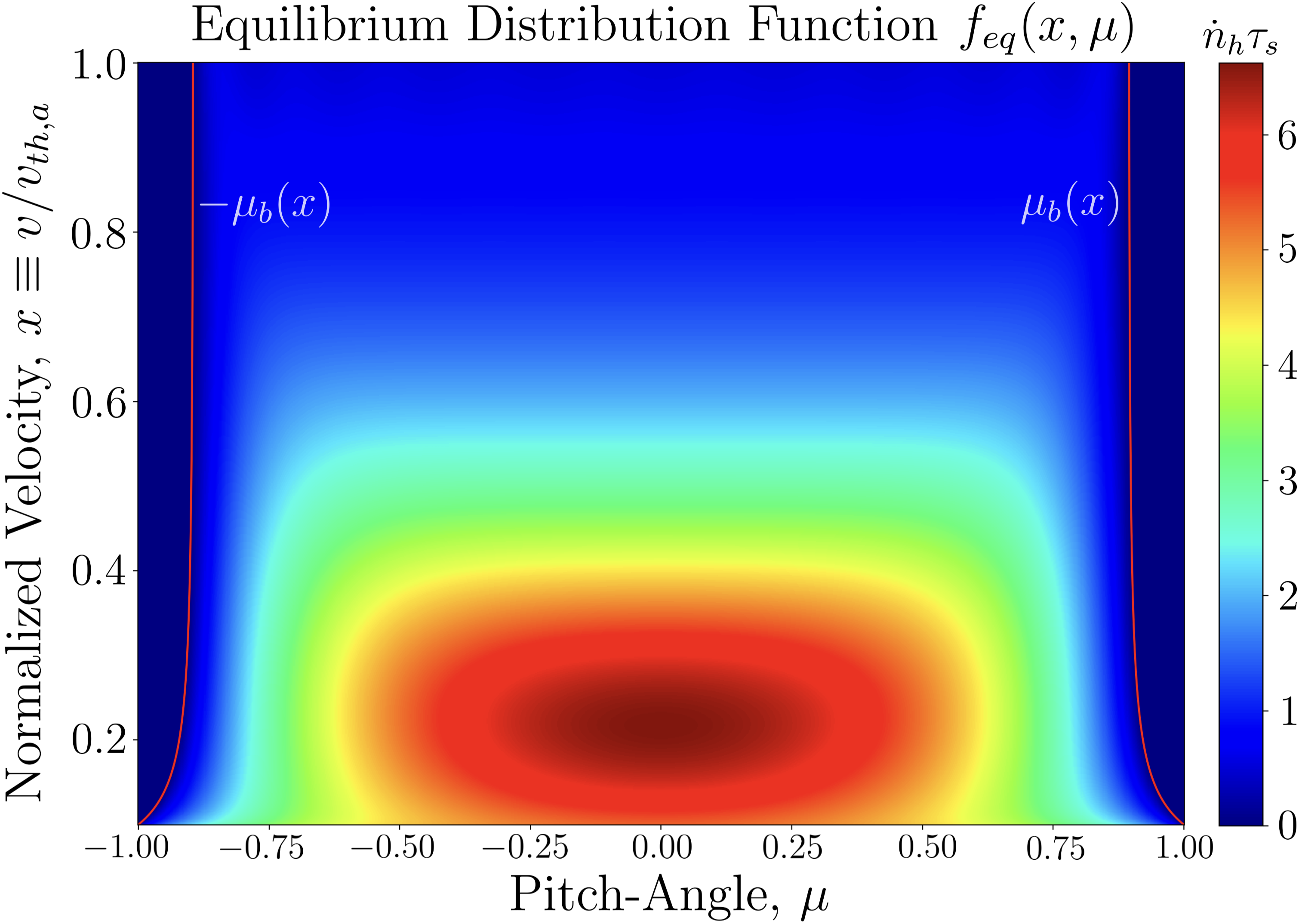}
\caption{Equilibrium Distribution Function $f_{eq}(x,\mu)$. Here we see the equilibrium distribution function (Eq.~(\ref{Eq:feq})\!) in $(x,\mu)$ coordinates for an isotropic confined delta source at $x_{0} = 1$ for $x_{a} = 0.1, R = 5$.\label{Fig:feq-square}}
\end{figure}

\begin{figure}
\centering
\noindent\includegraphics[height = 0.55\columnwidth,width=\columnwidth]{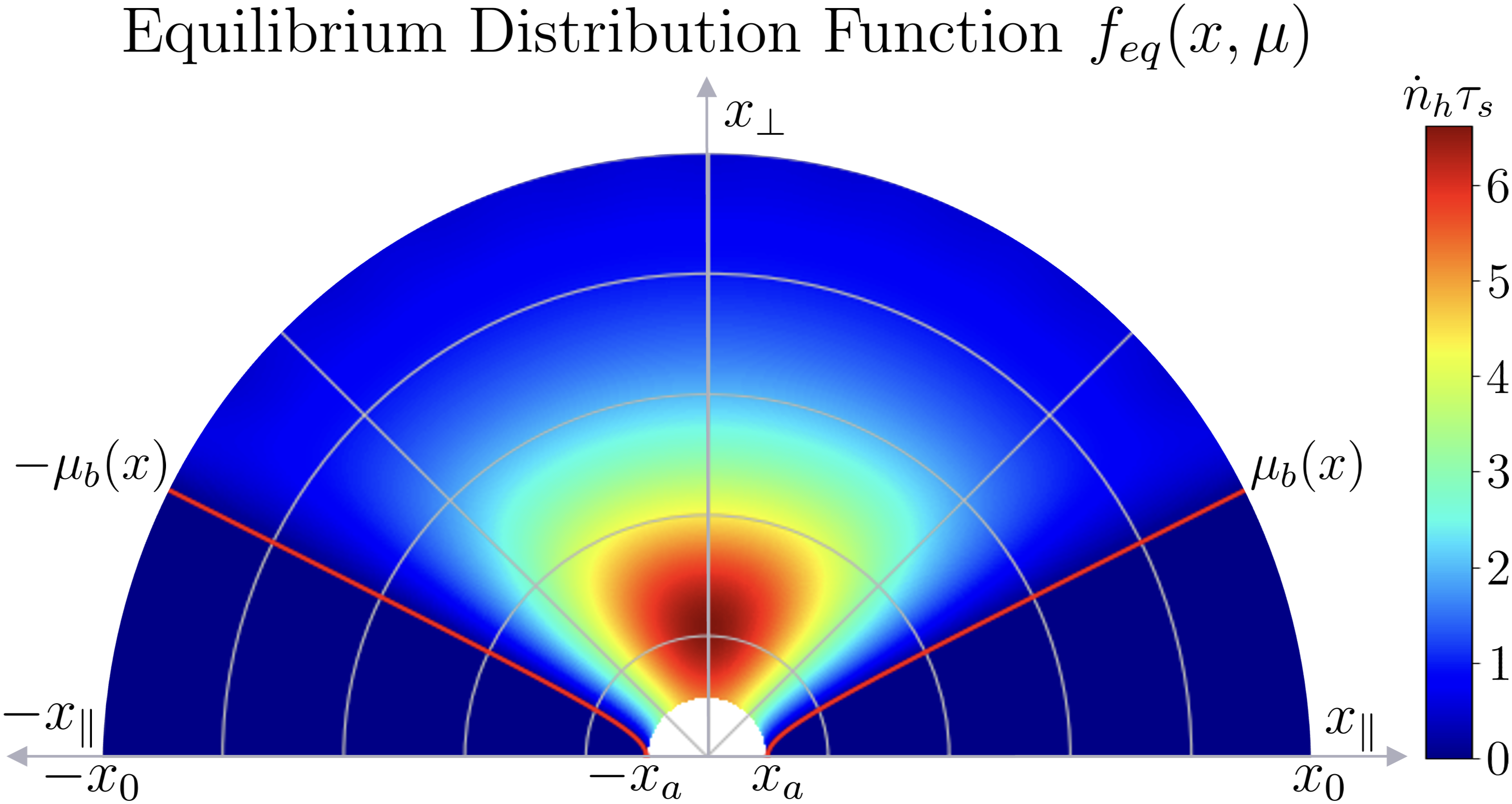}
\caption{Equilibrium Distribution $f_{eq}(x,\mu)$ in Phase Space. Here we see the equilibrium distribution function (Eq.~(\ref{Eq:feq})\!) in $(x_{\parallel},x_{\perp})$ coordinates for an isotropic confined delta source at $x_{0} = 1$ for $x_{a} = 0.1, R = 5$.\label{Fig:feq-round}}
\end{figure}

Our time-dependent solution for the loss spectra of an arbitrary initial distribution of alpha particles is certainly instructive. However, the more realistic situation in a fusion reactor is alpha particles being constantly produced by reactions until some steady-state is reached. This corresponds to isotropic delta shells being produced at $x = x_{0}$ at general start time $t = t_{0}$ with rate $\nu_{h}$, and can be obtained directly from our time-dependent Green's function solution (Eq.~(\ref{Eq:f(t,x,mu)})\!) by integrating from the distant past up to the present time,
\begin{align}f_{eq}(x,\mu) = \!\!\!\int\limits_{-\infty}^{t}\!\!\!f(t-t_{0},x,\mu)\nu_{h}dt_{0}
\end{align}
which as shown in Appendix \hyperref[App:I]{I.1} yields,
\begin{align}\label{Eq:feq}&f_{eq}(x,\mu) \!=\!\dot{n}_{h}\tau_{s}\frac{\!H\!(x_{0}\!-\!x)}{x^{3}\!+\!\eta^{3}}\frac{H(\mu_{b}\!-\!|\mu|)}{2\mu_{b}}\sum\limits_{k=0}^{\infty}\Biggl[\!\frac{2(-1)^{k}}{(k\!+\!1/2)\pi}\nonumber\\
&\cdot\cos(\sqrt{\lambda_{k}}\sin^{\!-1}\!\!\mu)\exp\!\!\left(\!\!-\!\!\!\int\limits_{x}^{x_{0}}\!\!\!\lambda_{k}\!\frac{Z_{\perp}\!(x)}{Z_{\parallel}(x)}\!\frac{dx}{2x}\!\right)\!\Biggr]\end{align}
where $H(x)$ is the Heaviside step function, $\lambda_{k}$ are our eigenvalues from Eq.~(\ref{Eq:eigenfunctions}), and $\dot{n}_{h}\equiv\hat{n}_{0}\nu_{h}$ is the rate of particle production. The equilibrium distribution (Eq.~(\ref{Eq:feq})\!) is shown in Figures~\ref{Fig:feq-square},\ref{Fig:feq-round}. \\
\indent This is entirely equivalent to considering our advection-diffusion equation (Eq.~(\ref{Eq:advection-diffusion})\!) in steady-state with a constant source outputting $\dot{n}_{h}$ confined particles per second isotropically at $x = x_{0}$,
\begin{equation}\begin{aligned}&\frac{1}{x^{2}}\frac{\partial}{\partial x}(x^{2}v(x)f_{eq}) = \nu(x)\frac{\partial}{\partial\mu}\!\!\left[\!(1\!-\!\mu^{2})\frac{\partial f_{eq}}{\partial\mu}\!\right]\!+S_{h}\\
&S_{h} \!\equiv \dot{n}_{h}\frac{\delta(x\!-\!x_{0})}{x^{2}}\frac{H(\mu_{b}\!-\!|\mu|)}{2\mu_{b}}
\end{aligned}\end{equation}
and considering $f_{eq} = g(x)h(\mu|\mu_{b}(x)\!)$ to solve by separation of variables, as shown in Appendix \hyperref[App:I]{I.2}. The time-dependent method is more powerful, however, since $\nu_{h}$ can be amended to depend on time.
\vspace{-8pt}
\section{XII. Applications}
\label{sec:relativity}
\vspace{-2pt}
The primary application of this work is modeling alpha particles in centrifugal mirror fusion reactors. Projected parameters for the DT scenario from Table \hyperref[Tab:transport]{1} in a reactor based on CMFX~\cite{Schwartz} are $(x_{a},R) \sim (0.16,6)$ as shown in Appendix \hyperref[App:J]{J}. However, our model should in principle apply to any highly energetic charged species across standard, tandem, and centrifugal mirror devices.\\
\indent For example, Ochs et al. previously derived a set of equations describing a similar model for relativistic tail electrons, see Eq.~(16-24)~\cite{Munirov}. It is worth noting that their Eq.~(16) is of the same form as our scalar collision operator (Eq.~(\ref{Eq:scalarcollisionoperator})\!) and can be rewritten in terms of our transport coefficients for normalized velocity $x\equiv v/v_{th,e}$ as,
\begin{equation}\begin{aligned}&\tau_{0}^{i}\frac{\partial f_{e}}{\partial t} = \frac{1}{x^{2}}\frac{\partial}{\partial x}\!\left(\!\gamma(x)^{2}Z_{\parallel}^{i}f_{a}\!+\!\frac{\gamma(x)^{3}}{2x}\frac{\partial f_{e}}{\partial x}\!\right)\\
&+\!\frac{\gamma(x)}{2x^{3}}\!\left(\!Z_{\perp}^{i}\!-\!\frac{1}{2x^{2}}\!\right)\!\!\frac{\partial}{\partial\mu}\!\!\left[\!(1\!-\!\mu^{2})\frac{\partial f_{e}}{\partial\mu}\!\right]\end{aligned}\end{equation}
where the Lorentz factor is given by,
\begin{align}\gamma(x) \equiv \sqrt{1\!+\!2\chi x^{2}},\quad\chi \equiv k_{B}T_{e}/m_{e}c^{2}\end{align}
Considering gyrotropicity and neglecting parallel diffusion, we obtain precisely our advection-diffusion equation (Eq.~(\ref{Eq:advection-diffusion})\!), except that now we have,
\begin{align}Z_{\parallel}(x)\equiv\gamma(x)^{2}Z_{\parallel}^{i},\quad Z_{\perp}\!(x) \equiv \gamma(x)\!\left(\!Z_{\perp}^{i}\!-\!\frac{1}{2x^{2}}\!\right)\end{align}
and our trapping boundary is amended to be \cite{Munirov},
\begin{align}\label{Eq:relhyperbola}\mu_{b}(x)\equiv\sqrt{1\!-\!\frac{1}{R}\!\left(\!1\!-\!\frac{x_{e}^{2}}{x^{2}}\!\left(\!\gamma(x)\!-\frac{\chi x_{e}^{2}}{2}\right)\!\!\right)}\end{align}
where $x_{e}\equiv\sqrt{\Phi_{e}/k_{B}T_{e}}$ and $\Phi_{e}$ is the potential experienced by electrons. The model is otherwise equivalent and a complete analytical description readily follows.\\
\indent In fact, a solution can be found following our method for any collision operator that can be placed in the form of Eq.~(\ref{Eq:advection-diffusion}). This approach will be most accurate when parallel diffusion and sheared rotation loss are negligible, and the trapping boundary varies slowly relative to pitch-angle scattering. For instance, if one wanted to consider relativistic effects for alpha particles, one could simply amend the collision operator in the appropriate limit~\cite{Karney}, employ the relativistic trapping boundary (Eq.~(\ref{Eq:relhyperbola}) with $x_{e}\rightarrow x_{a}$), and neglect parallel diffusion. A variation of this model could even be used for mirror-trapped high-energy particles in a stellarator~\cite{Ho,Bader}.
\vspace{-10pt}
\section{XIII. Conclusions}
\label{sec:conclusions}
\vspace{-5pt}
We have thus developed a framework for analyzing the losses of energetic alpha particles in a magnetic mirror subject to any applied potential $\Phi_{a} < k_{B}T_{a}$ with mirror ratio $R > 1$. We have identified the various confinement possibilities that an alpha particle might experience after being produced in fusion reactions and obtained analytical expressions for the fractions of particles that meet each of these fates. The only nontrivial possibility is becoming gradually deconfined through a series of pitch-angle scattering collisions, which we have analyzed using a Fokker-Planck model of the non-relativistic Landau collision operator. In solving this Fokker-Planck model, we have discovered a set of orthogonal asymptotic eigenfunctions of the Legendre operator for homogeneous boundary conditions on symmetric intervals with amplitude less than unity. We have used these eigenfunctions to obtain the first ever closed-form solutions for the time-dependent and steady-state distribution functions, the remaining number density at a given normalized velocity and time, and the PDFs of the loss velocities, energies, and times of alpha particles. Using these PDFs, one can compute the loss statistics of all related quantities for any initial velocity and angular distributions. We have verified that all of these calculations are valid in the proposed regime by comparison with Monte-Carlo simulations. The general approach presented here should work for any fast-ion species produced by any fusion reaction in a magnetic mirror subject to any applied potential between the midplane and ends. It can be trivially amended to work for relativistic alphas and tail electrons as well. Our model should provide a solid basis for analyzing ion-beam experiments at existing mirror facilities and eventually the loss spectra of highly energetic particles in magnetic mirror fusion reactors.
\section{Acknowledgments}
We would like to thank T. Rubin, M. W. Kunz, T. Foster, M. Q. May, H. Fetsch, R. Nies, A. Robbins, L. David, J. M. Molina, P. Karypis, and W. Chu for helpful discussions throughout the development of this paper.\\
\indent This research was supported by U.S. DOE Grant Number DE-AC02-09CH11466 and ARPA-E Grant Number DE-AR0001554.
\bibliography{mirror}
\bibliographystyle{abbrv}
\bibliographystyle{unsrt}
\nocite{*}
\section{Appendix A: Rosenbluth Potentials}
\label{App:A}
\begin{figure}
\centering
\noindent\includegraphics[width=0.95\columnwidth]{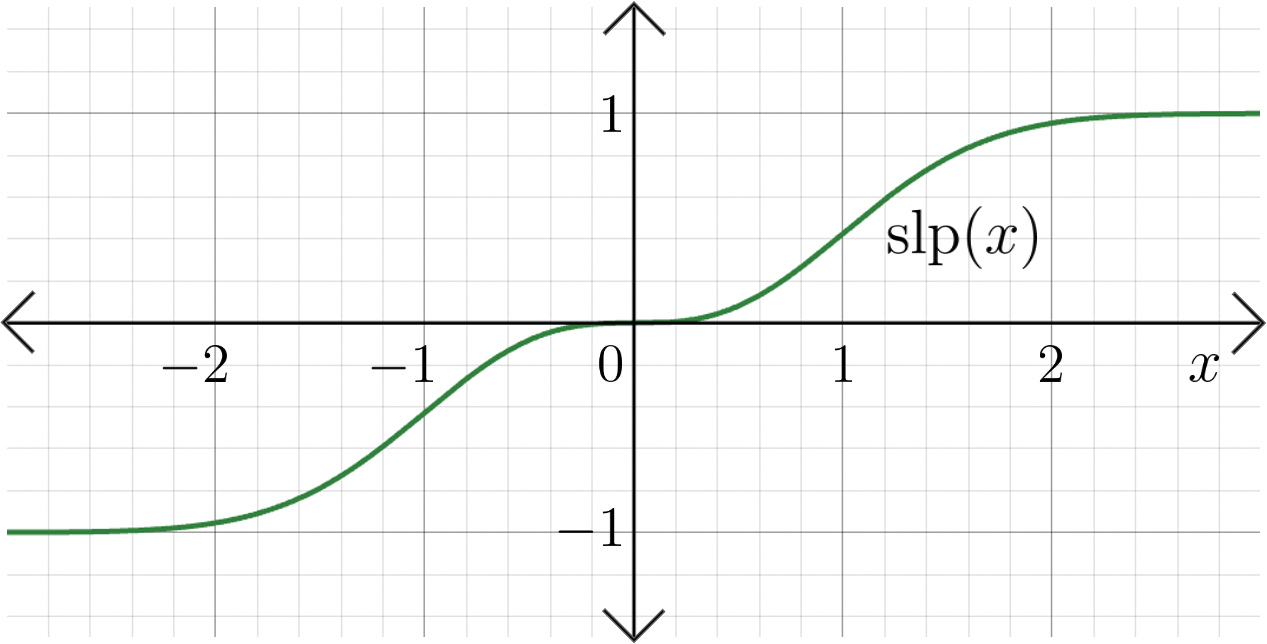}
\caption{Slope Function slp($x$). Here we see the slope function slp($x$) plotted vs $x$. Its domain is the real numbers $\mathbb{R}$ and its range is $(-1,1)$. It approaches horizontal asymptotes at $y=\pm 1$ as $x\rightarrow \pm \infty$ and has a saddle point at the origin. It is the numerator of the Chandrasekhar function $G(x) \equiv \text{slp}(x)/2x^{2}$.}\label{Fig:slope}
\end{figure}

\noindent We present a non-relativistic model of the collision operator for a highly energetic ion species (alpha particles) colliding with Maxwellian distributions of other less energetic, slower ions (protons and Boron-11/deuterium and tritium) and less energetic, albeit faster, electrons. From the Rosenbluth formulation of the collision operator we have,
\begin{equation*}
\begin{aligned}
&\left(\!\frac{\partial f_{a}}{\partial t}\!\right)_{\!\!coll} \!\!\!\!\!\!\!= -\!\sum_{b}\vec{\nabla}_{\vec{v}}\cdot\vec{J}_{ab}, \,\,\, C_{ab} \equiv\! \frac{4\pi}{m_{a}^{2}}\!\!\left(\!\frac{Z_{a}Z_{b}e^{2}}{4\pi\epsilon_{0}}\!\right)^{\!\!2}\!\!n_{b}\lambda_{ab}\\
&\quad\vec{J}_{ab} \!=\! \frac{C_{ab}}{n_{b}}\!\!\left[\!\biggl(\!1\!+\!\frac{m_{a}}{m_{b}}\!\biggr)f_{a}\vec{\nabla}_{\vec{v}}H_{b}\!-\!\frac{1}{2}\vec{\nabla}\cdot(f_{a}\vec{\nabla}_{\vec{v}}\vec{\nabla}_{\vec{v}}G_{b})\!\right]
\end{aligned}\end{equation*}
where the Rosenbluth potentials $G(\vec{v}),H(\vec{v})$ are given by,
\begin{align*}
&G(\vec{v}) \!=\!\! \!\int\!\!f_{b}(\vec{v}')|\vec{v}-\vec{v}'|d\vec{v}'\!, \quad\!\! H(\vec{v}) \!=\! \!\!\int\!\!\frac{f_{b}(\vec{v}')}{|\vec{v}-\vec{v}'|}\,\!d\vec{v}'\!\!
\end{align*}
And our Maxwellian bulk distributions are given by,
$$f_{b}(\vec{v}) = \frac{n_{b}}{\pi^{3/2}v_{th,b}^{3}}e^{-v^{2}\!/v_{th,b}^{2}}, \quad v_{th,b}\equiv\sqrt{\frac{2k_{B}T_{b}}{m_{b}}}\quad $$
Our first Rosenbluth potential is then given by,
\begin{align*}G(\vec{v}) = \frac{n_{b}}{\pi^{3/2}v_{th,b}^{3}}\!\int\!|\vec{v}-\vec{v}'|e^{-v'^{2}\!/v_{th,b}^{2}}\,d\vec{v}'\end{align*}
Choosing spherical coordinates such that $\vec{v} = v\mathbf{\hat{z}}$,
\begin{align*}&|\Vec{v}'\!-\!\vec{v}| \!=\! \sqrt{v'^{2}\!+\!v^{2}\!-\!2v'v\cos\theta} = v\!\left(\!1\!-\!\frac{2v'}{v}\cos\theta\!+\!\frac{v'^{2}}{v^{2}}\right)^{\!\!1/2}\end{align*}
Substituting into our integral we have for $x_{b}\!\equiv\! v/v_{th,b}$,
\begin{align*}&G(\vec{v}) \!=\!\frac{n_{b}}{\pi^{3/2}v_{th,b}^{3}}\!\!\!\iint\limits_{\!\Omega\,\,0}^{\,\,\,\,\,\,\,\,\,\,\,\,\,\,\infty}\!\!\!\!\left(\!v'^{2}\!\!-\!2vv'\!\!\cos\theta\!+\!v^{2}\right)^{\!1/2}\!\!\!e^{-v'^{2}\!/v_{th,b}^{2}}v'^{2}\!dv'\!d\Omega\\
&\boxed{G(\vec{v}) = n_{b}v_{th,b}\Biggl[\!\frac{1}{\sqrt{\pi}}e^{-x_{b}^{2}}\!+\!\left(\!x_{b}\!+\!\frac{1}{2x_{b}}\!\right)\!\text{erf}(x_{b})\!\Biggr]\!}\end{align*}
Similarly, we compute our second Rosenbluth potential,
\begin{align*}
&H(\vec{v}) =  \frac{n_{b}}{\pi^{3/2}v_{th,b}^{3}}\!\int\! \frac{e^{-v'^{2}\!/v_{th,b}^{2}}}{|\vec{v}-\vec{v}'|}\,\!d\vec{v}'
\end{align*}
Substituting our expression for $|\vec{v}-\vec{v}'|$ as before,
\begin{align*}
&\!H(\vec{v}) \!=\!\frac{n_{b}}{\pi^{3/2}v_{th,b}^{3}}\!\!\!\iint\limits_{\!\Omega\,\,0}^{\,\,\,\,\,\,\,\,\,\,\,\,\,\,\infty}\!\!\!\!\left(\!v'^{2}\!\!\!-\!2vv'\!\!\cos\theta\!+\!v^{2}\right)^{\!\!-1/2}\!\!\!e^{-v'^{2}\!\!/v_{th,b}^{2}}v'^{2}\! dv'\!d\Omega\\
&\quad\quad\quad\quad\quad\quad\quad\quad\boxed{H(\vec{v}) = \frac{n_{b}}{v}\text{erf}(x_{b})\!}\end{align*}
Now that we have our Rosenbluth potentials, we can evaluate the following related quantities,
\begin{align*}
&\vec{\nabla}_{\vec{v}}G(\vec{v})\!=\!\vec{\nabla}_{\vec{v}}\Biggl(\!n_{b}v_{th,b}\biggl[\!\frac{1}{\sqrt{\pi}}e^{-x_{b}^{2}}\!+\!\left(\!x_{b}+\frac{1}{2x_{b}}\!\right)\text{erf}(x_{b})\biggr]\!\Biggr)\\
&\boxed{\vec{\nabla}_{\vec{v}}G(\vec{v}) = \frac{n_{b}}{v}\!\left[\frac{1}{\sqrt{\pi}}\frac{e^{-x_{b}^{2}}\!\!}{x_{b}}+\!\left(\!1\!-\!\frac{1}{2x_{b}^{2}}\!\right)\!\text{erf}(x_{b})\!\right]\!\vec{v}}\\
&\vec{\nabla}_{\vec{v}}\vec{\nabla}_{\vec{v}}G(\vec{v})\!=\!\vec{\nabla}_{\vec{v}}\!\left\{\!\frac{n_{b}}{v}\!\left[\!\frac{1}{\sqrt{\pi}}\frac{e^{-x_{b}^{2}}\!\!}{x_{b}}\!+\!\left(\!1\!-\!\frac{1}{2x_{b}^{2}}\!\right)\!\text{erf}\!\left(x_{b}\right)\!\right]\!\!\vec{v}\!\right\}\\
&\vec{\nabla}_{\vec{v}}\vec{\nabla}_{\vec{v}}G(\vec{v}) = \frac{n_{b}}{v^{3}}\!\Biggl[\!\left[-\frac{3}{\sqrt{\pi}}\frac{e^{-x_{b}^{2}}\!\!}{x_{b}}\!+\!\left(\!\frac{3}{2x_{b}^{2}}\!-\!1\!\right)\text{erf}(x_{b})\!\right]\!\vec{v}\vec{v}\\
&+\left[\frac{1}{\sqrt{\pi}}\frac{e^{-x_{b}^{2}}\!\!}{x_{b}}\!+\!\left(\!1-\frac{1}{2x_{b}^{2}}\!\right)\!\text{erf}(x_{b})\!\right]\!v^{2}\overline{\overline{\mathbf{1}}}\Biggr]\\
&\boxed{\vec{\nabla}_{\vec{v}}\vec{\nabla}_{\vec{v}}G(\vec{v}) \!=\! n_{b}\!\!\left[\!\!\left(\!\text{erf}(x_{b})\!-\!\frac{\text{slp}(x_{b})}{2x_{b}^{2}}\!\right)\!\frac{v^{2}\overline{\overline{\mathbf{1}}}\!-\!\vec{v}\vec{v}}{v^{3}}\!+\!\frac{\text{slp}(x_{b})}{x_{b}^{2}}\frac{\vec{v}\vec{v}}{v^{3}}\right]\!}\!\!\!\!\!\!\!\!\\
&\vec{\nabla}_{\vec{v}}\cdot\vec{\nabla}_{\vec{v}}\vec{\nabla}_{\vec{v}}G(\vec{v}) = \vec{\nabla}_{\vec{v}}\cdot\!\Biggl\{\!\frac{n_{b}}{v^{3}}\!\Biggl[\!\left[\!-\frac{3}{\sqrt{\pi}}\frac{e^{-x_{b}^{2}}\!\!}{x_{b}}\!+\!\!\left(\!\frac{3}{2x_{b}^{2}}\!-\!1\!\right)\!\text{erf}(x_{b})\!\right]\!\vec{v}\vec{v}\\
&\!\!+\!\!\left[\!\frac{e^{-x_{b}^{2}}\!\!}{\sqrt{\pi}x_{b}}\!\!+\!\!\left(\!1\!-\!\frac{1}{2x_{b}^{2}}\!\right)\!\text{erf}(x_{b})\!\right]\!\!v^{2}\overline{\overline{\mathbf{1}}}\!\Biggr]\!\!\Biggr\}\!\!\\
&\quad\quad\quad\quad\quad\quad\boxed{\!\vec{\nabla}_{\vec{v}}\!\cdot\!\!\vec{\nabla}_{\vec{v}}\vec{\nabla}_{\vec{v}}G(\vec{v}) \!=\! -\frac{2n_{b}}{v^{3}}\text{slp}(x_{b})\vec{v}\!}\\
&\!\!\vec{\nabla}_{\vec{v}}H(\vec{v}) \!=\! \!\vec{\nabla}_{\vec{v}}\!\left(\!\frac{n_{b}}{v}\text{erf}(x_{b})\!\!\right)\!\!\rightarrow\!\boxed{\!\vec{\nabla}_{\vec{v}}H(\vec{v}) \!= \!-\frac{n_{b}}{v^{3}}\text{slp}(x_{b})\vec{v}\!}\end{align*}
Now we can substitute in our expression for $\vec{J}^{a/b}$,
\begin{align*}&\!\!\Vec{J}_{ab} \!=\!\! -\frac{C_{ab}}{n_{b}}\!\Biggl[\!f_{a}\!\frac{n_{b}\vec{v}}{v^{3}}\!\!\left(\!\!1\!+\!\frac{m_{a}}{m_{b}}\!\!\right)\!\text{slp}(x_{b})\!\!+\!\!\frac{1}{2}\!\Biggl\{\!\!f_{a}\!\!\left(\!\!\frac{-2n_{b}\vec{v}}{v^{3}}\text{slp}(x_{b})\!\!\right)\!\!+\!\vec{\nabla}f_{a}\!\cdot\!\vec{\nabla}\vec{\nabla}G\!\Biggr\}\!\!\Biggr]\\
&\!\!\!\!\!\!\!\!\!=\!\!-C_{ab}\!\Biggl[\!\!\left\{\!\!\frac{m_{a}\vec{v}}{m_{b}v^{3}}\,\!\text{slp}(x_{b})\!\!\right\}\!\!f_{a}\!\!+\!\!\frac{1}{2}\!\Biggl\{\!\frac{v^{2}\overline{\overline{\mathbf{1}}}\!-\!\vec{v}\vec{v}}{v^{3}}\!\!\left[\!\text{erf}(x_{b}\!)\!-\!\frac{\text{slp}(x_{b})}{2x_{b}^{2}}\!\right]\!\!+\!\!\frac{\vec{v}\vec{v}}{v^{3}}\frac{\text{slp}(x_{b})}{x_{b}^{2}}\!\!\Biggr\}\!\!\cdot\!\!\vec{\nabla}\!f_{a}\!\Biggr]\\
&\frac{\partial f_{a}}{\partial \vec{v}}= \frac{\partial}{\partial\vec{v}}\cdot C_{ab}\!\Biggl[\!\left\{\!\frac{m_{a}\vec{v}}{m_{b}v^{3}}\,\text{slp}(x_{b})\!\right\}f_{a}(\vec{v})\\
&+\frac{1}{2}\Biggl\{\frac{v^{2}\overline{\overline{\mathbf{1}}}-\vec{v}\vec{v}}{v^{3}}\!\left[\text{erf}(x_{b})\!-\!\frac{\text{slp}(x_{b})}{2x_{b}^{2}}\!\right]\!+\frac{\vec{v}\vec{v}}{v^{3}}\frac{\text{slp}(x_{b})}{x_{b}^{2}}\!\Biggr\}\!\cdot\!\frac{\partial f_{a}}{\partial\vec{v}}\Biggr]\end{align*}
\begin{align*}
&\text{where throughout we've defined the ``slope" function $\text{slp}(x)$ by,}\\
&\quad\quad\quad\boxed{\text{slp}(x)\equiv \text{erf}(x)\!-\!\frac{2}{\sqrt{\pi}}xe^{-x^{2}} \!\!\!\equiv\text{erf}(x)\!-\!x\,\text{erf}\,'\!(x)\!}\\
&\sim \frac{4}{\sqrt{\pi}}\!\sum\limits_{k=0}^{\infty}\frac{(-1)^{k}x^{2k+3}}{k!(2k+3)}\sim\frac{4}{\sqrt{\pi}}\!\!\left(\!\frac{x^{3}}{3}\!-\!\frac{x^{5}}{5}\!+\!\frac{x^{7}}{14}\!-\hdots\!\right)\,\, (x << 1)\\
&\sim 1\!-\!\frac{e^{-x^{2}}\!}{x\sqrt{\pi}}\!\!\sum\limits_{k=-1}^{\infty}\!\!\frac{(-1)^{k}(2k\!-\!1)!!}{(2x^{2})^k}\!\sim\! 1\!-\!\frac{e^{-x^{2}}\!\!}{\sqrt{\pi}}\!\!\left(\!2x\!+\!\frac{1}{x}\!-\!\hdots \!\!\right)\!(x \!>>\! 1)
\end{align*}
\newpage
\noindent It is displayed in Figure~\ref{Fig:slope}. This special function can curiously also be written in the following forms,
\begin{align*}\text{slp}(x) = \!\frac{1}{\sqrt{\pi}}\!\!\left[\,\int\limits_{-x}^{x}\!e^{-t^{2}}dt\!+\!\frac{d}{dx}\!\!\left(e^{-x^{2}}\right)\!\right] \!= -x^{2}\frac{d}{dx}\!\left(\!\frac{\text{erf}(x)}{x}\!\right)\end{align*}
\noindent Then we have in general,
\begin{adjustwidth}{-4in}{-4in}
\begin{empheq}[box=\widefbox]{align*}
&\!\!C_{ab} \equiv \frac{4\pi}{m_{a}^{2}}\!\!\left(\!\frac{Z_{a}Z_{b}e^{2}}{4\pi\epsilon_{0}}\!\right)^{\!\!2}\!\!n_{b}\lambda_{ab}, \quad x_{b} \equiv v/v_{th,b}\\
&\!\!^{\vec{v}\!}\vec{A}_{a} \equiv \sum_{b}C_{ab}\!\!\left\{\frac{m_{a}\vec{v}}{m_{b}v^{3}}\text{slp}(x_{b})\!\!\right\}\\
&\!\!\!\!\!\!\!\!\!\!\!^{\vec{v}}\overline{\overline{D}}_{a} \!\equiv \!\sum_{b}\!\frac{C_{ab}}{2}\!\!\left\{\!\frac{v^{2}\overline{\overline{\mathbf{1}}}\!-\!\vec{v}\vec{v}}{v^{3}}\!\!\left[\text{erf}(x_{b})\!-\!\frac{\text{slp}(x_{b})}{2x_{b}^{2}}\right]\!\!+\!\frac{\vec{v}\vec{v}}{v^{3}}\frac{\text{slp}(x_{b})}{x_{b}^{2}}\!\!\right\}\!\!\!\!\!\!\!\!\!\!\!\!\\
&\!\!\left(\frac{\partial f_{a}}{\partial t}\right)_{\!\!coll}\!\!\!\!\!\!= \frac{\partial}{\partial\vec{v}}\cdot\!\left(\!^{\vec{v}}\!\vec{A}_{a}f_{a}+^{\vec{v}}\!\overline{\overline{D}}_{a}\cdot\frac{\partial f_{a}}{\partial\vec{v}}\!\right)\end{empheq}
\end{adjustwidth}

\newpage
\section{Appendix B: Useful Vector Identities}
\label{App:B}
The following vector identities were useful in evaluating the Rosenbluth potential objects in Appendix \hyperref[App:A]{A}.
\begin{align*}&\vec{\nabla}_{\vec{v}}\vec{v} = \frac{\partial\vec{v}}{\partial\vec{v}} = \overline{\overline{\mathbf{1}}} \rightarrow \boxed{\vec{\nabla}_{\vec{v}}\vec{v} = \overline{\overline{\mathbf{1}}}}\\
&\vec{\nabla}_{\vec{v}}v = \vec{\nabla}_{\vec{v}}\sqrt{\vec{v}\cdot\vec{v}} = \frac{1}{2}\left(\vec{v}\cdot\vec{v}\right)^{-1/2}(\vec{\nabla}_{\vec{v}}\vec{v}\cdot\vec{v}+\vec{\nabla}_{\vec{v}}\vec{v}\cdot\vec{v})\\
&=\frac{1}{2v}(2\overline{\overline{\mathbf{1}}}\cdot\vec{v}) = \frac{\vec{v}}{v} \rightarrow \boxed{\vec{\nabla}_{\vec{v}}v = \frac{\vec{v}}{v}}\\
&\vec{\nabla}_{\vec{v}}\left(\frac{1}{v}\right) = \vec{\nabla}_{\vec{v}}(\vec{v}\cdot\vec{v})^{-1/2} = -\frac{1}{2}\left(\vec{v}\cdot\vec{v}\right)^{-3/2}(\vec{\nabla}_{\vec{v}}\vec{v}\cdot\vec{v}+\vec{\nabla}_{\vec{v}}\vec{v}\cdot\vec{v})\\
&=-\frac{1}{2v^{3}}2\overline{\overline{\mathbf{1}}}\cdot\vec{v} = -\frac{\vec{v}}{v^{3}} \rightarrow \boxed{\vec{\nabla}_{\vec{v}}\left(\frac{1}{v}\right) = -\frac{\vec{v}}{v^{3}}}\\
&\boxed{\vec{\nabla}_{\vec{v}}f(v) = \frac{df}{dv}\mathbf{\hat{v}} = \frac{df}{dv}\frac{\vec{v}}{v}}\\
&\vec{\nabla}_{\vec{v}}(f\vec{v}) = \mathbf{\hat{e}}_{i}\frac{\partial}{\partial v_{i}}\left(fv_{k}\mathbf{\hat{e}}_{k}\right) = \mathbf{\hat{e}}_{i}\left(\frac{\partial f}{\partial v_{i}}v_{k}+f\frac{\partial v_{k}}{\partial v_{i}}\right)\mathbf{\hat{e}}_{k}\\
&\boxed{\vec{\nabla}_{\vec{v}}(f\vec{v}) = \vec{\nabla}_{\vec{v}}f\vec{v}+\!f\overline{\overline{\mathbf{1}}}}\\
&\vec{\nabla_{v}}\cdot(f\overline{\overline{T}}) = \mathbf{\hat{e}}_{i}\frac{\partial}{\partial v_{i}}\cdot(fT_{kl}\mathbf{\hat{e}}_{k}\mathbf{\hat{e}}_{l}) = \left(\frac{\partial f}{\partial v_{i}}T_{kl}+f\frac{\partial T_{kl}}{\partial v_{i}}\right)\delta_{ik}\mathbf{\hat{e}}_{l}\\
&\boxed{\vec{\nabla}_{\vec{v}}\cdot(f\overline{\overline{T}})=\vec{\nabla}_{\vec{v}}f\cdot\overline{\overline{T}}+\!f(\vec{\nabla}_{\vec{v}}\cdot\overline{\overline{T}})}\\
&\vec{\nabla}_{\vec{v}}\cdot(f\vec{v}\vec{v}) = \mathbf{\hat{e}}_{i}\frac{\partial}{\partial v_{i}}\!\cdot\!(fv_{k}\mathbf{\hat{e}}_{k}v_{l}\mathbf{\hat{e}}_{l})\\
&=\!\! \left(\!\frac{\partial f}{\partial v_{i}}v_{k}v_{l}\!+\!f\frac{\partial v_{k}}{\partial v_{i}}v_{l}\!+\!fv_{k}\frac{\partial v_{l}}{\partial v_{i}}\!\right)\!\delta_{ik}\mathbf{\hat{e}}_{l}\\
&=\vec{\nabla}_{\vec{v}}f\cdot\vec{v}\vec{v}+\!f(\vec{\nabla}_{\vec{v}}\cdot\vec{v})\vec{v}+\!f\vec{v}\cdot\overline{\overline{\mathbf{1}}}\\
&\boxed{\vec{\nabla}_{\vec{v}}\cdot(f\vec{v}\vec{v}) = \vec{\nabla}_{\vec{v}}f\cdot\vec{v}\vec{v}+4f\vec{v}}\\
&\vec{w}\cdot\overline{\overline{T}} = w_{i}\mathbf{\hat{e}}_{i}\cdot T_{kl}\mathbf{\hat{e}}_{k}\mathbf{\hat{e}}_{l} = w_{i}T_{kl}\delta_{ik}\mathbf{\hat{e}}_{l} = w_{k}T_{kl}\mathbf{\hat{e}}_{l}\\
&\overline{\overline{T}}\cdot\vec{w} = T_{kl}\mathbf{\hat{e}}_{k}\mathbf{\hat{e}}_{l}\cdot w_{i}\mathbf{\hat{e}}_{i}\cdot = T_{kl}w_{i}\delta_{li}\mathbf{\hat{e}}_{k} = T_{kl}w_{l}\mathbf{\hat{e}}_{k} = w_{k}T_{lk}\mathbf{\hat{e}}_{l}\\
&\boxed{T_{lk} = T_{kl} \longrightarrow \vec{w}\cdot\overline{\overline{T}} = \overline{\overline{T}}\cdot\vec{w}}\end{align*}
\section{Appendix C: Tensor Transformation to Mirror Coordinates}
\label{App:C}
The coordinate transformation from Cartesian velocity coordinates $\vec{v} = (v_{x},v_{y},v_{z})$ to standard mirror coordinates $\vec{x} = (x,\mu,\varphi)$ was performed as follows. Our coordinates are defined by,
\begin{align*}x \equiv v/v_{_{th,a}}, \,\,\mu \equiv v_{z}/v,\,\,\text{tan}\,\varphi \equiv v_{y}/v_{x}, \end{align*}
which can be rewritten as,
\begin{align*}x = \frac{\sqrt{v_{x}^{2}\!+\!v_{y}^{2}\!+\!v_{z}^{2}}}{v_{th,a}}, \,\,\mu = \frac{v_{z}}{\!\sqrt{v_{x}^{2}\!+\!v_{y}^{2}\!+\!v_{z}^{2}}},\,\,\varphi = \tan^{\!-1}\!\!\left(\!\frac{v_{y}}{v_{x}}\!\right)\end{align*}
Then we can observe,
\begin{align*}\frac{v_{x}}{v_{th,a}} \!= x\sqrt{1\!-\!\mu^{2}}\cos\varphi, \,\frac{v_{y}}{v_{th,a}} \!= x\sqrt{1\!-\!\mu^{2}}\sin\varphi, \, \frac{v_{z}}{v_{th,a}} = x\mu\end{align*}
Through straightforward differentiation, we obtain the Jacobian matrix,
\begin{align*}\!\frac{\partial \vec{v}}{\partial \vec{x}} = v_{th,a}\!\begin{pmatrix}\sqrt{1\!-\!\mu^{2}}\cos\varphi&-\frac{x\mu\cos\varphi}{\sqrt{1-\mu^{2}}}&-x\sqrt{1\!-\!\mu^{2}}\sin\varphi\\\sqrt{1\!-\!\mu^{2}}\sin\varphi&-\frac{x\mu\sin\varphi}{\sqrt{1-\mu^{2}}}&x\sqrt{1\!-\!\mu^{2}}\cos\varphi\\\mu&x&0\end{pmatrix}\end{align*}
And since $\partial \vec{x}/\partial \vec{v} = (\partial \vec{v}/\partial \vec{x})^{-1}$,
\begin{align*}\!\frac{\partial \vec{x}}{\partial \vec{v}} = \frac{1}{v_{th,a}}\!\begin{pmatrix}\sqrt{1\!-\!\mu^{2}}\cos\varphi&\sqrt{1\!-\!\mu^{2}}\sin\varphi&\mu\\-\frac{\mu\sqrt{1-\mu^{2}}}{x}\cos\varphi&-\frac{\mu\sqrt{1-\mu^{2}}}{x}\sin\varphi&\frac{1-\mu^{2}}{x}\\\frac{-\sin\varphi}{x\sqrt{1-\mu^{2}}}&\frac{\cos\varphi}{x\sqrt{1-\mu^{2}}}&0\end{pmatrix}\end{align*}
The transformation rules for contravariant and covariant indices dictate that,
\begin{align*}&^{\vec{x}}T^{mn} = \frac{\partial x^{m}}{\partial v^{i}}\frac{\partial x^{n}}{\partial v^{j}}\,^{\vec{v}}T^{ij}, \quad ^{\vec{x}}T_{mn} = \frac{\partial v^{i}}{\partial x^{m}}\frac{\partial v^{j}}{\partial x^{n}}\,^{\vec{v}}T_{ij}\end{align*}
Then,
$$^{\vec{x}}\!A^{m} \!=\! \frac{\partial x^{m}}{\partial v^{k}}\,\!^{\vec{v}}\!A^{k}, \,\! ^{\vec{x}}\!D^{mn} \!=\! \frac{\partial x^{m}}{\partial v^{i}}\frac{\partial x^{n}}{\partial v^{j}}\,\!^{\vec{v}}\!D^{ij}, \,\! ^{\vec{x}}g_{mn} \!=\! \frac{\partial v^{i}}{\partial x^{m}}\frac{\partial v^{j}}{\partial x^{n}}\!\,^{\vec{v}}g_{ij}$$
where we call $\overline{\overline{g}}\leftrightarrow g_{mn}$ and $\overline{\overline{g}}^{-1}\leftrightarrow g^{mn}$. Taking $^{\vec{v}}\overline{\overline{g}} = \overline{\overline{\mathbf{1}}} \rightarrow ^{\vec{v}}\!\!g_{ij} = \delta^{i}_{j}$ we obtain,
\begin{align*}&^{\vec{x}}g_{mn} \!=\! \frac{\partial v^{i}}{\partial x^{m}}\frac{\partial v^{i}}{\partial x^{n}} = \left(\frac{\partial \vec{v}}{\partial \vec{x}}\right)^{\!\!T}\!\!\!\!\left(\frac{\partial \vec{v}}{\partial \vec{x}}\right), \quad ^{\vec{x}}g^{mn} = (^{\vec{x}}g_{mn})^{-1}\\
&^{\vec{x}\!}g_{mn}\!\!=\!v_{th,a}^{2}\!\!\begin{pmatrix}1&\!\!0&\!\!\!\!\!0\\0&\!\frac{x^{2}}{1-\mu^{2}}&\!\!\!\!0\\0&\!\!0&\!\!\!\!\!\!x^{2}\!(1\!-\!\mu^{2})\!\end{pmatrix}\!\!,\!^{\vec{x}\!}g^{mn}\!\!=\!\!\frac{1}{v_{th,a}^{2}\!}\!\!\begin{pmatrix}1&0&0\\0&\!\!\frac{1-\mu^{2}}{x^{2}}\!\!\!\!\!\!&0\\0&0&\!\!\!\!\frac{1}{x^{2}\!(1-\mu^{2})}\!\end{pmatrix}\end{align*}
where $g\equiv \text{det}(^{\vec{x}}g_{_{m\!n}}) = v_{th,a}^{6}x^{4} \rightarrow \sqrt{g} = v_{th,a}^{3}x^{2}$.
\newpage
\noindent We previously obtained the following exact expressions in Cartesian velocity coordinates,
\begin{align*}^{\vec{v}\!}\vec{A}_{a} = \sum_{b}C_{ab}\!\left\{\!\frac{m_{a}\vec{v}}{m_{b}v^{3}}\,\text{slp}\!\left(x_{b}\right)\!\right\}, \!\!\quad \text{slp}(x) \equiv \text{erf}(x)-\frac{2}{\sqrt{\pi}}xe^{-x^{2}}\\^{\vec{v}}\overline{\overline{D}}_{a} = \sum_{b}\frac{C_{ab}}{2}\!\left\{\!\frac{v^{2}\overline{\overline{1}}-\vec{v}\vec{v}}{v^{3}}\!\left(\!\text{erf}(x_{b})-\frac{\text{slp}(x_{b})}{2x_{b}^{2}}\!\right)\!+\frac{\vec{v}\vec{v}}{v^{3}}\frac{\text{slp}(x_{b})}{x_{b}^{2}}\!\right\}\end{align*}
To make our lives easier,
\begin{align*}
&^{\vec{v}\!}\vec{A}_{a} \equiv \gamma\vec{v}, \,\, ^{\vec{v}}\overline{\overline{D}}_{a} \equiv \frac{v^{2}\overline{\overline{\mathbf{1}}}\!-\!\vec{v}\vec{v}}{\alpha} \!+\! \frac{\vec{v}\vec{v}}{\beta}, \,\, \gamma \equiv \sum_{b}\!C_{ab}\!\left\{\!\!\frac{m_{a}}{m_{b}v^{3}}\text{slp}(x_{b})\!\!\right\}\\
&\alpha^{\!-1}\!\equiv\!\sum_{b}\frac{C_{ab}}{2v^{3}}\!\left(\!\text{erf}(x_{b})-\frac{\text{slp}(x_{b})}{2x_{b}^{2}}\!\right), \, \beta^{-1}\!\equiv\!\sum_{b}\frac{C_{ab}}{2v^{3}}\frac{\text{slp}(x_{b})}{x_{b}^{2}}\end{align*}
Performing our transformations we obtain,
\begin{align*}&\begin{pmatrix}^{\vec{x}}\!A_{a}^{1}\\^{\vec{x}}\!A_{a}^{2}\\^{\vec{x}}\!A_{a}^{3}\end{pmatrix} \!=\! \frac{\gamma}{v_{th,a}}\!\!\begin{pmatrix}\sqrt{1\!-\!\mu^{2}}\cos\varphi&\sqrt{1\!-\!\mu^{2}}\sin\varphi&\mu\\\!\!-\frac{\mu\sqrt{1-\mu^{2}}}{x}\cos\varphi&\!\!-\frac{\mu\sqrt{1-\mu^{2}}}{x}\sin\varphi&\!\!\frac{1-\mu^{2}}{x}\!\!\\\frac{-\sin\varphi}{x\sqrt{1-\mu^{2}}}&\!\!\frac{\cos\varphi}{x\sqrt{1-\mu^{2}}}&0\end{pmatrix}\!\!\!\begin{pmatrix}\!v_{x}\!\\\!v_{y}\!\\\!v_{z}\!\end{pmatrix}\\
&\!=\!\gamma\!\!\begin{pmatrix}\!x(1\!-\!\mu^{2})(\cos^{2}\!\varphi\!+\!\sin^{2}\!\varphi)+x\mu^{2}\!\\\!\!\mu(1\!-\!\mu^{2})\!-\!\mu(1\!-\!\mu^{2})(\cos^{2}\!\varphi\!+\!\sin^{2}\!\varphi)\!\!\!\\\!-\cos\varphi\sin\varphi\!+\!\sin\varphi\cos\varphi\!\end{pmatrix}\!\!=\!\gamma\!\!\begin{pmatrix}\!x\!\\\!0\!\\\!0\!\end{pmatrix}\!\!\rightarrow ^{\vec{x}\!}\!\!\vec{A}_{a} \!=\! \gamma\vec{x}\end{align*}
which makes sense since $\vec{x} = \partial\vec{x}/\partial\vec{v}\,\cdot\,\vec{v}$. With the diffusion tensor we must be a bit more delicate,
\begin{align*}&^{\vec{x}}\overline{\overline{D}}_{a} \!=\!\! \left(\!\frac{\partial\vec{x}}{\partial\vec{v}}\right)\!\cdot^{\vec{v}}\!\overline{\overline{D}}_{a}\!\cdot\!\left(\!\frac{\partial\vec{x}}{\partial\vec{v}}\right)^{\!\!\!T} \!\!\!=\!\!\left(\!\frac{\partial\vec{x}}{\partial\vec{v}}\right)\!\cdot\!\left(\!\!\frac{v^{2}\overline{\overline{\mathbf{1}}}\!-\!\vec{v}\vec{v}}{\alpha}\!+\!\frac{\vec{v}\vec{v}}{\beta}\!\right)\!\cdot\!\left(\!\frac{\partial\vec{x}}{\partial\vec{v}}\right)^{\!\!\!T}\\
&\text{where we can identify, using what we learned above,}\\
&\!\left(\!\frac{\partial\vec{x}}{\partial\vec{v}}\right)\!\overline{\overline{\mathbf{1}}}\!\left(\!\frac{\partial\vec{x}}{\partial\vec{v}}\right)^{\!\!\!T}\!\!\!=\!\!\left(\!\frac{\partial\vec{x}}{\partial\vec{v}}\right)\!\!\!\left(\!\frac{\partial\vec{x}}{\partial\vec{v}}\right)^{\!\!\!T} \!\!\!=\!\!\left[\!\!\left(\!\frac{\partial\vec{v}}{\partial\vec{x}}\right)^{\!\!\!T}\!\!\!\!\left(\!\frac{\partial\vec{v}}{\partial\vec{x}}\right)\!\!\right]^{-1}\!\!\!\!\!\!\!\!\!=\! (\overline{\overline{g}}_{mn})^{\!-1}\!\!= \overline{\overline{g}}^{mn}\\
&\!\left(\!\frac{\partial\vec{x}}{\partial\vec{v}}\right)\!\cdot\!\vec{v}\vec{v}\!\cdot\!\left(\!\frac{\partial\vec{x}}{\partial\vec{v}}\right)^{\!T}\!\!\!\!\!=\vec{x}\cdot\!\left[\!\left(\frac{\partial\vec{x}}{\partial\vec{v}}\right)\!\cdot\!\vec{v}\right]^{\!T} \!\!\!\!= \vec{x}\cdot\vec{x}^{T} \!\!= \vec{x}\vec{x}\end{align*}
Then in terms of our defined constants,
\begin{align*}&^{\vec{x}}\!\vec{A}_{a} = \gamma\vec{x}, \quad ^{\vec{x}}\overline{\overline{D}}_{a} = \frac{v_{th,a}^{2}x^{2}\overline{\overline{g}}^{-1}\!\!\!\!-\vec{x}\vec{x}}{\alpha}\!+\!\frac{\vec{x}\vec{x}}{\beta}\end{align*}
Substituting them, we obtain,
\begin{align*}\frac{\partial f_{a}}{\partial t} = \frac{1}{\sqrt{g}}\frac{\partial}{\partial\vec{x}}\cdot\left[\sqrt{g}\!\left(\!^{\vec{x}}\!\vec{A}f_{a}+^{\vec{x}}\!\overline{\overline{D}}_{a}\cdot\frac{\partial f_{a}}{\partial\vec{x}}\!\right)\right]\end{align*}
where the advection vector $^{\vec{x}}\!\vec{A}$, and diffusion tensor $^{\vec{x}}\overline{\overline{D}}_{a}$ are given for $x_{b}\equiv v/v_{th,b}$ by,
\begin{align*}\!^{\vec{x}}\!\vec{A} &= \sum_{b}\frac{C_{ab}}{v_{th,a}^{3}\!\!}\!\left\{\!\frac{m_{a}\vec{x}}{m_{b}x^{3}}\,\text{slp}(x_{b})\!\!\right\}, \,\,\,\text{slp}(x) \equiv \text{erf}(x)-\frac{2}{\sqrt{\pi}}xe^{-x^{2}}\\
\!\!\!^{\vec{x}}\overline{\overline{D}}_{a}\!&=\!\sum_{b}\!\frac{C_{ab}}{2v_{th,a}^{3}\!}\!\!\left\{\!\frac{v_{th,a}^{2}x^{2}\overline{\overline{g}}^{-1}\!\!\!\!\!\!\!-\!\vec{x}\vec{x}}{x^{3}}\!\!\left(\!\!\text{erf}(x_{b})\!-\!\frac{\text{slp}(x_{b})}{2x_{b}^{2}}\!\!\right)\!\!+\!\frac{\vec{x}\vec{x}}{x^{3}}\frac{\text{slp}(x_{b})}{x_{b}^{2}}\!\!\right\}\end{align*}
\section{Appendix D: From Tensors to the PDE}
\label{App:D}
\vspace{-\baselineskip}
\vspace{-\baselineskip}
\begin{align*}&\text{Our tensorial advection-diffusion equation is given by,}\\
&\frac{\partial f_{a}}{\partial t} = \frac{1}{\sqrt{g}}\frac{\partial}{\partial\vec{x}}\cdot\left[\sqrt{g}\!\left(\!^{\vec{x}}\!\vec{A}_{a}f_{a}+^{\vec{x}}\!\overline{\overline{D}}_{a}\cdot\frac{\partial f_{a}}{\partial\vec{x}}\!\right)\right]\\
&\text{where for simplicity, we can once again take,}\\
&^{\vec{x}}\!\vec{A}_{a}^{m} \equiv \gamma\vec{x}, \quad ^{\vec{x}}\overline{\overline{D}}_{a}^{mn} \equiv \frac{v_{th,a}^{2}x^{2}\overline{\overline{g}}^{-1}\!\!\!\!-\vec{x}\vec{x}}{\alpha} \!+\! \frac{\vec{x}\vec{x}}{\beta}\\
&\text{Then,}\\
&\frac{\partial f_{a}}{\partial t} = \frac{1}{\sqrt{g}}\frac{\partial}{\partial\vec{x}}\!\cdot\!\left[\!\sqrt{g}\!\left(\!\gamma\vec{x}f_{a}\!+\!\!\left(\frac{v_{th,a}^{2}x^{2}\overline{\overline{g}}^{-1}\!\!\!\!-\vec{x}\vec{x}}{\alpha} \!+\! \frac{\vec{x}\vec{x}}{\beta}\right)\!\cdot\frac{\partial f_{a}}{\partial\vec{x}}\!\right)\!\right]\\
&\text{And since $v_{th,a}^{2}x^{2}\overline{\overline{g}}^{-1} = \vec{x}\vec{x}+(1-\mu^{2})\mathbf{\hat{\mu}}\mathbf{\hat{\mu}}+\mathbf{\hat{\varphi}}\mathbf{\hat{\varphi}}/(1-\mu^{2})$},\\
&\!\!\!\!\frac{\partial f_{a}}{\partial t} \!\!=\!\! \frac{1}{\sqrt{g}}\frac{\partial}{\partial\vec{x}}\!\cdot\!\!\left[\!\sqrt{g}\!\left(\!\gamma\vec{x}f_{a}\!\!+\!\!\left(\!\frac{(1\!\!-\!\!\mu^{2})\mathbf{\hat{\mu}}\mathbf{\hat{\mu}}\!+\!\mathbf{\hat{\varphi}}\mathbf{\hat{\varphi}}/(1\!\!-\!\!\mu^{2})}{\alpha} \!+\! \frac{\vec{x}\vec{x}}{\beta}\right)\!\!\cdot\!\frac{\partial f_{a}}{\partial\vec{x}}\!\right)\!\!\right]\\
&\!\!\!\!\frac{\partial f_{a}}{\partial t} \!\!=\!\! \frac{1}{x^{2}}\frac{\partial}{\partial\vec{x}}\!\cdot\!\!\left[x^{2}\!\!\left(\!\!\gamma\vec{x}f_{a}\!+\!\frac{1}{\alpha}\!\!\left(\!\!(1\!-\!\mu^{2})\mathbf{\hat{\mu}}\frac{\partial f_{a}}{\partial\mu}\!+\!\frac{\mathbf{\hat{\varphi}}}{1\!-\!\mu^{2}}\frac{\partial f_{a}}{\partial\varphi}\!\right) \!\!+\! \frac{\vec{x}x}{\beta}\frac{\partial f_{a}}{\partial x}\!\right)\!\right]\\
&\!\!\!\!\frac{\partial f_{a}}{\partial t} \!\!=\!\! \frac{1}{x^{2}}\frac{\partial}{\partial x}\!\!\left[x^{3}\!\!\left(\!\!\gamma f_{a}\!+\!\frac{x}{\beta}\frac{\partial f_{a}}{\partial x}\!\right)\!\right]\!\!+\!\frac{1}{\alpha}\!\!\left[\!\frac{\partial}{\partial\mu}\!\!\left[(1\!-\!\mu^{2})\frac{\partial f_{a}}{\partial\mu}\!\right]\!\!+\!\frac{1}{1\!-\!\mu^{2}}\frac{\partial^{2}f_{a}}{\partial\varphi^{2}}\!\right]\!\!\\
&\text{Substituting our constants $\alpha$ and $\beta$,}\\
&\!\!\!\!\!\frac{\partial f_{a}}{\partial t}\!=\!\frac{1}{x^{2}}\frac{\partial}{\partial x}\!\!\left[\!\left\{\!\sum_{b}\frac{C_{ab}}{v_{th,a}^{3}}\frac{m_{a}}{m_{b}}\text{slp}(x_{b})\!\!\right\}\!\!f_{a}\!\!+\!\!\left\{\!\sum_{b}\frac{C_{ab}x}{2v_{th,a}^{3}}\frac{\text{slp}(x_{b})}{x_{b}^{2}}\!\right\}\!\!\frac{\partial f_{a}}{\partial x}\!\right]\!\!\\
&\!\!\!\!\!\!\!\!\!\!\!+\!\!\frac{1}{x^{3}}\!\!\left\{\!\sum_{b}\frac{C_{ab}}{2v_{th,a}^{3}\!}\!\!\left(\!\!\text{erf}(x_{b})\!-\!\frac{\text{slp}(x_{b})}{2x_{b}^{2}}\!\right)\!\!\right\}\!\!\left[\!\frac{\partial}{\partial\mu}\!\!\left[\!(1\!-\!\mu^{2})\frac{\partial f_{a}}{\partial\mu}\!\right]\!\!+\!\!\frac{1}{1\!-\!\mu^{2}}\frac{\partial^{2}f_{a}}{\partial\varphi^{2}}\!\right]\!\!\end{align*}
Until now our theory has been completely general and exact. Now let us consider the case where our ion species $a$ is faster than the other ion species, but slower than electrons, i.e. $v >> v_{th,b}$ for $b\neq e$ and $v << v_{th,e}$. To explore this regime, we'll need the following asymptotic expansions of our special functions,
\begin{align*}
&\underline{x_{b} >> 1}\\
&\text{erf}(x_{b})\sim 1 \!-\!\frac{e^{-x_{b}^{2}}}{\sqrt{\pi}}\!\left(\!\frac{1}{x_{b}}\!-\!\frac{1}{2x_{b}^{3}}\!+\!\frac{3}{4x_{b}^{5}}\!-\!\frac{15}{8x_{b}^{7}}\!\right)\!\sim 1 \!+\! \mathcal{O}\!\left(\!\frac{e^{-x_{b}^{2}}}{x_{b}}\!\right)\\
&\text{slp}(x_{b})\sim 1 \!-\! \frac{e^{-x_{b}^{2}}}{\sqrt{\pi}}\!\left(\!2x_{b}\!+\!\frac{1}{x_{b}}\!-\!\frac{1}{2x_{b}^{3}}\!+\!\frac{3}{4x_{b}^{5}}\!\right)\!\sim 1 \!+\! \mathcal{O}\!\left(x_{b}e^{-x_{b}^{2}}\right)\\
&\text{erf}(x_{b})\!-\!\frac{\text{slp}(x_{b})}{2x_{b}^{2}}\sim 1 \!-\!\frac{1}{2x_{b}^{2}}\!+\!\frac{e^{-x_{b}^{2}}}{\sqrt{\pi}}\!\!\left(\!\frac{1}{x_{b}^{3}}\!\right)\!\sim 1 \!-\! \frac{1}{2x_{b}^{2}}\!+\!\mathcal{O}\!\left(\!\frac{e^{-x_{b}^{2}}}{x_{b}^{3}}\!\right)\\
&\underline{x_{b} << 1}\\
&\text{erf}(x_{b})\sim\frac{2}{\sqrt{\pi}}\left(x_{b}-\frac{x_{b}^{3}}{3}+\frac{x_{b}^{5}}{5\cdot2!}-\frac{x_{b}^{7}}{7\cdot3!}\right)\!\sim\frac{2x_{b}}{\sqrt{\pi}}+\mathcal{O}(x_{b}^{3})\\
&\text{slp}(x_{b})\sim \frac{4}{\sqrt{\pi}}\left(\frac{x_{b}^{3}}{3}-\frac{x_{b}^{5}}{5}+\frac{x_{b}^{7}}{14}-\frac{x_{b}^{9}}{54}\right)\!\sim\frac{4x_{b}^{3}}{3\sqrt{\pi}}+\mathcal{O}(x_{b}^{5})\\
&\text{erf}(x_{b})\!-\!\frac{\text{slp}(x_{b})}{2x_{b}^{2}}\sim \frac{4}{\sqrt{\pi}}\!\left(\frac{x_{b}}{3}-\frac{x_{b}^{3}}{15}+\frac{x_{b}^{5}}{70}\right)\!\sim \frac{4x_{b}}{3\sqrt{\pi}}+\mathcal{O}(x_{b}^{3})\end{align*}
\vspace{-\baselineskip}
\\ \\Expanding in the appropriate limits for each species,\\
\vspace{-0.7\baselineskip}
\vspace{-0.55\baselineskip}
\begin{align*}&\frac{\partial f_{a}}{\partial t} = \frac{1}{x^{2}}\frac{\partial}{\partial x}\!\!\Biggl[\!\left\{\!\sum_{b\neq e}\frac{C_{ab}}{v_{th,a}^{3}}\frac{m_{a}}{m_{b}}\!+\!\frac{C_{ae}}{v_{th,a}^{3}}\frac{m_{a}}{m_{e}}\frac{4x_{e}^{3}}{3\sqrt{\pi}}\!\!\right\}\!f_{a}\\
&\!\!\!\!+\!\left\{\sum_{b\neq e}\frac{C_{ab}x}{2v_{th,a}^{3}}\frac{1}{x_{b}^{2}}\!+\!\frac{C_{ae}x}{2v_{th,a}^{3}}\frac{4x_{e}}{3\sqrt{\pi}}\!\right\}\!\frac{\partial f_{a}}{\partial x}\!\Biggr]\\
&\!\!\!\!+\!\!\frac{1}{x^{3}}\!\!\left\{\!\sum_{b\neq e}\frac{C_{ab}}{2v_{th,a}^{3}\!}\!\!\left(\!1\!-\!\frac{1}{2x_{b}^{2}}\!\right)\!\!+\!\frac{C_{ae}}{2v_{th,a}^{3}\!}\frac{4x_{e}}{3\sqrt{\pi}}\!\!\right\}\\
&\cdot\left[\!\frac{\partial}{\partial\mu}\!\left[(1\!-\!\mu^{2})\frac{\partial f_{a}}{\partial\mu}\!\right]\!\!+\!\frac{1}{1\!-\!\mu^{2}}\frac{\partial^{2}f_{a}}{\partial\varphi^{2}}\!\right]\!\!\end{align*}
\vspace{-0.7\baselineskip}

\noindent where we have neglected same species collisions since we have trace alpha particles $n_{a} << n_{i},n_{e}$,
\vspace{-0.5\baselineskip}
\begin{align*}
&\quad\quad\frac{\partial f_{a}}{\partial t} \!=\! \frac{1}{x^{2}}\frac{\partial}{\partial x}\!\!\Biggl[\Biggl[\!\left\{\!\sum_{b\neq e}\frac{C_{ab}}{v_{th,a}^{3}}\frac{m_{a}}{m_{b}}\!\right\}\!\!+\!\!\left\{\frac{4}{3\sqrt{\pi}}\frac{C_{ae}}{v_{th,a}^{3}}\frac{m_{a}v_{th,a}^{3}}{m_{e}v_{th,e}^{3}}\!\!\right\}\!x^{3}\!\Biggr]\!f_{a}\\
&\quad\quad\!\!\!\!\!+\!\frac{1}{2}\!\Biggl[\!\left\{\!\!\sum_{\,b\neq e}\!\frac{C_{ab}}{v_{th,a}^{3}}\frac{v_{th,b}^{2}}{v_{th,a}^{2}}\!\!\right\}\!\!\frac{1}{x}\!+\!\left\{\!\!\frac{4}{3\sqrt{\pi}}\frac{C_{ae}}{v_{th,a}^{3}\!}\frac{v_{th,a}}{v_{th,e}}\!\!\right\}\!x^{2}\!\Biggr]\!\frac{\partial f_{a}}{\partial x}\!\Biggr]\\
&\quad\quad\!\!\!\!\!+\!\frac{1}{2x^{3}}\!\!\!\left[\!\left\{\!\!\sum_{\,b\neq e}\!\!\frac{C_{ab}}{v_{th,a}^{3}}\!\right\}\!\!-\!\!\left\{\!\!\sum_{\,b\neq e}\!\!\frac{C_{ab}}{v_{th,a}^{3}}\!\frac{v_{th,b}^{2}}{v_{th,a}^{2}}\!\!\right\}\!\!\frac{1}{2x^{2}}\!+\!\left\{\!\!\frac{4}{3\sqrt{\pi}}\frac{C_{ae}}{v_{th,a}^{3}}\frac{v_{th,a}}{v_{th,e}}\!\!\right\}\!x\!\right]\\
&\quad\quad\!\!\!\cdot\left[\!\frac{\partial}{\partial\mu}\!\!\left[\!(1\!-\!\mu^{2})\frac{\partial f_{a}}{\partial\mu}\right]\!\!\!+\!\!\frac{1}{1\!-\!\mu^{2}}\frac{\partial^{2}f_{a}}{\partial\varphi^{2}}\!\right]\!\!\end{align*}
And dividing through by the coefficient in front of the $(\partial f_{a}/\partial x)/2x$ term we have,
\vspace{-0.7\baselineskip}
\begin{align*}
&\quad\tau_{0}^{i}\frac{\partial f_{a}}{\partial t} \!=\! \frac{1}{x^{2}}\frac{\partial}{\partial x}\!\!\left[\left(\!Z_{\parallel}^{i}\!+\!Z_{\parallel}^{e}x^{3}\!\right)\!f_{a}\!+\!\frac{1}{2}\!\left(\frac{1}{x}\!+\!Z_{\perp}^{e}x^{2}\!\right)\!\frac{\partial f_{a}}{\partial x}\right]\\
&\quad\quad\!\!\!\!+\!\frac{1}{2x^{3}}\!\!\left(Z_{\perp}^{i}\!+\!Z_{\perp}^{e}x\!-\!\frac{1}{2x^{2}}\!\right)\!\!\left[\!\frac{\partial}{\partial\mu}\!\!\left[(1\!-\!\mu^{2})\frac{\partial f_{a}}{\partial\mu}\right]\!\!+\!\frac{1}{1\!-\!\mu^{2}}\frac{\partial^{2}f_{a}}{\partial\varphi^{2}}\!\right]\!\!
\end{align*}
where our transport coefficients are given by,
\begin{align*}&(\tau_{0}^{i})^{-1}\!\equiv\! \frac{4\pi e^{4}}{(4\pi\epsilon_{0})^{2}}\!\!\sum_{b\neq e}\frac{n_{b}Z_{a}^{2}Z_{b}^{2}\lambda_{ab}}{m_{a}^{2}v_{th,a}^{3}}\frac{m_{a}T_{b}}{m_{b}T_{a}}\\
&(\tau_{0}^{e})^{-1}\!\equiv \frac{4\pi e^{4}}{(4\pi\epsilon_{0})^{2}}\frac{\frac{4}{3\sqrt{\pi}}n_{e}Z_{a}^{2}\lambda_{ae}}{m_{a}^{2}v_{th,a}^{3}}\!\left(\!\frac{m_{e}T_{a}}{m_{a}T_{e}}\!\right)^{1/2}\!\!\!\!\!\!\!\!\\
&Z_{\parallel}^{i}\equiv \frac{\sum_{b\neq e}n_{b}Z_{b}^{2}\lambda_{ab}\frac{m_{a}}{m_{b}}}{\sum_{b\neq e}n_{b}Z_{b}^{2}\lambda_{ab}\frac{m_{a}T_{b}}{m_{b}T_{a}}} = \frac{\sum_{b\neq e}n_{b}Z_{b}^{2}\lambda_{ab}T_{a}/m_{b}}{\sum_{b\neq e}n_{b}Z_{b}^{2}\lambda_{ab}T_{b}/m_{b}}\\
&Z_{\parallel}^{e} \equiv \frac{\frac{4}{3\sqrt{\pi}}n_{e}\lambda_{ae}\frac{m_{a}}{m_{e}}\!\left(\!\frac{m_{e}T_{a}}{m_{a}T_{e}}\!\right)^{3/2}}{\sum_{b\neq e}n_{b}Z_{b}^{2}\lambda_{ab}\frac{m_{a}T_{b}}{m_{b}T_{a}}} = \frac{\frac{4}{3\sqrt{\pi}}n_{e}\lambda_{ae}\left(\!\frac{m_{e}T_{a}^{5}}{m_{a}^{3}T_{e}^{3}}\!\right)^{1/2}}{\sum_{b\neq e}n_{b}Z_{b}^{2}\lambda_{ab}T_{b}/m_{b}}\\
&Z_{\perp}^{i} \equiv \frac{\sum_{b\neq e}n_{b}Z_{b}^{2}\lambda_{ab}}{\sum_{b\neq e}n_{b}Z_{b}^{2}\lambda_{ab}\frac{m_{a}T_{b}}{m_{b}T_{a}}} = \frac{\sum_{b\neq e}n_{b}Z_{b}^{2}\lambda_{ab}T_{a}/m_{a}}{\sum_{b\neq e}n_{b}Z_{b}^{2}\lambda_{ab}T_{b}/m_{b}}\\
&Z_{\perp}^{e}=\frac{\frac{4}{3\sqrt{\pi}}n_{e}\lambda_{ae}\!\left(\!\frac{m_{e}T_{a}}{m_{a}T_{e}}\!\right)^{1/2}}{\sum_{b\neq e}n_{b}Z_{b}^{2}\lambda_{ab}\frac{m_{a}T_{b}}{m_{b}T_{a}}} = \frac{\frac{4}{3\sqrt{\pi}}n_{e}\lambda_{ae}\!\left(\!\frac{m_{e}T_{a}^{3}}{m_{a}^{3}T_{e}}\!\right)^{1/2}}{\sum_{b\neq e}n_{b}Z_{b}^{2}\lambda_{ab}T_{b}/m_{b}} = \frac{\tau_{0}^{i}}{\tau_{0}^{e}}
\end{align*}
We compute the Coulomb logarithms\!\,\,$\lambda_{\alpha\beta}$\! in Appendix \hyperref[App:E]{E}.\\
We solve the PDE in the main text and Appendix \hyperref[App:F]{F}.
\section{Appendix E: Coulomb Logarithms - Classical Model of Collisions}
\label{App:E}
\begin{figure}[h]
\centering
\noindent\includegraphics[width = \columnwidth, height = 0.6\columnwidth]{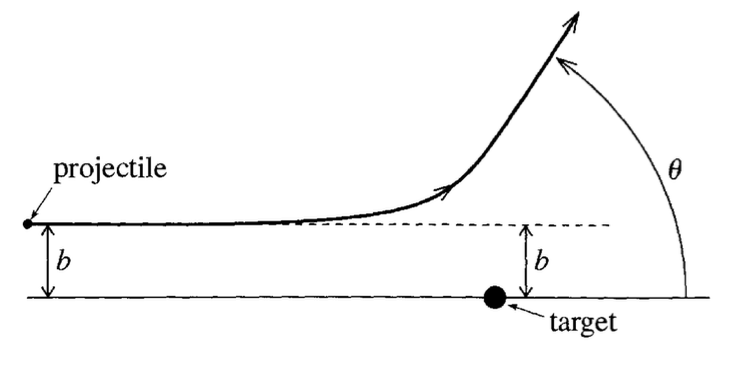}
\caption{Rutherford Scattering. Shown here is the deflection of one charged particle off another, with impact parameter $b$ and corresponding scattering angle $\theta$.}\label{Fig:Rutherford}
\end{figure}

The Coulomb logarithm for classical collisions of charged particles of species $\alpha,\beta$ is given by,
\begin{align*}
\boxed{\lambda_{\alpha\beta} \equiv \ln\Lambda_{\alpha\beta} = \ln\left(\frac{\lambda_{D}}{b_{\pi/2}^{\alpha\beta}}\right)}
\end{align*}
where the Debye length $\lambda_{D}$ is given by,
\begin{align*}
\boxed{\lambda_{D} = \left(\sum\limits_{\gamma}\frac{n_{\gamma}q_{\gamma}^{2}}{\epsilon_{0}k_{B}T_{\gamma}}\right)^{-1/2}}
\end{align*}
and the sum is taken over all species $\gamma$. The large angle impact parameter for collisions of species $\alpha$ with species $\beta$, namely $b_{\pi/2}^{\alpha\beta}$ can be derived as follows. Consider the setup shown in Figure~\ref{Fig:Rutherford} for one charged particle deflecting off another.
\\
The Lagrangian is given by,
\begin{align*}
\mathcal{L} = \frac{1}{2}M_{\alpha\beta}\dot{R}^{2}+\frac{1}{2}m_{\alpha\beta}\left(\dot{r}^{2}+r^{2}\dot{\theta}^{2}\right)-\frac{q_{\alpha}q_{\beta}}{4\pi\epsilon_{0}r}
\end{align*}
where,
\begin{align*}
&\vec{r} \equiv \vec{r}_{\alpha}-\vec{r}_{\beta}, \quad \vec{R} \equiv \frac{m_{\alpha}\vec{r}_{\alpha}+m_{\beta}\vec{r}_{\beta}}{m_{\alpha}+m_{\beta}}\\
&m_{\alpha\beta} \equiv \frac{m_{\alpha}m_{\beta}}{m_{\alpha}+m_{\beta}}, \quad M_{\alpha\beta} = m_{\alpha}+m_{\beta}
\end{align*}
Euler-Lagrange Equations yield,
\begin{align*}
&\frac{d}{dt}\!\left(\frac{\partial\mathcal{L}}{\partial\dot{r}}\right) = \frac{\partial\mathcal{L}}{\partial r} \rightarrow m_{\alpha\beta}\ddot{r} = \frac{q_{\alpha}q_{\beta}}{r^{2}}\\
&\frac{d}{dt}\!\left(\frac{\partial\mathcal{L}}{\partial\dot{\varphi}}\right) = \frac{\partial\mathcal{L}}{\partial \varphi} \rightarrow 
l \equiv m_{\alpha\beta}r^{2}\dot{\theta} = const.
\end{align*}
Defining $u \equiv b/r$ we obtain Binet's equation,
\begin{align*}\frac{d^{2}u}{d\theta^{2}} + u = -\cot\alpha, \quad \tan\alpha \equiv \frac{4\pi\epsilon_{0}\mu b v_{0}^{2}}{q_{\alpha}q_{\beta}}\end{align*}
whose general solution is given by,
\begin{align*}
u(\theta) = C_{1}\cos\theta + C_{2}\sin\theta - \cot\alpha
\end{align*}
Imposing boundary conditions $(u,u',\theta_{0}) = (0,-1,\pi) \rightarrow (u,\theta_{f}) = (0,\theta_{\infty})$ we find,
\begin{align*}&0 = -C_{1} - \cot\alpha \rightarrow C_{1} = -\cot\alpha\\
&-1 = -C_{2} \rightarrow C_{2} = 1\\
&0 = -\cot\alpha\cos\theta_{\infty}+C_{2}\sin\theta_{\infty} - \cot\alpha\\ &\tan\left(\!\frac{\theta_{\infty}}{2}\!\right) = \frac{\sin\theta_{\infty}}{1+\cos\theta_{\infty}} = \cot\alpha = \frac{q_{\alpha}q_{\beta}}{4\pi\epsilon_{0}\mu b^{\alpha\beta} v_{0}^{2}}\end{align*}
Then, for the case $\theta_{\infty} = \pi/2$ we have,
\begin{align*}
&\tan\left(\frac{\pi}{4}\right) = \frac{q_{\alpha}q_{\beta}}{4\pi\epsilon_{0}\mu b_{\pi/2}^{\alpha\beta} v_{0}^{2}} \rightarrow \boxed{b_{\pi/2}^{\alpha\beta} = \frac{q_{\alpha}q_{\beta}}{4\pi\epsilon_{0}m_{\alpha\beta} \langle v_{0}^{2}\rangle}}\end{align*}
The only remaining unknown is an estimate of the mean squared relative velocity $\langle v_{0}^{2} \rangle$, which we can compute as follows. The relative velocity is defined as,
\begin{align*}\vec{v}_{0} \equiv \vec{v}_{\alpha}-\vec{v}_{\beta} \rightarrow v_{0}^{2} = (\vec{v}_{\alpha}-\vec{v_{\beta}})\cdot(\vec{v}_{\alpha}-\vec{v_{\beta}})\end{align*}
If we choose some given direction for $\vec{v}_{\alpha}$ the direction of $\vec{v}_{\beta}$ can be described by a relative pair of random variable angles $(\theta,\varphi) \in [0,\pi]\times[0,2\pi)$. In that case,
\begin{align*}v_{0}^{2} = &v_{\alpha}^{2}+v_{\beta}^{2}-2v_{\alpha}v_{\beta}\cos\theta\\
&\langle v_{0}^{2}\rangle = \frac{1}{4\pi}\int\limits_{0}^{2\pi}\!\!\int\limits_{0}^{\pi}\!\left(v_{\alpha}^{2}+v_{\beta}^{2}-2v_{\alpha}v_{\beta}\cos\theta\right)\sin\theta\,d\theta\,d\varphi\\
&\langle v_{0}^{2}\rangle = \frac{1}{2}\left[\frac{1}{2}\left(v_{\alpha}^{2}+v_{\beta}^{2}-2v_{\alpha}v_{\beta}\cos\theta\right)^{2}\frac{1}{2v_{\alpha}v_{\beta}}\right|_{0}^{\pi}\\
&\langle v_{0}^{2}\rangle = \frac{1}{8v_{\alpha}v_{\beta}}\left[\left|v_{\alpha}+v_{\beta}\right|^{2}-|v_{\alpha}-v_{\beta}|^{2}\right]
\end{align*}
Assuming without loss of generality that $v_{\alpha} > v_{\beta}$, and taking $v_{\alpha},v_{\beta} \sim v_{th,\alpha},v_{th,\beta}$ where $v_{th,\gamma}\equiv\sqrt{2k_{B}T_{\gamma}/m_{\gamma}}$,
\begin{align*}
&\langle v_{0}^{2}\rangle = \frac{1}{8v_{\alpha}v_{\beta}}\left[8v_{\alpha}^{3}v_{\beta}+8v_{\alpha}v_{\beta}^{3}\right]\rightarrow \boxed{\langle v_{0}^{2}\rangle = v_{th,\alpha}^{2}+v_{th,\beta}^{2}}
\end{align*}
where the symmetry in $\alpha,\beta$ retroactively makes our assumption about $\alpha$ being faster irrelevant. The boxed equations now constitute a complete set sufficient to determine the Coulomb logarithm $\lambda_{\alpha\beta}$ for any pair of species $\alpha,\beta$ in SI units.\\ \\

\section{Appendix F: Method of Characteristics}
\label{App:F}
\noindent Consider the advection equation,
\begin{align*}\frac{\partial g}{\partial t} + \frac{1}{x^{2}}\frac{\partial}{\partial x}\!\left(x^{2}v(x)g\right) + \lambda\nu(x)g = 0\end{align*}
where,
\begin{align*}&v(x)\equiv-\frac{Z_{\parallel}(x)}{\tau_{0}^{i}x^{2}}, \quad \nu(x)\equiv\frac{Z_{\perp}\!(x)}{2\tau_{0}^{i}x^{3}}\end{align*}
Define $G(t,x)\equiv x^{2}v(x)g(t,x)$,
\begin{align*}&\frac{\partial G}{\partial t} + v(x)\frac{\partial G}{\partial x} = -\lambda\nu(x)G\end{align*}
Consider characteristic paths $X(T)$ satisfying,
\begin{align*}
&\frac{dX}{dT}=v(X) \rightarrow \int\frac{-\tau_{0}^{i}X^{2}}{Z_{\parallel}^{i}\!+\!Z_{\parallel}^{e}X^{3}}\,dX =\!\int \!dT\\
&\frac{-\tau_{0}^{i}}{3Z_{\parallel}^{e}}\ln\!\left(X^{3}\!\!+\eta^{3}\right) = T - C\\
&X(T;C) = \left[e^{\!-3Z_{\parallel}^{e}(T-C)/\tau_{0}^{i}}-\eta^{3}\right]^{1/3}\end{align*}
where $C\in\mathbb{R}$ is a constant of integration labeling one of infinitely many characteristic paths satifying $dX/dT = v(X)$. Let us choose the path $C_{\star}$ that passes through $(X,T) = (x,t)$, namely that for which $X(t) = x$,
\begin{align*}&X(t;C_{\star}) = \left[e^{-3Z_{\parallel}^{e}(t-C_{\star}\!)/\tau_{0}^{i}}-\eta^{3}\right]^{1/3} \!\!\!\!\!\!=\, x\\
&C_{\star} = t+\frac{\tau_{0}^{i}}{3Z_{\parallel}^{e}}\ln(x^{3}\!+\eta^{3})\end{align*}
Then our characteristic path of interest is,
\begin{align*}\boxed{X(T;t,x) = \left[(x^{3}\!+\eta^{3})e^{3Z_{\parallel}^{e}(t-T)/\tau_{0}^{i}}\!-\eta^{3}\right]^{1/3}}\end{align*}
Now let us define $\mathscr{G}(T)\equiv G(T,X(T))$ with $dt/dT = 1$,
\begin{align*}&\frac{d\mathscr{G}}{dT} = \frac{\partial G}{\partial T}+\frac{dX}{dT}\frac{\partial G}{\partial X} = -\lambda \nu(X)\mathscr{G}\\
&\mathscr{G}(T) \!=\! \mathscr{G}(0)\exp\!\!\left(\!\!-\!\!\!\int\limits_{0}^{T}\!\!\lambda(\mu_{b}(X(T)\!)\!)\nu(X(T)\!)\,dT\!\!\right)\!\!\equiv\! \mathscr{G}(0)e^{-\zeta(T)}\end{align*}
And by definition then,
\begin{align*}&G(T,\!X(T)\!) = G(0,\!X(0)\!)e^{-\zeta(T)}\end{align*}
And in turn,
\begin{align*}&g(T,\!X(T)\!) \!=\! \frac{X(0)^{2}}{X(T)^{2}}\frac{\!v(X(0)\!)}{v(X)}g(0,\!X\!(0)\!)e^{-\zeta(T)}\end{align*}
And at $(T,X(T)) = (t,x)$,
\begin{align*}g(t,x\!) \!=\! \frac{X(0)^{3}\!+\eta^{3}}{x^{3}\!+\eta^{3}}g(0,X(0)\!)e^{-\zeta(t)}\end{align*}
Consider the delta shell initial condition,
\begin{align*}\boxed{g(0,x) = \frac{\delta(x-x_{0})}{x^{2}}}\end{align*}
Then,
\begin{align*}g(t,x\!) \!=\! \frac{X(0)^{3}\!+\eta^{3}}{x^{3}\!+\eta^{3}}\frac{\delta(X(0)\!-\!x_{0})}{X(0)^{2}}e^{-\zeta(t)}\end{align*}
where we can observe that,
\begin{align*}X(0) = \left[(x^{3}\!+\eta^{3})e^{3Z_{\parallel}^{e}t/\tau_{0}^{i}}\!-\eta^{3}\right]^{1/3}\end{align*}
And that therefore,
\begin{align*}\frac{X(0)^{3}\!+\eta^{3}}{x^{3}\!+\eta^{3}}\frac{\delta(X(0)\!-\!x_{0})}{X(0)^{2}} = e^{3Z_{\parallel}^{e}t/\tau_{0}^{i}}\frac{\delta\left(z(x)-x_{0}\right)}{z(x)^{2}}\end{align*}
where,
\begin{align*}&z(x)\equiv\left[(x^{3}\!+\eta^{3})e^{3Z_{\parallel}^{e}t/\tau_{0}^{i}}\!-\eta^{3}\right]^{1/3}\!\!\!\!\rightarrow\,z'(x) = \frac{x^{2}}{z(x)^{2}}e^{3Z_{\parallel}^{e}t/\tau_{0}^{i}}\end{align*}
Then we can use the property of the delta function,
\begin{align*}\delta(f(x)) = \!\!\sum\limits_{\substack{i:f(x_{i})=0}}\frac{\delta(x-x_{i})}{|f'(x_{i})|}\end{align*}
where
$\!z(x(t)) \!=\! x_{0}$. Then our trajectory is given by,
\begin{align*}\boxed{x(t) \!\equiv\! \left[(x_{0}^{3}\!+\!\eta^{3})e^{-3Z_{\parallel}^{e}t/\tau_{0}^{i}}\!-\!\eta^{3}\right]^{1/3}}\end{align*}
And our delta function can be rewritten,
\begin{align*}e^{3Z_{\parallel}^{e}t/\tau_{0}^{i}}\frac{\delta\left(z(x)-x_{0}\right)}{z(x)^{2}} = \frac{e^{3Z_{\parallel}^{e}t/\tau_{0}^{i}}}{z(x)^{2}}\frac{\delta(x-x(t)\!)}{|z'(x(t))|} = \frac{\delta(x-x(t)\!)}{x(t)^{2}}\end{align*}
Putting it all together we are left with,
\begin{align*}g(t,x) = \frac{\delta(x-x(t)\!)}{x(t)^{2}}\exp\!\left(\!-\!\!\int\limits_{0}^{t}\!\lambda(\mu_{b}(X(T)\!)\!)\nu(x(T)\!)\,dT\!\!\right)\end{align*}
And finally, we use the delta function and observe that $X(T;t,x(t)\!)=x(T)$ to perform a routine change of variables in the integral and obtain,
\begin{align*}\boxed{g(t,x) = \frac{\delta(x-x(t)\!)}{x(t)^{2}}\exp\!\left(\!-\!\!\int\limits_{0}^{t}\!\lambda(\mu_{b}(x(t))\!)\!)\nu(x(t)\!)\,dt\!\!\right)}\end{align*}
which makes perfect sense since our original equation is conservative in the absence of pitch-angle scattering $\nu(x)=0$, and our solution should therefore be a moving delta function, with velocity $v(x)$, appropriately scaled by the Jacobian of our space $x^{2}$ to conserve particles.

\section{Appendix G: Physical Optics (PO) Approximation}
\label{App:G}
\noindent Our general distribution in pitch-angle is given by,
\begin{align*}h(\mu|\mu_{b})\!\sim\!\frac{1}{(1\!-\!\mu^{2})^{1/4}}\!\left(\!a\cos(\sqrt{\lambda}\sin^{\!-1}\!\mu)\!+\!b\sin(\sqrt{\lambda}\sin^{\!-1}\!\mu)\!\right)\end{align*}
Our boundary conditions $h(\pm\mu_{b}|\mu_{b}) = 0$ allow two possible solutions,
\begin{align*}&a\!\neq\!0,b\!=\!0 \rightarrow\! \lambda_{k}^{(i)} \!\!=\! \frac{\!(k\!+\!1/2)^{2}\pi^{2}\!}{(\sin^{\!-1}\!\!\mu_{b})^{2}},\,\!h_{k}^{(i)} \!\sim\! \frac{\cos\!\left(\!\!\sqrt{\!\lambda_{k}^{(i)}}\!\sin^{\!-1}\!\!\mu\!\!\right)}{(1\!-\!\mu^{2})^{1/4}}\\
&a\!=\!0,b\!\neq\!0 \rightarrow\!\lambda_{k}^{(ii)} \!\!=\! \frac{\!k^{2}\pi^{2}\!}{(\sin^{\!-1}\!\!\mu_{b})^{2}},\,\!h_{k}^{(ii)} \!\sim\! \frac{\sin\!\left(\!\!\sqrt{\!\lambda_{k}^{(ii)}}\!\sin^{\!-1}\!\!\mu\!\!\right)}{(1\!-\!\mu^{2})^{1/4}}\end{align*}
Then, our general solution for $f(t,x,\mu)$ is a linear combination of these two solutions,
\begin{align*}&\!f(t,x,\mu)\!\sim\!\!\frac{\delta(x\!-\!x(t)\!)}{x(t)^{2}}\!\sum\limits_{k=0}^{\infty}\biggl[a_{k}h_{k}^{\!(i)}\!e^{\!\!-\!\!\int\limits_{0}^{t}\!\!\lambda_{k}^{\!(i)}\!\nu dt}\!\!\!\!\!\!\!\!\!\!\!+\!b_{k}h_{k}^{\!(ii)}\!e^{\!\!-\!\!\int\limits_{0}^{t}\!\!\lambda_{k}^{\!(ii)}\! \nu dt}\!\biggr]\!\!\!\!\end{align*}
If we start with a delta shell in velocity, uniformly distributed in angle, then our initial condition is,
\begin{align*}
f(0,x,\mu) \!=\! \hat{n}_{0}\frac{\delta(x\!-\!x_{0})\!}{x_{0}^{2}}\frac{H(\mu_{b}\!-\!|\mu|)}{2\mu_{b}}
\end{align*}
where $H(x)$ is the Heaviside step function. However since $a$ and $b$ are functions, we will need more information to completely define them. If we further impose that in the absence of pitch-angle scattering, particles are conserved and the distribution remains uniform,
\begin{align*}\lim_{\nu\rightarrow0}f(t,x,\mu) = \hat{n}_{0}\frac{\delta(x\!-\!x(t))\!}{x(t)^{2}}\frac{H(\mu_{b}\!-\!|\mu|)}{2\mu_{b}}\end{align*}
Then, defining $z \!\equiv\! \sin^{\!-1}\!\!\mu/\sin^{\!-1}\!\!\mu_{b},\zeta\equiv\sin^{\!-1}\!\!\mu_{b}$ and renormalizing $a_{k},b_{k} \rightarrow \hat{a}_{k},\hat{b}_{k}$ by $\hat{n}_{0}/2\mu_{b}$ we can match our coefficients via,
\begin{align*}\sum_{k=0}^{\infty}\left[\hat{a}_{k}\cos\!\left(\!(k\!+\!1/2)\pi z\right)+\hat{b}_{k}\sin\!\left(k\pi z\right)\right] = \sqrt{|\cos(\zeta z)|}\phi(z)\end{align*}
where since we do not have a constant term available for matching on the LHS, we periodically extend the RHS beyond our interval to be the even square wave $\phi(z)$ with average value zero,
\begin{align*}
\phi(z)\!\equiv\! \begin{cases}1, \,z\!\in\![-1,1]\quad \phi(z\!+\!4n) = \phi(z)\\
\!-1, \,z\!\in\!(1,3]\quad\quad\quad\forall\,\,n\in\mathbb{Z}\end{cases}\!\!\!\!\!\!\!\!\!\!\!\!\!\!\!\!\!\!\!\!\!\!\!\!\!\!\!\!\!\!\!\!\!\!\!\!\!\!\!\!\!\!\!\!\!\!\!\!\!\!\!\!\!\!,\quad\quad\quad\quad\quad\quad\quad
\end{align*}
Since our $\phi(z)$ is now even, we can immediately discard all the $b_{k}$. The $a_{k}$ can be found via Fourier's trick,
\begin{align*}\hat{a}_{k} =\! \int_{-2}^{2}\!\!\!\!\!\sqrt{|\cos(\zeta z)|}\phi(z)\cos\left(\!(k\!+\!1/2)\pi z\right)dz,\,\,\,\, b_{k} = 0\end{align*}
which can be rewritten,
\begin{align*}\hat{a}_{k}\!=\!\!\int\limits_{0}^{1}\!\!w(z)dz\!-\!\!\!\int\limits_{1}^{2}\!\!w(z)dz,\, w(z)\!\equiv\!\!\sqrt{|\!\cos(\zeta z)|}\cos\!\left(\!(k\!+\!\!1/2)\pi z\right)\end{align*}
Putting everything together we obtain,
\begin{align*}&f(t,x,\mu)\!\sim\!\frac{\hat{n}_{0}}{2\mu_{b}}\frac{\delta(x\!-\!x(t)\!)}{x(t)^{2}}\sum\limits_{k=0}^{\infty}\hat{a}_{k}\!\cos\!\left(\!\!\sqrt{\!\lambda_{k}}\sin^{\!\!-1}\!\!\mu\!\right)\nonumber\\
&\cdot \exp\!\left(\!\!-\!\!\int\limits_{0}^{t}\!\!\lambda_{k}(\mu_{b}(x(t)\!)\!)\nu(x(t)\!)dt\!\!\right)\!\!,\, |\mu| < \mu_{b}(x)\!\end{align*}
\noindent Then at any given time our remaining density is simply,
\begin{align*}n(t) \!\equiv\! \!\smashoperator[r]{\int\limits_{0}^{\infty}}\!\!\!\smashoperator[r]{\int\limits_{\,-\mu_{b}(x)}^{\,\mu_{b}(x)}} \!\!f(t,x,\mu)x^{2}d\mu\,dx
\end{align*}
which yields,
\begin{align*}n(t) \!\sim\frac{\hat{n}_{0}}{2\mu_{b}}\sum\limits_{k=0}^{\infty}\hat{a}_{k}\!\!\!\!\!\!\!\smashoperator[r]{\int\limits_{\,-\mu_{b}(x)}^{\,\mu_{b}(x)}} \!\!\cos\!\left(\!\!\sqrt{\lambda_{k}}\sin^{\!-1}\!\!\mu\!\right)\!d\mu\exp\!\!\left(\!\!-\!\!\int\limits_{0}^{t}\!\!\lambda_{k}\nu(x)dt\!\right)\end{align*}
where our eigenvalues of interest have, just as in the geometrical optics approximation, turned out to be $\lambda_{k}\equiv\lambda_{k}^{(i)}$, and we've written $x\equiv x(t), \mu_{b}\equiv\mu_{b}(x)$, $\lambda_{k}\equiv\lambda_{k}(\mu_{b})$ for brevity. Carrying out our integral in $\mu$,
\begin{align*}n(t|x_{0}) \!\sim \hat{n}_{0}\frac{\mu_{b}(x_{0})}{\mu_b}\frac{\sum\limits_{k=0}^{\infty}\hat{a}_{k}I_{\mu}(x)\exp\!\left(\!\!-\!\!\int\limits_{0}^{t}\!\lambda_{k}\nu(x)dt\!\right)}{\sum\limits_{k=0}^{\infty}\hat{a}_{k}I_{\mu}(x_{0})}\end{align*}
where,
\begin{align*}I_{\mu}(\mu_{b})\equiv\!\!\!\int\limits_{-\mu_{b}}^{\mu_{b}}\!\!\frac{\cos\left(\sqrt{\lambda_{k}}\sin^{\!-1}\!\!\mu\right)}{(1-\mu^{2})^{1/4}}d\mu\end{align*}
And using our bijection between $x(t)\leftrightarrow t(x)$,
\begin{align*}n(x|x_{0}) \!\sim \hat{n}_{0}\frac{\mu_{b}(x_{0})}{\mu_b}\frac{\sum\limits_{k=0}^{\infty}\hat{a}_{k}I_{\mu}(\mu_{b})\exp\!\left(\!\!-\!\!\int\limits_{x}^{x_{0}}\!\lambda_{k}\frac{Z_{\perp}\!(x)}{Z_{\parallel}(x)}\frac{dx}{2x}\!\right)}{\sum\limits_{k=0}^{\infty}\hat{a}_{k}(\mu_{b}(x_{0})\!)I_{\mu}(\mu_{b}(x_{0})\!)}\end{align*}
In the zero potential limit $\Phi_{a} = 0$,
\begin{align*}\!n(x|x_{0}) \!\sim \hat{n}_{0}\frac{\sum\limits_{k=0}^{\infty}\!\hat{a}_{k}(\mu_{b0})I_{\mu}(\mu_{b0})\!\left(\!\frac{x^{3}}{x_{0}^{3}}\!\frac{x_{0}^{3}+\eta^{3}}{x^{3}+\eta^{3}}\!\right)^{\!\beta_{k}/6}\!\!\!\!\!\!\!\!}{\sum\limits_{k=0}^{\infty}\!\hat{a}_{k}(\mu_{b0})I_{\mu}(\mu_{b0})},\,\beta_{k}\!\equiv\!\lambda_{k}(\mu_{b0})\frac{Z_{\perp}^{i}}{Z_{\parallel}^{i}}\!\!\end{align*}
Correspondingly,
\begin{align*}n(t|x_{0})\sim \hat{n}_{0}\frac{\sum\limits_{k=0}^{\infty}\!\hat{a}_{k}(\mu_{b0})I_{\mu}(\mu_{b0})\!\left(\!1\!-\!\frac{\eta^{3}}{x_{0}^{3}}\!\left(e^{3Z_{\parallel}^{e}t/\tau_{0}^{i}}\!-\!1\right)\!\!\right)^{\!\beta_{k}/6}\!\!\!\!\!\!\!\!}{\sum\limits_{k=0}^{\infty}\hat{a}_{k}(\mu_{b0})I_{\mu}(\mu_{b0})}\!\!\end{align*}
\section{Appendix H: Probability Distributions}
\label{App:H}
We compute the probability distribution functions (PDFs) of normalized velocities $x$ at which alpha particles become deconfined. We are concerned here only with those particles that leave gradually via pitch angle scattering, not the substantial fraction that are born detrapped, nor those that ultimately makes it to the circular region around the origin and are trapped by the potential. We therefore consider $x \in [x_{a},x_{0}]$. The distribution of $x$ values that particles leave at is of course proportional to how many particles are leaving when the distribution is at a given value of $x$.
\begin{align*}&\int\limits_{x_{a}}^{x_{0}}\!p_{x}(x|x_{0})dx \propto\!\!\int\limits_{x_{a}}^{x_{0}}\!\frac{dn}{dx}\,dx = n_{0}\mu_{b}(x_{0}) \!-\! n(x_{a}|x_{0})\end{align*}
We chose to normalize it to unity as a proper PDF,
\begin{align*}
&\boxed{p_{x}(x|x_{0}) = \frac{dn/dx}{\hat{n}_{0}\!-\!n(x_{a}|x_{0})}, \quad x \in [x_{a},x_{0}]}
\end{align*}
For the basic scaling (S), we obtained the loss distribution on $x\in[x_{a},x_{0}]$ in Eq,~(\ref{Eq:px-S}), rewritten here for convenience,
\begin{align*}\boxed{p_{x}(x|x_{0})\sim\sum_{k=0}^{\infty}\frac{n_{0}^{\!(k)}\!(\mu_{b0}\!)}{\hat{n}_{0}\!-\!n(x_{a}|x_{0})}\frac{\beta_{k}\eta^{3}\!/2x}{x^{3}+\eta^{3}}\!\left(\!\frac{x^{3}}{x_{0}^{3}}\frac{x_{0}^{3}\!+\!\eta^{3}}{x^{3}\!+\!\eta^{3}}\!\right)^{\!\beta_{k}/6}}\end{align*}
Performing a change of variables $v = v_{th,a}x \rightarrow p_{v}(v) = p_{x}(x)|\frac{dx}{dv}| = p_{x}(v/v_{th,a})/v_{th,a}$ and defining $v_{0} \equiv v_{th,a}x_{0}$ and $v_{a} \equiv v_{th,a}x_{a}$ we have on $v\in[v_{a},v_{0}]$,
\begin{align*}\boxed{p_{v}(v|x_{0}) \!\sim\! \sum_{k=0}^{\infty}\frac{n_{0}^{\!(k)}\!(\mu_{b0}\!)}{\hat{n}_{0}\!-\!n(x_{a}|x_{0})}\frac{\beta_{k}\eta^{3}v_{th,a}^{3}/2v}{v^{3}\!+\!\eta^{3}v_{th,a}^{3}}\!\!\left(\!\frac{v^{3}}{v_{0}^{3}}\frac{v_{0}^{3}\!+\!\eta^{3}v_{th,a}^{3}}{v^{3}\!+\!\eta^{3}v_{th,a}^{3}}\!\right)^{\!\!\beta_{k}/6}\!}\end{align*}
Performing a change of variables $E \!=\! E_{th,a}x^{2} \rightarrow p_{E}(E) \!=\! p_{x}(x)|\frac{dx}{dE}| = \frac{1}{2E_{th,a}}\!\sqrt{\!\frac{E_{th,a}}{E}}p_{x}\!\!\left(\!\sqrt{\!\frac{E}{E_{th,a}}}\right)$ where $E_{th,a}\!\equiv\! k_{B}T_{a} = m_{a}v_{th,a}^{2}/2$ and defining $E_{0} \equiv E_{th,a}x_{0}^{2}$ we have on $E\in[\Phi_{a},E_{0}]$,
\begin{align*}\!\!\boxed{\!p_{E}(\!E|x_{0}\!)\!\sim\!\!\!\sum_{k=0}^{\infty}\frac{n_{0}^{\!(k)}\!(\mu_{b0}\!)}{\hat{n}_{0}\!-\!n(x_{a}|x_{0})}\frac{\beta_{k}\eta^{3}E_{th,a}^{3/2}/4E}{E^{3/2}\!+\!\eta^{3}E_{th,a}^{3/2}}\!\left(\!\frac{E^{3/2}\!}{E_{0}^{3/2}}\frac{E_{0}^{3/2}\!\!\!+\!\eta^{3}E_{th,a}^{3/2}}{E^{3/2}\!+\!\eta^{3}E_{th,a}^{3/2}}\!\right)^{\!\!\beta_{k}/6}}\!\end{align*}
Performing a change of variables $p_{t}(t) = p_{x}(x)|\frac{dx}{dt}|$ where $\frac{dx}{dt} = \!-Z_{\parallel}(x)/\tau_{0}^{i}x^{2}$ we have on $t\in[0,t_{a}]$
\begin{align*}\boxed{p_{t}(t|x_{0})\!\sim\frac{Z_{\parallel}(x)}{2x^{3}}\!\sum_{k=0}^{\infty}\frac{n_{0}^{\!(k)}\!(\mu_{b0}\!)}{\hat{n}_{0}\!-\!n(x_{a}|x_{0})}\frac{\beta_{k}\eta^{3}}{x^{3}\!+\!\eta^{3}}\!\!\left(\!\frac{x^{3}}{x_{0}^{3}}\frac{x_{0}^{3}\!+\!\eta^{3}}{x^{3}\!+\!\eta^{3}}\!\right)^{\!\beta_{k}/6}}\end{align*}
where we write $x \equiv x(t)$ for brevity and again,
\begin{align*}x(t) \!=\! \left[\!\left(x_{0}^{3}+\eta^{3}\right)\!e^{-3Z_{\parallel}^{e}t/\tau_{0}^{i}}\!-\!\eta^{3}\right]^{1/3}\!\!\!\!\!\!\!\!, \quad \eta\equiv\!\left(\!Z_{\parallel}^{i}/Z_{\parallel}^{e}\!\right)^{1/3}\end{align*}
and we've defined,
\begin{align*}&x(t_{a}) \equiv x_{a} \rightarrow t_{a} \equiv \frac{\tau_{0}^{i}}{3Z_{\parallel}^{e}}\ln\!\left(\!\frac{x_{0}^{3}+\eta^{3}}{x_{a}^{3}+\eta^{3}}\!\right)\end{align*}
In shorthand, we can write in full generality,
\begin{align*}
p_v(v|x_{0}) = \frac{p_{x}(v/v_{th,a}|x_{0})}{v_{th,a}}, \, v \in [v_{a},v_{0}]\\
p_E(E|x_{0}) = \frac{p_{x}(\sqrt{E/E_{th,a}}|x_{0})}{2\sqrt{EE_{th,a}}}, \, E \in [\Phi_{a},E_{0}]\\
p_{t}(t|x_{0}) = \frac{Z_{\parallel}(x)}{\tau_{0}^{i}x^{2}}p_{x}(x|x_{0}), \, t \in [0,t_{a}]
\end{align*}
where again $x \equiv x(t)$ for the PDF in time.

\section{Appendix I: Steady-State Model}
\label{App:I}
Our steady-state distribution can be derived by two methods. The first is generalizing our time-dependent Green's function solution for the distribution function in the long-time limit, which we will refer to as the time-dependent solution. The second is once again solving the Fokker-Planck PDE by separation of variables in steady-state ($\partial/\partial t = 0$) with an appropriately defined source-function $S_{h}$. We will refer to this as the source function solution.
\subsection{1. Time-Dependent Solution}
Our steady-state scenario corresponds to isotropic delta shells being constantly produced at $x = x_{0}$ for start times $t_{0}\in(-\infty,t\,]$ with rate $\nu_{h}$. It can therefore be obtained directly from our time-dependent Green's function solution (Eq.~(\ref{Eq:f(t,x,mu)})\!) by integrating from the distant past up to the present time,
\begin{align*}f_{eq}(x,\mu) = \!\!\!\int\limits_{-\infty}^{t}\!\!\!f(t-t_{0},x,\mu)\nu_{h}dt_{0}
\end{align*}
Let us once define the angular distribution $\hat{f}(t,\mu)\equiv x^{2}f(t,x,\mu)/\delta(x-x(t)\!)$ to express,
\begin{align*}f_{eq}(x,\mu) = \!\!\!\int\limits_{-\infty}^{t}\!\!\!\frac{\delta(x-x(t\!-\!t_{0})\!)}{x(t\!-\!t_{0})^{2}}\hat{f}(t-t_{0},\mu)\nu_{h}dt_{0}
\end{align*}
We will have to use the property of the delta function,
\begin{align*}\delta(z(x)) = \!\!\sum\limits_{\substack{i:z(x_{i})=0}}\frac{\delta(x-x_{i})}{|z'(x_{i})|}\end{align*}
so let us define $z(t_{0})\equiv x-x(t-t_{0})$. Then,
\begin{align*}z(t_{0}^{*}) \equiv 0 \rightarrow t_{0}^{*} \equiv t - \frac{\tau_{s}}{3}\ln\!\left(\frac{x_{0}^{3}\!+\!\eta^{3}}{x^{3}\!+\!\eta^{3}}\right)\end{align*}
and correspondingly,
\begin{align*}
z'(t_{0}^{*})  = -\frac{x(t-t_{0})^{3}\!+\!\eta^{3}}{\tau_{s}x(t\!-\!t_{0})^{2}}
\end{align*}
Then our delta function becomes, 
\begin{align*}\frac{\delta(x\!-\!x(t\!-\!t_{0})\!)}{x(t\!-\!t_{0})^{2}} = \tau_{s}\frac{\delta(t_{0}\!-\!t_{0}^{*})}{x(t\!-\!t_{0})^{3}\!+\!\eta^{3}}\end{align*}
And we can easily observe that for any function $q(t_{0})$,
\begin{align*}\int\limits_{-\infty}^{t}\!\!\delta(t_{0}\!-\!t_{0}^{*})q(t_{0})dt_{0} = q(t_{0}^{*})H(t\!-\!t_{0}^{*})\end{align*}
Observing finally that,
\begin{align*}H(t\!-\!t_{0}^{*}) = H(x_{0}\!-\!x),\quad x(t\!-\!t_{0}^{*}) = x\end{align*}
and defining the confined particle production rate $\dot{n}_{h}\equiv \hat{n}_{0}\nu_{h}$, we obtain the equilibrium distribution,
\begin{equation*}\boxed{\begin{aligned}f_{eq}(x,\mu) \!=\! \dot{n}_{h}\tau_{s}\frac{\!H\!(x_{0}\!-\!x)}{x^{3}\!+\!\eta^{3}}\frac{H(\mu_{b}\!-\!|\mu|)}{2\mu_{b}}\sum\limits_{k=0}^{\infty}\Biggl[\!\frac{2(-1)^{k}}{\pi(k\!+\!1/2)}\\
\cdot\cos(\sqrt{\lambda_{k}}\sin^{\!-1}\!\!\mu)\exp\!\!\left(\!\!-\!\!\!\int\limits_{x}^{x_{0}}\!\!\!\lambda_{k}\!\frac{Z_{\perp}\!(x)}{Z_{\parallel}\!(x)}\!\frac{dx}{2x}\!\right)\!\Biggr]\end{aligned}}\end{equation*}
Now let us see if the source function solution agrees.
\subsection{2. Source Function Solution}
\noindent Recall our advection-diffusion equation (Eq.~(\ref{Eq:advection-diffusion})\!),
\begin{align*}\!\!\!\!\!\frac{\partial f_{a}}{\partial t} \!+\! \frac{1}{x^{2}}\frac{\partial}{\partial x}\!\left(x^{2}v(x)f_{a}\right)\!=\!\nu(x)\frac{\partial}{\partial\mu}\!\!\left[\!(1\!-\!\mu^{2})\frac{\partial f_{a}}{\partial\mu}\!\right]\end{align*}
where $v(x)\!\equiv\! -Z_{\parallel}(x)/\tau_{0}^{i}x^{2}$, and $\nu(x)\!\equiv\! Z_{\perp}(x)/2\tau_{0}^{i}x^{3}$.
Then in steady-state, with a constant source outputting $\dot{n}_{h}$ confined particles per second isotropically at $x = x_{0}$,
\begin{equation*}\begin{aligned}&\frac{1}{x^{2}}\frac{\partial}{\partial x}(x^{2}v(x)f_{eq}) = \nu(x)\frac{\partial}{\partial\mu}\!\!\left[\!(1\!-\!\mu^{2})\frac{\partial f_{eq}}{\partial\mu}\!\right]\!+S_{h}\\
&S_{h} \!\equiv \dot{n}_{h}\frac{\delta(x\!-\!x_{0})}{x^{2}}\frac{H(\mu_{b}\!-\!|\mu|)}{2\mu_{b}}\rightarrow\! \int\limits_{\!x_{a}}^{\infty}\!\!\!\int\limits_{\,-\mu_{b}(x)}^{\mu_{b}(x)}\!\!\!\!\!\!S_{h}d\mu dx = \dot{n}_{h}
\end{aligned}\end{equation*}
Then let us proceed by separation of variables $f_{eq}(x,\mu) = g(x)h(\mu|\mu_{b}(x)\!)$. For $x \neq x_{0}$,
\begin{align*}
&\frac{1}{\nu(x)x^{2}g}\frac{\partial}{\partial x}(x^{2}v(x)g) \!=\! \frac{1}{h}\frac{\partial}{\partial\mu}\!\!\left[\!(1\!-\!\mu^{2})\frac{\partial h}{\partial\mu}\!\right] \!=\! -\lambda(\mu_{b})\end{align*}
We separate this into the following two equations,
\begin{align*}&\frac{\partial}{\partial x}(x^{2}v(x)g) = -\lambda(\mu_{b})\nu(x)x^{2}g\\
&\frac{\partial}{\partial\mu}\!\!\left[\!(1\!-\!\mu^{2})\frac{\partial h}{\partial\mu}\!\right] = -\lambda(\mu_{b})h\end{align*}
Our equation in $g(x)$ is a separable ODE. Let us define $G(x)\equiv x^{2}v(x)g$. Then,
\begin{align*}\frac{dG}{dx} = -\lambda\frac{\nu(x)}{v(x)}G \,\rightarrow\, G(x) = G(x_{0})\exp\!\!\left(\!\!-\!\!\int\limits_{x}^{x_{0}}\!\!\lambda\frac{\nu(x)}{v(x)}dx\!\!\right)\end{align*}
which after transforming back to $g(x)$ becomes,
\begin{align*}
&g(x) = g(x_{0})\frac{x_{0}^{3}\!+\!\eta^{3}}{x^{3}\!+\!\eta^{3}}\exp\!\!\left(\!\!-\!\!\int\limits_{x}^{x_{0}}\!\!\lambda(\mu_{b})\frac{Z_{\perp}(x)}{Z_{\parallel}(x)}\frac{dx}{2x}\!\!\right)\end{align*}
where we have used the definitions of $\nu(x)$ and $v(x)$ in Eq.~(\ref{Eq:advection-diffusion}).
We can figure out what $g(x_{0})\equiv g(x_{0}^{-})$ is by integrating over our source function in our original steady-state equation,
\begin{align*}&\int\limits_{x_{0}-\epsilon}^{x_{0}+\epsilon}\!\!\!\frac{\partial}{\partial x}(x^{2}v(x)f_{eq})dx = \!\!\!\!\int\limits_{x_{0}-\epsilon}^{x_{0}+\epsilon}\!\!\!\!\nu(x)\mathcal{L}(f_{eq})x^{2}dx+\!\!\!\!\int\limits_{x_{0}-\epsilon}^{x_{0}+\epsilon}\!\!\!\!S_{h}x^{2}dx\end{align*}
In the limit as $\epsilon\rightarrow 0$,
\begin{align*}
&\left[x^{2}v(x)g\right|_{x_{0}-\epsilon}^{x_{0}+\epsilon} = \frac{\dot{n}_{h}}{2\mu_{b}(x_{0})} \rightarrow g(x_{0}^{-}) = \frac{\dot{n}_{h}\tau_{0}^{i}}{2\mu_{b}(x_{0})Z_{\parallel}^{e}}\end{align*}
Since this only holds for $x < x_{0}$, we have in total,
\begin{align*}&\boxed{g(x) = \frac{\dot{n}_{h}\tau_{0}^{i}}{2\mu_{b}(x_{0})Z_{\parallel}^{e}}\frac{H(x\!-\!x_{0})}{x^{3}\!+\!\eta^{3}}\exp\!\!\left(\!\!-\!\!\int\limits_{x}^{x_{0}}\!\!\lambda_{k}\frac{Z_{\perp}(x)}{Z_{\parallel}(x)}\frac{dx}{2x}\!\right)}\end{align*}
where $H(x)$ is the Heaviside step function. Our equation in $h(\mu|\mu_{b})$ is the same eigenvalue problem of the Legendre Operator with homogeneous boundary conditions presented in the main text (Eq.~(\ref{Eq:Legendre})\!). In the GO approximation,
\begin{align*}h(\mu|\mu_{b})\sim a(\mu_{b})\cos(\sqrt{\lambda}\sin^{\!-1}\!\mu)+b(\mu_{b})\sin(\sqrt{\lambda}\sin^{\!-1}\!\mu)\end{align*}
Our boundary conditions $h(\pm\mu_{b}|\mu_{b}) = 0$ allow two solutions,
\begin{align*}&a\!\neq\!0,b\!=\!0 \rightarrow\! \lambda_{k}^{(i)} \!\!=\! \frac{\!(k\!+\!1/2)^{2}\pi^{2}\!\!\!}{\!(\sin^{\!-1}\!\!\mu_{b})^{2}\!},\,\!h_{k}\!\sim\! \cos\!\left(\!\!\sqrt{\!\lambda_{k}^{(i)}}\!\sin^{\!-1}\!\!\mu\!\!\right)\nonumber\\
&a\!=\!0,b\!\neq\!0 \!\rightarrow\!\lambda_{k}^{(ii)} \!\!=\! \frac{\!k^{2}\pi^{2}\!\!}{\!(\sin^{\!-1}\!\!\mu_{b})^{2}\!},\,\!h_{k}\!\sim\! \sin\!\!\left(\!\!\sqrt{\!\lambda_{k}^{(ii)}}\!\sin^{\!-1}\!\!\mu\!\!\right)\!\!\end{align*}
Then our general equilibrium solution is,
\begin{align*}&f_{eq}(x,\mu) \!=\! \frac{\dot{n}_{h}\tau_{s}}{2\mu_{b}(x_{0})}\frac{\!H\!(x_{0}\!-\!x)}{x^{3}\!+\!\eta^{3}}\!\sum\limits_{k=0}^{\infty}\biggl[\!a_{k}h_{k}^{\!(i)}\!e^{\!-\!\!\int\limits_{x}^{x_{0}}\!\!\lambda_{k}^{\!(i)}\!\!\frac{Z_{\perp}}{Z_{\parallel}}\!\frac{dx}{2x}}\!\!\!\!\!\!\!\\
&\quad\quad\quad\quad\quad\quad\quad\quad\quad\quad\quad\quad+\!b_{k}h_{k}^{\!(ii)}\!e^{\!-\!\!\int\limits_{x}^{x_{0}}\!\!\lambda_{k}^{\!(ii)}\!\!\frac{Z_{\perp}}{Z_{\parallel}}\!\frac{dx}{2x}}\!\biggr]\end{align*}
Let us impose initial condition,
\begin{align*}\lim_{x\rightarrow x_{0}^{-}}\!\!f(x,\mu) \!=\!\frac{\dot{n}_{h}\tau_{s}}{x_{0}^{3}\!+\!\eta^{3}}\frac{H(\mu_{b}(x_{0})\!-\!|\mu|)}{2\mu_{b}(x_{0})}
\end{align*}
such that the angular distribution does not affect our normalization due to the source. However since our $a_{k}$ and $b_{k}$ are functions of $\mu_{b}$, we will need more information to completely define them. If we further impose that in the absence of pitch-angle scattering, particles are conserved and the distribution remains uniform, we have,
\begin{align*}\lim_{\nu\rightarrow0}f(t,x,\mu)\!=\!\dot{n}_{h}\tau_{s}\frac{\!H\!(x_{0}\!-\!x)}{x^{3}\!+\!\eta^{3}}\frac{H(\mu_{b}\!-\!|\mu|)}{2\mu_{b}}\end{align*}
Then, defining $z \!\equiv\! \sin^{\!-1}\!\!\mu/\sin^{\!-1}\!\!\mu_{b}$ and renormalizing $a_{k},b_{k} \rightarrow \hat{a}_{k},\hat{b}_{k}$ by $\dot{n}_{h}\tau_{s}H(x_{0}\!-\!x)/2\mu_{b}(x^{3}\!+\!\eta^{3})$ we can match our coefficients via,
\begin{align*}\sum_{k=0}^{\infty}\left[\hat{a}_{k}\cos\!\left(\!(k\!+\!1/2)\pi z\right)+\hat{b}_{k}\sin\!\left(k\pi z\right)\right] = \phi(z)\end{align*}
where since we do not have a constant term available for matching on the LHS, we periodically extend the RHS beyond our interval to be the even square wave $\phi(z)$ with average value zero,
\begin{align*}
\phi(z)\!\equiv\! \begin{cases}1, \,z\!\in\![-1,1]\quad \phi(z\!+\!4n) = \phi(z)\\
\!-1, \,z\!\in\!(1,3]\quad\quad\quad\forall\,\,n\in\mathbb{Z}\end{cases}\!\!\!\!\!\!\!\!\!\!\!\!\!\!\!\!\!\!\!\!\!\!\!\!\!\!\!\!\!\!\!\!\!\!\!\!\!\!\!\!\!\!\!\!\!\!\!\!\!\!\!\!\!\!,\quad\quad\quad\quad\quad\quad\quad
\end{align*}
Since our $\phi(z)$ is now even, we can immediately discard all the $b_{k}$. The $a_{k}$ can be found via Fourier's trick,
\begin{align*}\label{Eq:coefficients}a_{k} =\! \frac{\dot{n}_{h}\tau_{s}}{2\mu_{b}}\frac{H(x_{0}\!-\!x)}{x^{3}\!+\!\eta^{3}}\!\int_{0}^{2}\!\!\!\phi(z)\cos\left(\!(k\!+\!1/2)\pi z\right)dz,\,\,\,\, b_{k} = 0\end{align*}
Then we obtain for $\lambda_{k}\equiv\lambda_{k}^{(i)}$,
\begin{equation*}\boxed{\begin{aligned}f_{eq}(x,\mu) \!=\! \dot{n}_{h}\tau_{s}\frac{\!H\!(x_{0}\!-\!x)}{x^{3}\!+\!\eta^{3}}\frac{H(\mu_{b}\!-\!|\mu|)}{2\mu_{b}}\sum\limits_{k=0}^{\infty}\Biggl[\!\frac{2(-1)^{k}}{\pi(k\!+\!1/2)}\\
\cdot\cos(\sqrt{\lambda_{k}}\sin^{\!-1}\!\!\mu)\exp\!\!\left(\!\!-\!\!\!\int\limits_{x}^{x_{0}}\!\!\!\lambda_{k}\!\frac{Z_{\perp}\!(x)}{Z_{\parallel}\!(x)}\!\frac{dx}{2x}\!\right)\!\Biggr]\end{aligned}}\end{equation*}
which is precisely the same result obtained by the time-dependent solution method. The time-dependent solution is more powerful, however, since $\nu_{h}$ can be amended to depend on time.
\subsection{3. Particle Confinement Time}
Now that we have confirmed our expression for the equilibrium distribution, our confinement time can be estimated via,
\begin{align*}\tau_{c}\!\sim\!\frac{\int \!f_{a}d\Vec{v}}{-\int\!\!\left(\!\frac{\partial f_{a}}{\partial t}\!\right)_{\!coll}\!\!\!\!\!\!\!\!d\Vec{v}}\sim\frac{\int \!f_{eq}d\Vec{v}}{\int\!S_{h}d\Vec{v}}\sim\frac{1}{\dot{n}_{h}}\! \int\limits_{\!x_{a}}^{\infty}\!\!\!\int\limits_{\,-\mu_{b}(x)}^{\mu_{b}(x)}\!\!\!\!\!\!f_{eq}\,x^{2}d\mu dx
\end{align*}
Substituting our expression for $f_{eq}(x,
\mu)$,
\begin{align*}
&\tau_{c}\!\sim\tau_{s}\sum\limits_{k=0}^{\infty}\Biggl[\frac{2(-1)^{k}}{\pi(k\!+\!1/2)}\!\int\limits_{\!x_{a}}^{x_{0}}\!\!\frac{x^{2}}{x^{3}\!+\!\eta^{3}}\frac{1}{2\mu_{b}}\\
&\cdot\exp\!\!\left(\!\!-\!\!\!\int\limits_{x}^{x_{0}}\!\!\!\lambda_{k}\!\frac{Z_{\perp}\!(x)}{Z_{\parallel}\!(x)}\!\frac{dx}{2x}\!\right)\!\!\!\!\!\int\limits_{\,-\mu_{b}(x)}^{\mu_{b}(x)}\!\!\!\!\!\!\!\cos(\sqrt{\lambda_{k}}\sin^{\!-1}\!\!\mu)d\mu dx\Biggr]\end{align*}
Performing the integral in pitch-angle we have,
\begin{align*}
&\tau_{c}\!\sim\tau_{s}\sum\limits_{k=0}^{\infty}\Biggl[\frac{2(-1)^{k}}{\pi(k\!+\!1/2)}\!\int\limits_{\!x_{a}}^{x_{0}}\!\!\frac{x^{2}}{x^{3}\!+\!\eta^{3}}\frac{1}{2\mu_{b}}\\
&\cdot\exp\!\!\left(\!\!-\!\!\!\int\limits_{x}^{x_{0}}\!\!\!\lambda_{k}\!\frac{Z_{\perp}\!(x)}{Z_{\parallel}\!(x)}\!\frac{dx}{2x}\!\right)\!\!\!\left(\!\frac{2\pi(k\!+\!1/2)\sin^{\!-1}\!\!\mu_{b}\sqrt{1\!-\!\mu_{b}^{2}}(-1)^{k}}{\pi^{2}(k\!+\!1/2)^{2}\!-\!(\sin^{\!-1}\!\!\mu_{b})^{2}}\!\right)\!\!dx\!\Biggr]\end{align*}
In general then, our confinement time becomes,
\begin{align*}\tau_{c}\!\sim\tau_{s}\!\sum\limits_{k=0}^{\infty}\int\limits_{\!x_{a}}^{x_{0}}\!\!\frac{x^{2}}{x^{3}\!\!+\!\eta^{3}}\exp\!\!\left(\!\!-\!\!\!\int\limits_{x}^{x_{0}}\!\!\!\lambda_{k}\!\frac{Z_{\perp}\!(x)}{Z_{\parallel}\!(x)}\!\frac{dx}{2x}\!\right)\\
\cdot\frac{2\sin^{\!-1}\!\!\mu_{b}\sqrt{1\!-\!\mu_{b}^{2}}/\mu_{b}}{\pi^{2}(k\!+\!1/2)^{2}\!-\!(\sin^{\!-1}\!\!\mu_{b})^{2}}dx\end{align*}
We can rewrite this using Eqs.~(\ref{Eq:n(t|x0)},\ref{Eq:n(x|x0)}), \begin{align*}
&\tau_{c}\!\sim\tau_{s}\!\int\limits_{\!x_{a}}^{x_{0}}\!\!\frac{x^{2}}{x^{3}\!\!+\!\eta^{3}}\sum\limits_{k=0}^{\infty}\frac{n_{0}^{(k)}}{\hat{n}_{0}}\exp\!\!\left(\!\!-\!\!\int\limits_{x}^{x_{0}}\!\!\lambda_{k}\frac{Z_{\perp}(x)}{Z_{\parallel}(x)}\frac{dx}{2x}\!\right)\!dx\\
&\tau_{c}\!\sim\tau_{s}\!\int\limits_{\!x_{a}}^{x_{0}}\!\!\frac{x^{2}}{x^{3}\!\!+\!\eta^{3}}\frac{n(x|x_{0})}{\hat{n}_{0}}dx\end{align*}
And finally, integrating by parts,
\begin{align*}
&\tau_{c}\!\sim\tau_{s}\!\!\left[\!\left[\frac{\ln(x^{3}\!\!+\!\eta^{3})}{3}\frac{n(x|x_{0})}{\hat{n}_{0}}\right|_{x_{a}}^{x_{0}}\!\!\!\!-\!\!\int\limits_{x_{a}}^{x_{0}}\frac{\ln(x^{3}\!\!+\!\eta^{3})}{3}\frac{dn/dx}{\hat{n}_{0}}dx\!\right]\end{align*}
We can recognize the loss velocity PDF $p_{x}(x|x_{0})$ (Eq.~(\ref{Eq:p(x|x0)})\!),
\begin{align*}
&\tau_{c}\!\sim\tau_{s}\Biggl[\!\frac{\ln(x_{0}^{3}\!\!+\!\eta^{3})}{3}\!-\!\frac{n(x_{a}|n_{0})}{\hat{n}_{0}}\frac{\ln(x_{a}^{3}\!\!+\!\eta^{3})}{3}\\
&\quad-\frac{\hat{n}_{0}\!-\!n(x_{a}|x_{0})}{\hat{n}_{0}}\!\!\int\limits_{x_{a}}^{x_{0}}\frac{\ln(x^{3}\!\!+\!\eta^{3})}{3}p_{x}(x|x_{0})dx\!\Biggr]\end{align*}
Using the definition of our expectation value (Eq.~(\ref{Eq:expectations})\!),
\begin{align*}
&\tau_{c}\!\sim\!\frac{\tau_{s}}{3}\ln(x_{0}^{3}\!\!+\!\eta^{3})\!-\!\frac{n(x_{a}|n_{0})}{\hat{n}_{0}}\frac{\tau_{0}^{i}}{3Z_{\parallel}^{e}}\ln(x_{a}^{3}\!\!+\!\eta^{3})\\
&\quad-\frac{\hat{n}_{0}\!-\!n(x_{a}|x_{0})}{\hat{n}_{0}}\left\langle\!\frac{\tau_{s}}{3}\ln(x^{3}\!+\!\eta^{3})\!\!\right\rangle^{\!(ii)}\\
&\tau_{c}\!\sim\!\frac{\hat{n}_{0}\!-\!n(x_{a}|x_{0})}{\hat{n}_{0}}\!\left\langle\!\frac{\tau_{s}}{3}\!\ln\!\left(\!\frac{x_{0}^{3}\!+\!\eta^{3}}{x^{3}\!+\!\eta^{3}}\!\right)\!\!\right\rangle^{\!\!(ii)}\!\!\!\!\!\!\!+\!\frac{n(x_{a}|n_{0})}{\hat{n}_{0}}\frac{\tau_{s}}{3}\!\ln\!\left(\!\frac{x_{0}^{3}\!+\!\eta^{3}}{x_{a}^{3}\!\!+\!\eta^{3}}\!\right)\end{align*}
And finally recognizing $\langle t\rangle^{\!(ii)}$ (Eq.~(\ref{Eq:<t>ii})\!) and $t_{a}$ (Eq.~(\ref{Eq:ta})\!),
\begin{align*}
&\tau_{c}\sim \frac{\hat{n}_{0}\!-\!n(x_{a}|x_{0})}{\hat{n}_{0}}\langle t\rangle^{\!(ii)}\!\!+\!\frac{n(x_{a}|x_{0})}{\hat{n}_{0}}t_{a}\!\sim\!\frac{F_{l}^{(ii)}\!\langle t\rangle^{\!(ii)}\!+\!\cancel{F_{l}^{(iii)}t_{a}}}{F_{l}^{(ii)}\!\!+\!\cancel{F_{l}^{(iii)}}}\end{align*}
where we can understand that this calculation misinterprets those particles that make it to $x = x_{a}$ as being ``lost" as $t \!=\! t_{a}$ to eliminate our $F_{l}^{(iii)}$ contributions. Then, \begin{align*}&\boxed{\tau_{c}\sim \langle t\rangle^{\!(ii)}}\end{align*}
Interestingly enough, we recover the same result as the mean loss time even though we are here concerned with a steady-state source. This is because our source is equivalent to a cascade of density shells produced with frequency $\nu_{h}\equiv \dot{n}_{h}/\hat{n}_{0}$, each with mean confinement time $\langle t\rangle^{\!(ii)}$, so their ensemble also has confinement time $\langle t\rangle^{\!(ii)}$.
\subsection{4. Energy Confinement Time}
The energy confinement time is similarly given by,
\begin{align*}&\tau_{E}\!\sim\tau_{s}\!\!\int\limits_{\!x_{a}}^{x_{0}}\!\!\frac{x^{4}/x_{0}^{2}}{x^{3}\!+\eta^{3}}\frac{n(x|x_{0})}{\hat{n}_{0}}dx\end{align*}
Let us define,
\begin{align*}c(x)\equiv\!\int\limits_{x}^{x_{0}}\frac{x^{4}}{x^{3}\!+\eta^{3}}dx\end{align*}
Then we can once again integrate by parts to obtain,
\begin{align*}&\tau_{E}\sim \frac{\tau_{s}}{x_{0}^{2}}\frac{F_{l}^{(ii)}\langle c(x)\rangle^{\!(ii)}\!+\cancel{F_{l}^{(iii)}c(x_{a})}}{F_{l}^{(ii)}\!+\cancel{F_{l}^{(iii)}}}\end{align*}
which yields,
\begin{align*}&\boxed{\tau_{E}\sim \tau_{s}\frac{\langle c(x)\rangle^{\!(ii)}}{x_{0}^{2}}}\end{align*}
Correspondingly, we recognize  $\tau_{E} \sim E_{int}/P_{\text{fus}}^{\alpha}$ to find,
\begin{align*}\boxed{P_{\text{fus}}^{\alpha} \sim \dot{n}_{h}E_{th,a}x_{0}^{2},\,\,E_{int} \sim \dot{n}_{h}\tau_{s}E_{th,a}\langle c(x)\rangle^{\!(ii)}}\end{align*}
where $P_{\text{fus}}^{\alpha}$ is the power produced by the alpha particle source and $E_{int}$ is the confined internal energy between $x\in[x_{a},x_{0}]$ in steady-state.

\section{Appendix J: Sheared Rotation Loss}
\label{App:J}
In the main text we assumed that all particles of our fast ion species `$a$' experience the same potential $\Phi_{a}$. In a centrifugal magnetic mirror, this potential is a combination of centrifugal and ambipolar electric potentials.
\begin{align*}\Phi_{a} = \Phi_{a}^{\omega} + \Phi_{a}^{\text{ambi}}\end{align*}
The ambipolar electric potential is the result of differences in charge mobility between ions and electrons, and produces an outward electric field. If we neglect corrections from the increased rate of electron scattering, which should be relatively small in the high-Mach-number limit, then this process continues roughly until the bulk ion and electron potential-to-temperature ratios equilibrate \cite{Munirov,Schwartz}. For $T_{e}\sim T_{i}$,
\begin{align*}\Phi_{i}^{\omega} + \Phi_{i}^{\text{ambi}} = \Phi_{e}^{\omega} + \Phi_{e}^{\text{ambi}}\end{align*}
Since $m_{e}/m_{i} << 1$ we can neglect $\Phi_{e}^{\omega}$ and since $q_{e} = -q_{i}$, their ambipolar potentials are equal and opposite,
\begin{align*}\frac{q_{i}}{q_{a}}\Phi_{a}^{\text{ambi}} \sim \Phi_{i}^{\text{ambi}} \sim -\frac{1}{2}\Phi_{i}^{\omega} \sim -\frac{1}{2}\frac{m_{i}}{m_{a}}\Phi_{a}^{\omega}\end{align*}
Then the total potential is proportional to the centrifugal potential,
\begin{align*}\Phi_{a} \sim \left(\!1-\frac{1}{2}\frac{m_{i}q_{a}}{m_{a}q_{i}}\!\right)\Phi_{a}^{\omega}\end{align*}
The centrifugal potential depends on $r$, the radial position within the reactor, as follows,
\begin{align*}\Phi_{a}^{\omega}(r) = \frac{1}{2}m_{a}\omega(r)^{2}r^{2}\end{align*}
Typically the plasma in a centrifugal mirror will be approximately annular and concentrated near some radius $R_{\ast}\sim R_{0}/2$ where $R_{0}$ is the reactor radius, and experience a more or less uniform potential $\Phi_{a}(R_{\ast})$. It is easy to see from our expression, however, that the centrifugal potential drops to zero at $r = 0$, and also at $r = R_{0}$ if a no-slip condition $\omega(R_{0}) = 0$ is imposed at the walls due to charge exchange and neutral drag. Therefore, it is conceivable that particles could diffuse a distance $R_{0}/2$ across the magnetic field lines to either of these regions with lower confining potential, and exit the reactor more rapidly than our treatment in the main text suggests \cite{White}. We will argue that the timescale of this cross-field diffusion process is much too long to seriously impact our results, using elementary transport theory.\\
\indent Cross-field diffusion is characterized by the length scale $\Delta l \sim \rho_{L}$ where $\rho_{L}$ is the Larmor radius, and timescale $\Delta t \sim \tau_{\text{coll}}$ where $\tau_{\text{coll}}$ is the collision time. In this case, the collision time is precisely the timescale for pitch-angle diffusion, which we have found in the main text to be $\tau_{\perp}^{i}\sim \tau_{0}^{i}/Z_{\perp}^{i}$, by the definition of $Z_{\perp}^{i}$. We also know the timescale of our problem to be $t_{a}$ (Eq.~(\ref{Eq:ta})\!) which for reasonable parameter values is an $O(1)$ multiple of the slowing down time $\tau_{s} \sim \tau_{0}^{i}/Z_{\parallel}^{e}$. From the theory of Braginskii transport across the magnetic field then, we have that the timescale of radial diffusion $\tau_{RD}$ scales as,
\begin{align*}\tau_{\text{RD}} \sim \frac{(R_{0}/2)^{2}}{\rho_{L}^{2}}\tau_{\perp}^{i} \sim  \frac{R_{0}^{2}}{4\rho_{L}^{2}}\frac{Z_{\parallel}^{e}}{Z_{\perp}^{i}}\tau_{s} >> \tau_{s}\end{align*}
which far exceeds the timescale of our problem when there is a sufficient combination of the reactor radius being much greater than the Larmor radius, $R_{0} >> \rho_{L}$ and drag dominating over pitch-angle scattering, $Z_{\parallel}^{e} >> Z_{\perp}^{i}$. Both of these conditions should hold in a centrifugal mirror designed to confine a fusion plasma.\\
\indent Sheared rotation loss and other non-uniform potential effects can also be explicitly considered through careful extension of the Fokker-Planck equation (Eq.~(\ref{Eq:advection-diffusion})\!), distribution function (Eq.(\ref{Eq:f(t,x,mu)})\!), and loss boundary (Eq.~(\ref{Eq:hyperbola})\!) to depend on $r$, the radial position of particles within the reactor.
\vspace{-2pt}
\subsection*{Projected Confining Potential}
\vspace{5pt}
In Table 1 of Schwartz et al.~\cite{Schwartz}, we see projected parameters for a DT centrifugal mirror fusion reactor based on CMFX. They consider a Mach number of $M = 4.5$ and mirror ratio $R = 6$ for a 50/50 DT plasma. Based on their definition of the Mach number in Eq.~(2.2),
\begin{align*}
M^{2}\equiv \frac{m_{i}\omega^{2} R_{\text{max}}^{2}}{T_{e}}
\end{align*}
For a 50/50 DT plasma, $m_{i} \sim 2.5$ and $q_{i} = 1$, and our alpha particles have $m_{a} = 4$ and $q_{a} = 2$.
\begin{align*}
&\Phi_{a}^{\omega} = \frac{8}{5}\Phi_{i}^{\omega},\quad\Phi_{a}^{\text{ambi}} = 2\Phi_{i}^{\text{ambi}}\end{align*}
Based on our previous discussion we can relate the centrifugal and ambipolar potentials,
\begin{align*}
&\Phi_{i}^{\text{ambi}} \sim -\frac{1}{2}\Phi_{i}^{\omega}\end{align*}
Then our total potential is given by,
\begin{align*}
&\Phi_{a} = \Phi_{a}^{\omega} + \Phi_{a}^{\text{ambi}} = \frac{3}{5}\Phi_{i}^{\omega}\end{align*}
Recalling our definition $x_{a}\equiv\sqrt{\Phi_{a}/T_{a}}$ we have,
\begin{align*}
&x_{a}^{2} = \frac{\Phi_{a}}{T_{a}} = \frac{3\Phi_{i}^{\omega}}{5T_{a}} = \frac{3T_{e}}{10T_{a}}\frac{m_{i}\omega^{2}R_{\text{max}}^{2}}{T_{e}} = \frac{3T_{e}}{10T_{a}}M^{2}\end{align*}
Using our projected species temperatures from the DT scenario in Table \hyperref[Tab:transport]{1},
\begin{align*}
&x_{a}^{2} = \frac{3}{10}\!\left(\!\frac{15\text{keV}}{3500\text{keV}}\!\right)\!(4.5)^{2} = 0.026036 \rightarrow x_{a} \sim 0.16\end{align*}
Thus, our projected parameters are $(x_{a},R) \sim (0.16,6)$.

\include{texfiles/Appendix-K}
\end{document}